\documentclass[acmtog, authorversion]{acmart}
\acmSubmissionID{1022}

\usepackage{wrapfig}
\usepackage[ruled]{algorithm2e} 

\usepackage{xspace}
\renewcommand\UrlFont{\ttfamily\color{blue}}

\makeatletter
\DeclareRobustCommand\onedot{\futurelet\@let@token\@onedot}
\def\@onedot{\ifx\@let@token.\else.\null\fi\xspace}

\makeatother
\SetAlFnt{\small}
\SetAlCapFnt{\small}
\SetAlCapNameFnt{\small}
\SetAlCapHSkip{0pt}

\newcommand{\bgcolor}[2]{\setlength{\fboxsep}{0pt}\colorbox{#1}{\strut #2}}
\definecolor{cbad}{HTML}{EAC2C2}
\definecolor{cmeh}{HTML}{FFE1C9}
\definecolor{cok}{HTML}{FDF3D0}
\definecolor{cgood}{HTML}{B3D09F}
\newcommand{\coloredCell}[3]{\definecolor{mycolor}{HTML}{#2}\colorbox{mycolor}{#3}}

\newcommand{\cellCD}[3][24pt]{\coloredCell{#1}{#2}{#3}}

\newcommand{\BGcolor}[3][HTML]{\definecolor{mycolor}{HTML}{#2}\bgcolor{mycolor}{#3}}

\copyrightyear{2025}
\acmYear{2025}
\setcopyright{cc}
\setcctype{by}
\acmConference[SIGGRAPH Conference Papers '25]{Special Interest Group on Computer Graphics and Interactive Techniques Conference Conference Papers }{August 10--14, 2025}{Vancouver, BC, Canada}
\acmBooktitle{Special Interest Group on Computer Graphics and Interactive Techniques Conference Conference Papers (SIGGRAPH Conference Papers '25), August 10--14, 2025, Vancouver, BC, Canada}
\acmDOI{10.1145/3721238.3730727}
\acmISBN{979-8-4007-1540-2/2025/08}

\begin{document}
\title{Splat and Replace: 3D Reconstruction with Repetitive Elements}

\author{Nicolás Violante}
\orcid{0009-0000-3169-8075}
\affiliation{%
 \institution{Inria \& Université Côte d’Azur, France and Adobe}
 \country{USA}}
\email{nicolas.violante@inria.fr}

\author{Andreas Meuleman}
\orcid{0000-0002-9899-6365}
\affiliation{%
 \institution{Inria \& Université Côte d’Azur}
 \country{France}
 }
\email{andreas.meuleman@inria.fr}

\author{Alban Gauthier}
\orcid{0000-0002-3710-0879}
\affiliation{%
 \institution{Inria \& Université Côte d’Azur}
 \country{France}
 }
\email{alban.gauthier@inria.fr}

\author{Fredo Durand}
\orcid{0000-0001-9919-069X}
\affiliation{%
 \institution{MIT}
 \country{USA}
 }
\email{fredo@mit.edu}

\author{Thibault Groueix}
\orcid{0000-0002-7984-8252}
\affiliation{%
 \institution{Adobe}
 \country{USA}
 }
\email{groueix@adobe.com}

\author{George Drettakis}
\orcid{0000-0002-9254-4819}
\affiliation{%
 \institution{Inria \& Université Côte d’Azur}
 \country{France}
 }
\email{george.drettakis@inria.fr}

\renewcommand{\shortauthors}{N. Violante et al.}

\begin{abstract}
We leverage repetitive elements in 3D scenes to improve novel view synthesis. Neural Radiance Fields (NeRF) and 3D Gaussian Splatting (3DGS) have greatly improved novel view synthesis but renderings of unseen and occluded parts remain low-quality if the training views are not exhaustive enough.
Our key observation is that our environment is often full of repetitive elements. 
We propose to leverage those repetitions to improve the reconstruction of low-quality parts of the scene due to poor coverage and occlusions. 
We propose a method that segments each repeated instance in a 3DGS reconstruction, registers them together, and allows information to be shared among instances. Our method improves the geometry while also accounting for appearance variations across instances.
We demonstrate our method on a variety of synthetic and real scenes with typical repetitive elements, leading to a substantial improvement in the quality of novel view synthesis.
\end{abstract}



\begin{teaserfigure}
    \includegraphics[width=\textwidth]{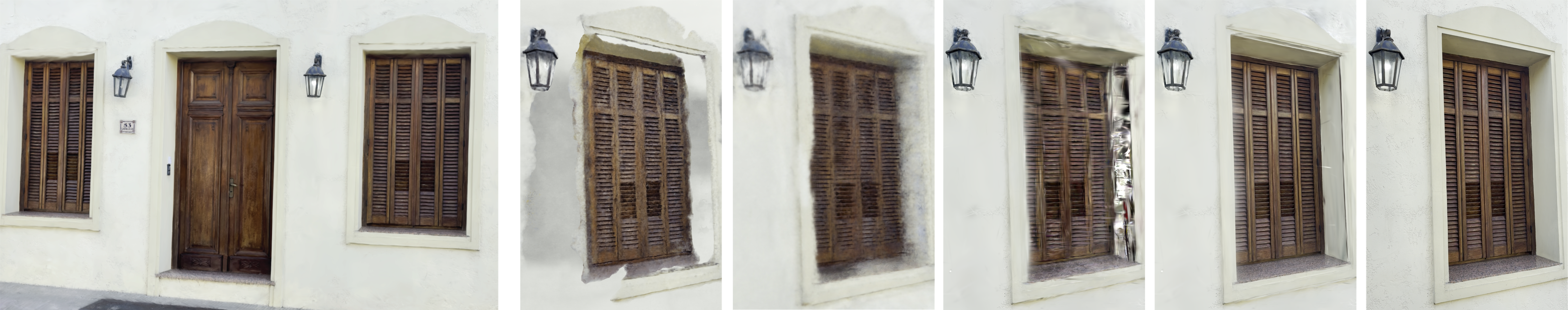}
    \makebox[\textwidth][l]{\hspace{182pt}Nerfbusters\hspace{25pt}Bayes' Rays\hspace{35pt}3DGS*\hspace{47pt}Ours\hspace{33pt}Ground Truth}
  \caption{Our method improves 3D reconstruction in  unseen views, by leveraging the multi-view information contained in repetitive elements (the two windows in this example). From left to right, we compare Nerfbusters~\cite{warburg_nerfbusters_2023}, Bayes Rays~\cite{goli_BayesRays_2023}, an improved version of 3D Gaussian Splatting~\cite{kerbl_3Dgaussians_2023} described in Section~\ref{sec:results}, our method, and the ground truth novel test view of a real scene.
  }
  \label{fig:teaser}
\end{teaserfigure}

\maketitle

\section{Introduction}

Capturing real scenes from photos and allowing 3D navigation has become widely accessible thanks to progress in novel view synthesis using Neural Radiance Fields (NeRF)~\cite{mildenhall_nerf_2020,barron2022mipnerf360,mueller2022instant} and  3D Gaussian Splatting (3DGS)~\cite{kerbl_3Dgaussians_2023}.
However, good quality novel view synthesis  requires careful and exhaustive capture of multiple views of a scene, and the visual quality of renderings from unseen or occluded regions is typically very low. 
We focus on scenes that exhibit \emph{repetitions}:
repetitions can be found everywhere, from urban scenes (pillars), building facades (windows) and furniture (tables and chairs) to decorative elements. 
We present a method that exploits the information provided by a set of \emph{instances} of the same object in a multi-view capture to improve the visual quality of novel views.

Repetitions have received little attention in novel view synthesis. Symmetry has been used to improve 3D reconstruction~\cite{mitra_symmetry_2013}, and repetitions have been used for single-image reconstruction~\cite{cheng_structure_2023}. Rodriguez et al.~\shortcite{rodriguez_exploiting_2018} presented a  solution to improve novel view synthesis for the restricted case of instances in a plane.
In addition, previous methods do not handle the inevitable differences in appearance between instances. In contrast, we aim for a more general solution where repetitive elements can be any 3D object, and appearance variations are handled. 
This involves three important challenges. First, we need high-quality 3D segmentation of the instances.  
Second, we need high-quality 3D registration between instances.
Finally, each instance usually presents differences in illumination and appearance that need to be accounted for.

Given a base 3DGS reconstruction, our goal is to segment repetitive objects in the scene, i.e. \emph{instances}, and fuse them together into a common and improved \textit{shared representation}. 

To address the above challenges, we first exploit learning-based 2D masks, and a single user click per instance, and then segment each instance in 3D. For this we follow previous work that uses contrastive learning~\cite{gu_egolifter_2024, kim_garfield_2024, choi_clickGaussian_2024, cen_segment_2024, ying_omniseg3d_2024, choi_clickgaussian_2025, ye_gaussian_2023} but show that additional regularization and postprocessing are required to have cleanly segmented 3D instances from a 3DGS reconstruction. 

We next find per-instance rigid transformations to bring them all to a common coordinate frame. Standard 3D point-cloud matching methods are ill-suited to 3DGS data, which is noisy and does not correspond well to an underlying geometry because it optimizes image radiance rather than geometry. 
To overcome this problem we exploit the ability of 3DGS to render additional high-quality views, enabling the use of robust 2D image matching algorithms, which we lift to 3D  using depth provided by 3DGS. 

Finally, we build a shared representation of a template for all the instances by taking the union of all the 3D Gaussian primitives. A fine-tuning step allows gradients to flow from each instance to the shared representation, improving the overall reconstruction.
In particular, partially-occluded instances are completed based on the visible instances. In addition low-quality geometry and appearance caused by poor coverage in the capture are enhanced with information from better-covered instances.
We model differences in appearance between instances using an offset representation for Spherical Harmonics (SH); we also show that this simple solution is faster than more complex options such as a multi-layer perceptron (MLP). 

\noindent
In summary, our contributions are:
\begin{itemize}
\item A shared representation and a fine-tuning process that improves overall reconstruction and novel view synthesis quality by using information from all instances.
	\item An improved 3D segmentation based on constrastive learning, using additional regularization and post-processing.
	\item Introducing the use of novel view synthesis to enable robust 2D matching, which then allows 3D registration of the noisy 3D Gaussians of each primitive.
	
\end{itemize}
\noindent
We evaluate our method on synthetic and real data with independent test trajectories unseen during reconstruction. 
Our method improves the overall quality of poorly captured or occluded regions in scenes with repetitive elements, outperforming other generalization solutions.
\section{Related Work}

We build on novel view synthesis and in particular 3D Gaussian Splatting (3DGS), segmentation, 3D registration, symmetries and repetitions, and  multi-illumination multi-view capture. \\

\textit{Novel view synthesis and Generalization.}
Our method strives to improve the quality of novel views by exploiting repetitions. Neural Radiance Fields (NeRF)~\cite{mildenhall_nerf_2020,barron2022mipnerf360,mueller2022instant} revolutionized novel view synthesis, permitting high quality rendering of novel views, albeit at a high computation cost for scene optimization and slow rendering. More recently 3D Gaussian Splatting (3DGS)~\cite{kerbl_3Dgaussians_2023} has permitted much faster optimization and rendering, resulting in widespread adoption (see surveys~\cite{chen2024survey3dgs,fei20243d}).

Several methods attempt to improve novel view synthesis generalization by introducing \emph{priors}~\cite{warburg_nerfbusters_2023}, or quantifying uncertainty~\cite{goli_BayesRays_2023}.
Diffusion models have been used as priors, typically for single or few-view use cases ~\cite{wu_reconfusion_2024,gao_cat3d_2024}. For 3DGS, LongLRM~\cite{ziwen_longlrm_2024} achieves impressive reconstruction and novel view synthesis quality with as few as 32 views, but with significant resource requirements (80Gb GPU).
We restrict generalization to repetitive elements, sidestepping the need for expensive diffusion models.\\

\textit{Radiance Field Segmentation.}
The lack of large-scale segmented 3D datasets has inspired the use of 2D vision models to segment 3D scenes. This creates the challenge of 
multi-view consistency, even with recent video segmentation~\cite{ravi_sam2_2024}.

Various authors~\cite{Shuaifeng_semanticnerf_2021,tschernezki_N3F_2022,kobayashi_FFD_2022} lift 2D class labels, or deep {features} ~\cite{caron_dino_2021,radford_clip_2021},  by augmenting the 3D representation with auxiliary feature channels  and optimizing them to match the 2D signal via differentiable rendering. Segmentation is then obtained via nearest neighbors in 3D feature space~\cite{goel_isrf_2023}. Such feature distillation has been extended to 3DGS~\cite{qiu_featuresplatting_2024,Zhou_Feature3DGS_2024, qin_langsplat_2023, lee_rethinking_2024}. 

Recent approaches propose to directly distill \textit{2D masks}~\cite{bhalgat_contrastive_2023, fan_nerfsos_2023}. The lack of association between masks from different views is resolved in 3D via contrastive learning, pushing rendered features closer together when they belong to the same mask in an image and farther otherwise.  Many authors \cite{gu_egolifter_2024, kim_garfield_2024, choi_clickGaussian_2024, cen_segment_2024, ying_omniseg3d_2024, choi_clickgaussian_2025, ye_gaussian_2023} extend this idea to 3DGS segmentation.
Such methods often have  residual artifacts and we propose  additional regularization and postprocessing  that significantly improve results.
\\

\textit{Registration}.
We refer the reader to a recent survey~\cite{huang_surveyregistration_2021}. Our specific sub-problem is the  registration of two clouds of Gaussian primitives, with \textit{varying density}, \textit{partial overlap}, \textit{large} pose variation and \textit{noise}. Because 3D Gaussians are  optimized to reproduce images, they are not necessarily located on the surface; density variation occur across instances depending on their coverage and projected size in the training views.  
Since our Gaussians are not located on the geometric surface, 3D descriptors ~\cite{johnson_spinimage_1999, Rusu_histogram3DFPFH_2009,qi_pointnet_2017, wang2019deep} would yield poor features.

We will exploit the ability of 3DGS to render realistics images to enable the use of 2D image matching. We also use the ability of 3DGS to provide depth to lift the matches to 3D.
We build on recent work on 3D reconstruction and 2D matching, in particular DUSt3R~\cite{wang_dust3r_2023} and MASt3R~\cite{leroy_grounding_2024}.  Due to its robustness under extreme pose variations, we adopt MASt3R and demonstrate its usefulness in building a registration pipeline for 3DGS, alongside more traditional tools such as PnP-RANSAC~\cite{lepetit_epnp_2009, fischler_ransac_1981}. \\

\begin{figure*}[!th]
\includegraphics[width=\linewidth]{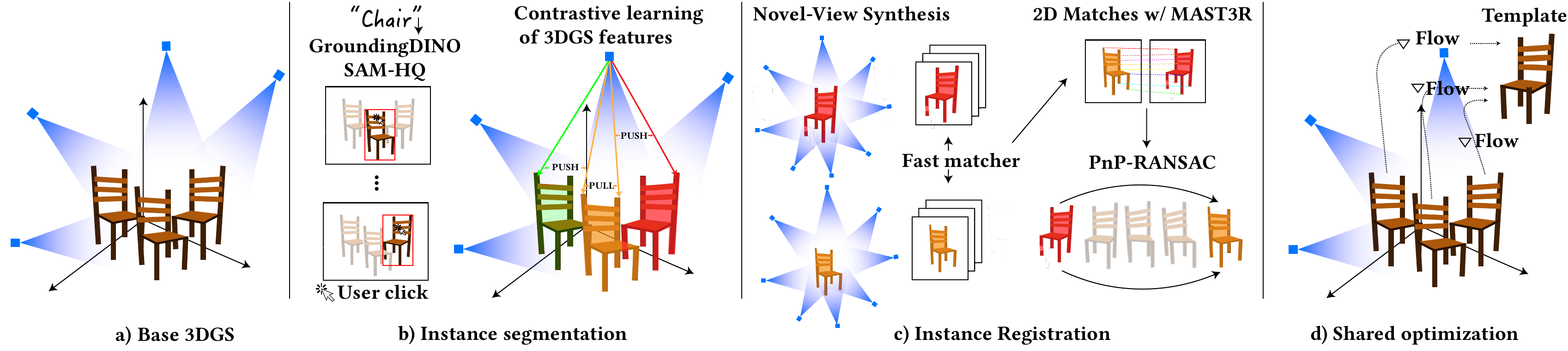}
\caption{
\label{fig:overview}
	\textbf{Overview of our method.} \textbf{(a)} We start with a base 3DGS reconstruction, then \textbf{(b)} use SAM-HQ masks and a user click to identify repetitive instances, and train contrastive features to perform instance segmentation in 3D. \textbf{(c)} To register the instances, we first render \emph{additional} views using 3DGS for each instance. After selecting the best pairs of views with a fast 2D matcher, we find robust 2D matches using MASt3R, which we lift to 3D with the depth obtained from 3DGS. Finally, we use a PnP-RANSAC solver to register the instances given the 2D-3D correspondences. \textbf{(d)} A finetuning step for our shared representation follows, allowing gradients to flow from the instances to the template. Once our method is complete, we can replace the instances with the optimized template, with significant improvement in visual quality. 
    }
\end{figure*}

\textit{Symmetry, self-similarities, repetitions}. 
The concept of symmetry and repetitions has received steady attention in Computer Graphics, with some approaches focusing on detecting symmetry~\cite{mitra_partial_2006, mitra_symmetry_2013, je_robust_2024} and others on exploiting it for downstream applications. In this paper, we offload the problem of symmetry detection to the user, who indicates repeating elements with a few clicks, 
and focus on how to best \textit{exploit} symmetry to improve novel-view synthesis. Previous work has shown the benefits of using symmetry for model compression~\cite{zheng_consolidation_2010}, scan consolidation~\cite{mitra_partial_2006},  symmetric triangulation~\cite{podolak_symmetry_2007} and model completion~\cite{thrun_completion_2005}. In the latter, the authors propose to use  partial symmetries to reconstruct missing regions in 3D scans. Repetitions have been used to compute structure and perform inverse rendering for a single image~\cite{cheng_structure_2023}, including in a generative context~\cite{zhang2023seeing}. 
Rodriguez et al.~\shortcite{rodriguez_exploiting_2018} use instances to improve reconstruction as we do, but are restricted to a planar configuration, and only test window instances on a facade. 
Similarly, we fill in missing details in one instance by borrowing from the other repetitions for 3DGS reconstructions.
Naively copying and pasting Gaussians would lead to artifacts because each repetition is observed under different illumination conditions. We thus tackle a multi-view multi-illumination reconstruction problem: geometry and surface properties are shared among instances, but illumination conditions vary.\\

\textit{Multi-view multi-illumination reconstruction.}
Most novel view synthesis methods assume that the scene has been captured under a single lighting condition. There have been several attempts to model changing conditions, for example NeRF-W~\cite{martinbrualla_nerfw_2020} that encodes varying appearance in an additional feature vector for NeRF. Specific methods have been developed to capture scenes under multiple illumination conditions, typically using specialized ``dome-like'' equipment~\cite{debevec2000acquiring}, but also low-end cellphone capture with a flash~\cite{bi2020deep}. 
Recently, 3DGS was extended for relighting ~\cite{poirier_GSrelighting_2024}, 
using a diffusion-model based prior to augment a single illumination capture to multi-illumination.

In our case, each instance can be seen as the same object illuminated under a different lighting configuration; however, the lighting variations are often less severe, since all instances are in the same overall scene. This allows us to represent appearance variation with spherical harmonics offsets.

\section{Overview}

We take as input a multi-view image dataset of a scene that contains multiple instances of the same element.  
Our method (Figure~\ref{fig:overview}) consists of three main stages. First, we perform 3D instance segmentation (Sec.~\ref{sec:3D_segmentation}), with an extended version of 3DGS, where an extra feature 
per Gaussian is trained using contrastive learning.
After segmentation, a user manually selects which instances should be grouped together by simply clicking on them. 
This  requires  one click per instance in only one image. 
Second, we propose a method to merge instances into a single coordinate system (Sec.~\ref{sec:instance_matching}). 
To deal with the geometric ambiguity of 3DGS, we  introduce a matching procedure that exploits the novel view synthesis capabilities of 3DGS to generate 
additional viewpoints. These new views  
enable robust 2D matching algorithms, which we exploit to do 2D-3D correspondences and 3D registration.
Third, we train a shared representation for all instances using common 
geometry but individual appearance parameters (Sec.~\ref{sec:scene_finetuning}).  
Each instance is seen from different viewpoints in the input images used during optimization. Our representation allows gradients to flow from each instance to the template using the rendering loss from each such viewpoint. This improves the overall reconstruction quality for each instance despite occlusions or lack of detail in different view of each instance used in training. 

\section{Method}
\label{sec:method}

\subsection{3D Instance Segmentation}
\label{sec:3D_segmentation}
Segmenting objects in 3D in a 3DGS scene is challenging. 3D Gaussian primitives do not strictly model object geometry: sometimes they  are placed in space around (but not on) a surface, or are large with low opacity, which allows 3DGS to model specific appearance in the input photos. Inspired by previous work~\cite{bhalgat_contrastive_2023, fan_nerfsos_2023}, we handle this task using contrastive learning~\cite{radford_clip_2021} with 2D masks, assigning a feature to each Gaussian.

To segment each training view, the user provides a text query such as ``\textit{pillar}'' to GroundingDINO~\cite{liu_groundingdino_2024}, which outputs a bounding box per instance, fed to SAM-HQ~\cite{ke_samhq_2023}, which in turn outputs a 2D mask per instance. We call ``\textit{background}'' the complement of the union of the masks in the training view, i.e. the union of all pixels \textit{not} containing the text query. We train using a cosine distance for the pull and push losses. The push loss includes a margin of 0.3. We add extra loss terms for the hard cases (distance between features larger than 0.5). The pull loss is weighted by the ratio of negative-to-positive pairs to counter the imbalance due to more negative pairs.

During contrastive learning, the trainable per-Gaussian features are rasterized into per-pixel features. To train the features, pairs of pixels are sampled in a given input image and undergo the following contrastive loss. If the pixels belong to the same 2D mask, their features are pulled together. If they belong to different masks, their features are pushed apart. 

However, we observe three main challenges in applying contrastive learning to 3DGS: 1) low-opacity Gaussians under the surface of objects do not receive enough gradients because they are occluded by other Gaussians, 2) large Gaussians that are shared among different objects get conflicting gradients and 3) several Gaussian primitives, typically at the border between an instance and the background are ``left behind'', resulting in visual artifacts and harming the subsequent optimization (Sec.~\ref{sec:scene_finetuning}). \\

\textit{Regularization and Optimization.}
To address the first two challenges, we add $\ell_1$ regularization terms on the opacity and the scaling during 3DGS training to discourage both phenomena 

\begin{equation}\label{eq:l1_reg_opacity_scale}
\lambda_{\text{opacity}}\frac{1}{N}\sum_{i=1}^N \mathbf{o}_i + \lambda_{\text{scale}}\frac{1}{N}\sum_{i=1}^N \mathbf{s}_i,
\end{equation}
\noindent
where $N$ is the number of Gaussian primitives. 

After 3DGS training is complete, we train the contrastive features. Contrastive methods are sensitive to the number of pixels sampled per mask to form positive and negative pairs. Therefore, we uniformly sample $M_u$ pixels in the image, so larger masks get more samples, and $M_s$ pixels per mask so even small masks get enough samples to form pairs (we set $M_u=M_s=4096$).

We include a regularization term to encourage the rendered features $\hat{\mathbf{f}}$ to be unit norm, similar to~\cite{cen_segment_2024}.\\

\textit{Interactive segmentation.}
Once the features are trained, different instances can be segmented based on the similarity of their features. 
For each instance $q$, to form a query feature $\mathbf{f}_q$, the user clicks inside the mask of the instance in any training view, and we select the rendered feature corresponding to the clicked pixel.
We perform the 3D segmentation by finding the closest Gaussians in the scene with a simple threshold in the contrastive feature space.  Thus, the segmented instance $\mathcal{I}_q$ for query $\mathbf{f}_q$ is simply the set of Gaussians with contrastive features $\mathbf{f}_i$ that verify $d(\mathbf{f}_i, \mathbf{f}_q) \leq \tau$. We set $\tau=0.1$ in our experiments. Since this only requires  one click per instance from the user, this interaction is fast and low-effort (please see supplementary video). At the end of this stage, we have a set of instances $\mathcal{I}_1, \dots \mathcal{I}_M$ formed by their corresponding Gaussian primitives for each object of interest. We compute corresponding 2D masks $\{ \mathcal{M}_{q,j} | 1\leq q \leq M, 1\leq j \leq N, \}$ of the $q$-th instance in  the $j$-th training view by comparing the rendered features in the training viewpoints with $\mathbf{f}_q$. The resulting masks are better than the ones produces by SAM-HQ because they aggregate information from all views~\cite{siddiqui_panoptic_2023}.\\

\textit{Post-processing.} This process can leave some residual Gaussians out of the segmented instances contributing to the appearance of the instance and the background Fig.~\ref{fig:ablation_seg} (left). It is important to remove them from the reconstruction of the background otherwise each instance will get slightly different gradients, hurting the sharing of information amongst them. We filter them out with a space carving approach.  To decide if a Gaussian primitive should be removed from the background, we check if at least one of the following three conditions is met in \emph{every} train image. First, the center projects inside the 2D masks $\{ \mathcal{M}_{q,j} \}$ of the instances.  Second, the center projects outside the image boundary. Third, the center is not visible in the current camera (this includes, for example, primitives behind the camera). If at least one of the conditions is met for all training views, the Gaussian affects the instance's rendering and is removed from the scene.

In Fig.~\ref{fig:ablation_seg}, we show how the regularization in Eq.~\ref{eq:l1_reg_opacity_scale} and the postprocessing step significantly improve overall visual quality.

\subsection{Novel View Synthesis for Instance Registration}\label{sec:instance_matching}

We now have $M$ segmented instances, which we need to register into a common coordinate system with a per-instance rigid transformation, i.e. a rotation $R$ and a translation $T$.
We choose the instance that has the largest number
of Gaussian primitives as the template instance $\mathcal{I}_T$. 

We have found purely  geometric registrations inapplicable because 3D Gaussians are optimized to reproduce the appearance of the scene but not necessarily its geometry. In particular, several Gaussians \textit{behind} the surface often contribute to the final appearance, and the instances have different spatial density of Gaussian primitives, e.g. very detailed regions have much smaller Gaussians than flat regions. Naive 3D point cloud registration does not work very well, as we show in Figure~\ref{fig:ablation_register_3d}.

On the other hand,  Gaussian primitives represent a radiance field that allows high quality novel view synthesis, which allows us to leverage
2D image matching. These methods can enable reliable instance registration in 3D if provided with suitable views of the instances. We use novel view synthesis to render such views. The estimation of the pose of a reference object from an image  typically has two steps: a coarse estimation stage, and a refinement step~\cite{Nguyen_gigapose_2024}. 

Our registration process for coarse pose estimation has four steps: 1) Rendering of a dense set of views of each instance from all directions 2) Fast matching of all pairs of all views to find a reduced set of $k$ pairs between each instance $\mathcal{I}_i$ and the template $\mathcal{I}_T$, 3) Dense matching of $k$ pairs of views of $\mathcal{I}_i$ and $\mathcal{I}_T$ to find the best such pair, 4) Lifting the 2D matches of the chosen pair to 3D and using  a PnP-RANSAC solver to find the transformation of $\mathcal{I}_i$ to $\mathcal{I}_T$.

\paragraph{Rendering dense set of views for each instance.}
We sample 25 cameras pointing towards the center, on a virtual sphere around each instance, by regularly sampling 5 azimuths and 5 elevations. Since we have a base 3DGS representation of the scene, we use these cameras to generate novel views of each instance. 

\paragraph{Fast matching to choose $k$ pairs of views.}
Next, we quickly find candidate pairs of views where the source instances $\mathcal{I}_i$ and target (template) $\mathcal{I}_T$ are seen from similar points of view. For this, we use a fast matcher~\cite{potje_xfeat_2024}, and select the top $k$ pair of source-target views with the most matches as candidates for the next stage. We set $k$ to 10 in our experiments.

\paragraph{Dense matching to find the best pair.}
Once we have the top $k$ pairs of views, we find \textit{dense} 2D matches with MASt3R~\cite{leroy_grounding_2024}, a transformer-based solution more robust than fast matching but of higher computational cost, which makes it prohibitively expensive to run directly on all pairs. We denote these matches
$P_{\text{2D}}=\{x_1 \dots x_N \}$ and $Q_{\text{2D}}=\{y_1 \dots y_N \}$. In pixel coordinates, the points $x_i$ and $y_i$ are of the form $(u_i, v_i)$.

\paragraph{Lifting to 3D and Perspective-n-Point solver.}
We backproject these 2D points using the depth $D_i$ obtained from 3DGS, (following~\cite{Kerbl_hierarchicalgaussians_2024}), the intrinsics  matrix $\mathbf{K}$ and extrinsics $\mathbf{M_{\text{cam}}}$. We obtain the corresponding 3D point matches $P_{\text{3D}}=\{p_1 \dots p_N \}$ and $Q_{\text{3D}}=\{q_1 \dots q_N \}$ expressed in world coordinates: 
\begin{equation}
p_i=D_i\mathbf{M_\text{cam}}\mathbf{K^{-1}}(u_i \ v_i \ 1)^T
\end{equation}
Back-projected depth tends to be closer to the geometric surface than the Gaussian primitives themselves because depth is computed as a weighted average of the Gaussian centers, some of which are in front of the surface and some of which are behind. 

Given one of the $k$ pairs of views, we find the rigid transformation between the source and the target Gaussians using Perspective-n-Point (PnP) on the 2D-3D correspondences between the source 3D points $P_{\text{3D}}$ and the target 2D keypoints $Q_{\text{2D}}$. The PnP-RANSAC solver outputs the camera-to-world transformation corresponding to the camera that sees the source 3D points $P_{\text{3D}}$ from the target camera viewpoint.
We run this robust PnP-RANSAC solver for each pair and keep the transformation with the largest number of inliers. 

Our pose refinement consist of two steps: first we run ICP~\cite{zhang_icp_1994} on the Gaussian centers, initialized with the coarse initialization of the pose, then we further refine the pose by adding its parameters to the subsequent 3DGS optimization.

For symmetric objects (e.g. the pawns in the \textsc{Chessboard} scene), the matching procedure finds correspondences based on appearance. For both symmetric and textureless objects, multiple rigid transformations are valid depending on the type of symmetry. In this case, finding one of them during registration is enough to finetune the shared representation.

\par

\subsection{Shared Representation and Optimization}\label{sec:scene_finetuning}

Now that we have the transformations of all the instances into a single template
space, we can create a shared representation for the repetitive scene elements. Our goal for this representation is twofold. First, we want a geometry representation that shares the information provided by each instance, thus significantly improving the 3DGS reconstruction quality for all instances. In particular, we want to complete instances with occluded parts using the information from the visible instances, and improve low-quality regions by propagating fine-grain details from high-quality parts. Second, we want our representation to handle the differences in appearance mainly due to different illumination.

Once we have all the transformations of instances $\mathcal{I}_i$ to the template $\mathcal{I}_T$, we create the initial shared representation by placing all the instances in a common coordinate system and taking their union. 

For each Gaussian primitive, all parameters are shared across instances of the same object, effectively constraining the geometry of all instances with multi-view information from all instances. To handle the differences in appearance, we decompose the SH coefficients of each primitive into a shared component and an offset term for each instance:
\begin{equation}
    \mathbf{c}^{\ell m} = \lambda \mathbf{c}_{\text{shared}}^{\ell m} + (1-\lambda)\mathbf{c}_{\text{offset}}^{\ell m}
\end{equation}
where $\ell$ is the degree, $m$ the order of the coefficient, and $\lambda$  a mixing weight set to $0.8$. The offset parameters $\mathbf{c}_{\text{offset}}^{\ell m}$ represent the individual differences between instances, typically due to illumination, and we encourage them to be small with an $\ell_1$ penalty.

\paragraph{Optimization} 
We replace each instance with a \textit{reference} to the shared representation that undergoes the per-instance geometric transform and SH offset. As a result, we can propagate gradients from all instances to the shared representation. 
We optimize the shared geometry and SHs with gradient descent. 
For example, if $\mathcal{I}_1$ and $\mathcal{I}_2$ are seen from opposite sides in template space, the shared representation will benefit from the information each instance provides and be well reconstructed from both sides. We jointly optimize over the pose to refine its estimation from the previous step.

\begin{table*}
    \centering
    \small
    \caption{\textbf{Quantitative evaluation.} We report average PSNR, SSIM, LPIPS, and KID across  test views on our synthetic and real dataset. We compute the metrics both in the complete images and in the masked regions where replacement of instances took place to isolate the effect of our method.
        For easier readability, we color-code results as a linear gradient between 
        \BGcolor{e0a4a4}{ w}\BGcolor{e7b2ac}{o}\BGcolor{eec0b5}{r}\BGcolor{f6cebe}{s}\BGcolor{fcdcc6}{t}\BGcolor{ffe4ca}{ }\BGcolor{fee8cc}{a}\BGcolor{feeccd}{n}\BGcolor{fdf0cf}{d}\BGcolor{f6f0cb}{ }\BGcolor{e3e6bd}{b}\BGcolor{cfdcae}{e}\BGcolor{bbd2a0}{s}\BGcolor{a8c992}{t }            }
    \addtolength{\tabcolsep}{-0.4em}
    \resizebox{\linewidth}{!}{
    \newdimen\wate \wate=15pt
    \newdimen\wperc \wperc=14pt
    \begin{tabular}{l|cccc|cccc|cccc|cccc}
    \toprule
        & \multicolumn{4}{c|}{Synthetic} & \multicolumn{4}{c}{Synthetic (masked)} & \multicolumn{4}{|c|}{Real}  & \multicolumn{4}{|c}{Real (masked)} \\
    &  PSNR$\uparrow$    &  SSIM$\uparrow$   &LPIPS $\downarrow$ &  KID$\downarrow$ 
    & PSNR$\uparrow$    &  SSIM$\uparrow$   &LPIPS $\downarrow$ &  KID$\downarrow$ 
    &  PSNR$\uparrow$    &  SSIM$\uparrow$   &LPIPS $\downarrow$ & KID$\downarrow$
    &  PSNR$\uparrow$    &  SSIM$\uparrow$   &LPIPS $\downarrow$ & KID$\downarrow$
    \\
    \midrule
    Nerfbusters & \cellCD[\wate]{e0a4a4}{19.48} & \cellCD[\wate]{e7b2ac}{0.690} & \cellCD[\wate]{e0a4a4}{0.464} & \cellCD[\wate]{e0a4a4}{0.4185} & \cellCD[\wate]{e0a4a4}{17.97} & \cellCD[\wate]{e7b2ac}{0.605} & \cellCD[\wate]{e7b2ac}{0.219} & \cellCD[\wate]{e7b2ac}{0.2098} & \cellCD[\wate]{e0a4a4}{19.26} & \cellCD[\wate]{e0a4a4}{0.757} & \cellCD[\wate]{e0a4a4}{0.422} & \cellCD[\wate]{e0a4a4}{0.3017} & \cellCD[\wate]{e0a4a4}{17.62} & \cellCD[\wate]{e0a4a4}{0.687} & \cellCD[\wate]{e0a4a4}{0.138} & \cellCD[\wate]{e0a4a4}{0.1827} \\
BayesRays & \cellCD[\wate]{e0a4a4}{19.50} & \cellCD[\wate]{e0a4a4}{0.655} & \cellCD[\wate]{eec0b5}{0.418} & \cellCD[\wate]{fcdcc6}{0.3018} & \cellCD[\wate]{e0a4a4}{18.20} & \cellCD[\wate]{e0a4a4}{0.573} & \cellCD[\wate]{e0a4a4}{0.223} & \cellCD[\wate]{e0a4a4}{0.2242} & \cellCD[\wate]{feeccd}{22.05} & \cellCD[\wate]{eec0b5}{0.779} & \cellCD[\wate]{f6cebe}{0.385} & \cellCD[\wate]{e3e6bd}{0.1215} & \cellCD[\wate]{fdf0cf}{21.97} & \cellCD[\wate]{fcdcc6}{0.743} & \cellCD[\wate]{f6cebe}{0.123} & \cellCD[\wate]{f6f0cb}{0.0837} \\
Nerfacto & \cellCD[\wate]{e7b2ac}{20.18} & \cellCD[\wate]{e7b2ac}{0.683} & \cellCD[\wate]{f6cebe}{0.405} & \cellCD[\wate]{feeccd}{0.2288} & \cellCD[\wate]{e7b2ac}{19.37} & \cellCD[\wate]{e7b2ac}{0.607} & \cellCD[\wate]{e7b2ac}{0.215} & \cellCD[\wate]{fcdcc6}{0.1694} & \cellCD[\wate]{fdf0cf}{22.43} & \cellCD[\wate]{ffe4ca}{0.803} & \cellCD[\wate]{ffe4ca}{0.368} & \cellCD[\wate]{f6f0cb}{0.1303} & \cellCD[\wate]{feeccd}{21.54} & \cellCD[\wate]{fee8cc}{0.771} & \cellCD[\wate]{ffe4ca}{0.114} & \cellCD[\wate]{e3e6bd}{0.0765} \\
3DGS & \cellCD[\wate]{fee8cc}{23.37} & \cellCD[\wate]{feeccd}{0.800} & \cellCD[\wate]{f6f0cb}{0.266} & \cellCD[\wate]{fdf0cf}{0.1808} & \cellCD[\wate]{fee8cc}{22.78} & \cellCD[\wate]{feeccd}{0.761} & \cellCD[\wate]{feeccd}{0.139} & \cellCD[\wate]{fcdcc6}{0.1607} & \cellCD[\wate]{fdf0cf}{22.57} & \cellCD[\wate]{e3e6bd}{0.849} & \cellCD[\wate]{bbd2a0}{0.273} & \cellCD[\wate]{cfdcae}{0.0974} & \cellCD[\wate]{feeccd}{21.87} & \cellCD[\wate]{f6f0cb}{0.798} & \cellCD[\wate]{e3e6bd}{0.089} & \cellCD[\wate]{e3e6bd}{0.0766} \\
3DGS* & \cellCD[\wate]{feeccd}{24.41} & \cellCD[\wate]{e3e6bd}{0.843} & \cellCD[\wate]{e3e6bd}{0.236} & \cellCD[\wate]{e3e6bd}{0.1389} & \cellCD[\wate]{fee8cc}{23.02} & \cellCD[\wate]{fdf0cf}{0.778} & \cellCD[\wate]{f6f0cb}{0.119} & \cellCD[\wate]{fdf0cf}{0.1076} & \cellCD[\wate]{fdf0cf}{22.59} & \cellCD[\wate]{f6f0cb}{0.842} & \cellCD[\wate]{bbd2a0}{0.273} & \cellCD[\wate]{cfdcae}{0.0994} & \cellCD[\wate]{feeccd}{21.91} & \cellCD[\wate]{f6f0cb}{0.799} & \cellCD[\wate]{f6f0cb}{0.090} & \cellCD[\wate]{e3e6bd}{0.0749} \\
Ours & \cellCD[\wate]{a8c992}{27.62} & \cellCD[\wate]{a8c992}{0.897} & \cellCD[\wate]{a8c992}{0.163} & \cellCD[\wate]{a8c992}{0.0316} & \cellCD[\wate]{a8c992}{27.54} & \cellCD[\wate]{a8c992}{0.887} & \cellCD[\wate]{a8c992}{0.063} & \cellCD[\wate]{a8c992}{0.0097} & \cellCD[\wate]{a8c992}{24.18} & \cellCD[\wate]{a8c992}{0.868} & \cellCD[\wate]{a8c992}{0.248} & \cellCD[\wate]{a8c992}{0.0483} & \cellCD[\wate]{a8c992}{24.63} & \cellCD[\wate]{a8c992}{0.847} & \cellCD[\wate]{a8c992}{0.068} & \cellCD[\wate]{a8c992}{0.0348} \\
    \bottomrule
    \end{tabular}
    }
    \vspace{5pt}
    \label{tab:quantitative}
\end{table*}

\section{Data \& Evaluation}

We implemented our method building on the original codebase of 3DGS~\cite{kerbl_3Dgaussians_2023}. Our source code and data are available at \url{https://repo-sam.inria.fr/nerphys/splat-and-replace}.

\subsection{Synthetic \& Real Scenes}
We evaluate our method on four synthetic scenes and four real scenes, the former allowing more precise quantitative evaluation. For each scene, we optimize the reconstruction over the training sequence and evaluate over a held-out test sequence. For synthetic scene, \textsc{Office}, \textsc{Temple}, \textsc{Chessboard} and \textsc{Classroom}, we render 200 images for training and 50 for testing. Our test views are far from the training views, and are not used in training. The image size is $1370\times 912$, rendered using Blender's Cycles~\cite{blender}, along with ground truth masks and depth maps, that we use in our quantitative evaluation.
We use the following repetitive elements for each scene: 3 tables and 9 chairs for \textsc{Office}, 13 columns for \textsc{Temple}, 5 desks for \textsc{Classroom}, 2 rooks, 2 knights, 2 bishops, and 8 pawns for \textsc{Chessboard}.

For each real scene, \textsc{House}, \textsc{MeetingRoom}, \textsc{Pillars} and \textsc{Facade},
 we run COLMAP~\cite{schonberger_structure--motion_2016} on both the train and test sequences together to obtain a coherent calibration, then exclude the SfM points for which the test views were used for triangulation from the 3DGS initialization. We compute monocular depth using Depth-Anything-v2~\cite{Yang_depthanythingv2_2024} for depth supervision, and the segmentation masks using GroundingDINO~\cite{liu_groundingdino_2024} and SAM-HQ~\cite{ke_samhq_2023}. 
We use the following repetitive elements for each scene: 2 windows for \textsc{House}, 4 windows and 4 railings for \textsc{Facade}, 6 chairs and 3 tables for  \textsc{MeetingRoom}, and 4 pillars for \textsc{Pillars}.

We also include one scene from ScanNet++~\cite{yeshwanth2023scannet++} and two scenes from DL3DV~\cite{ling2024dl3dv} with repetitive elements (chairs).
\subsection{Evaluation}\label{sec:results}

\textit{Baselines.} We compare with the vanilla 3DGS~\cite{kerbl_3Dgaussians_2023}, and an improved version, denoted 3DGS*, using monocular depth regularization, exposure adjustment from ~\cite{Kerbl_hierarchicalgaussians_2024} and our proposed opacity and scale regularization for the segmentation (see Sec.~\ref{sec:3D_segmentation}). We additionally compare with Nerfbusters~\cite{warburg_nerfbusters_2023} and Bayes' Rays~\cite{goli_BayesRays_2023}, which both aim at better generalization by measuring uncertainty in novel views and using it to remove floaters. Finally, Nerfacto~\cite{tancik2023nerfstudio} is a strong NeRF-based baseline that trades off speed and quality, and is also the underlying model that Nerfbusters and Bayes' Rays use.\\

\begin{figure*}
    \centering
    \begin{minipage}[b]{\linewidth}
        \rotatebox{90}{\hspace{0.5cm}Nerfbusters\phantom{y}}
        \begin{minipage}[b]{0.23\linewidth}
            \centering
            \textsc{Temple}\\
                \vspace{0.15cm}
            \includegraphics[width=\linewidth]{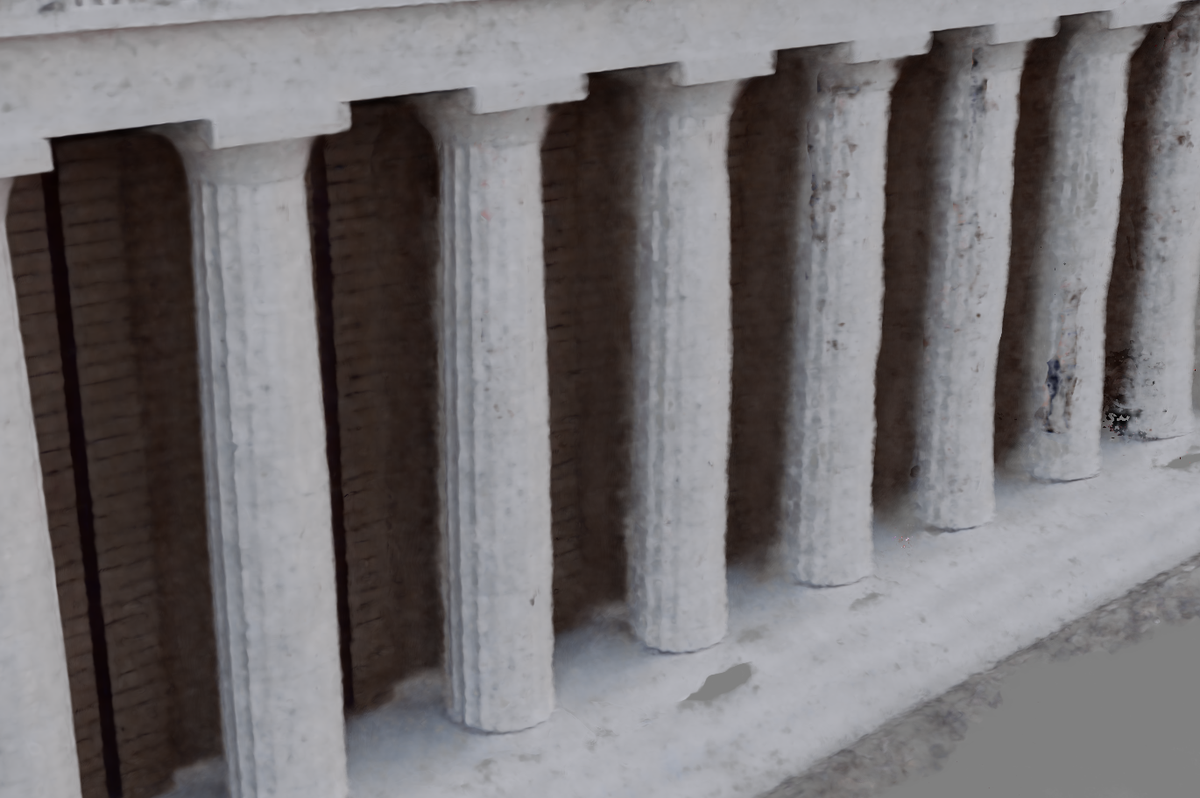}
        \end{minipage}%
        \hspace{0.01\linewidth}%
        \begin{minipage}[b]{0.23\linewidth}
            \centering
            \textsc{Classroom}\\
                \vspace{0.15cm}
            \includegraphics[width=\linewidth]{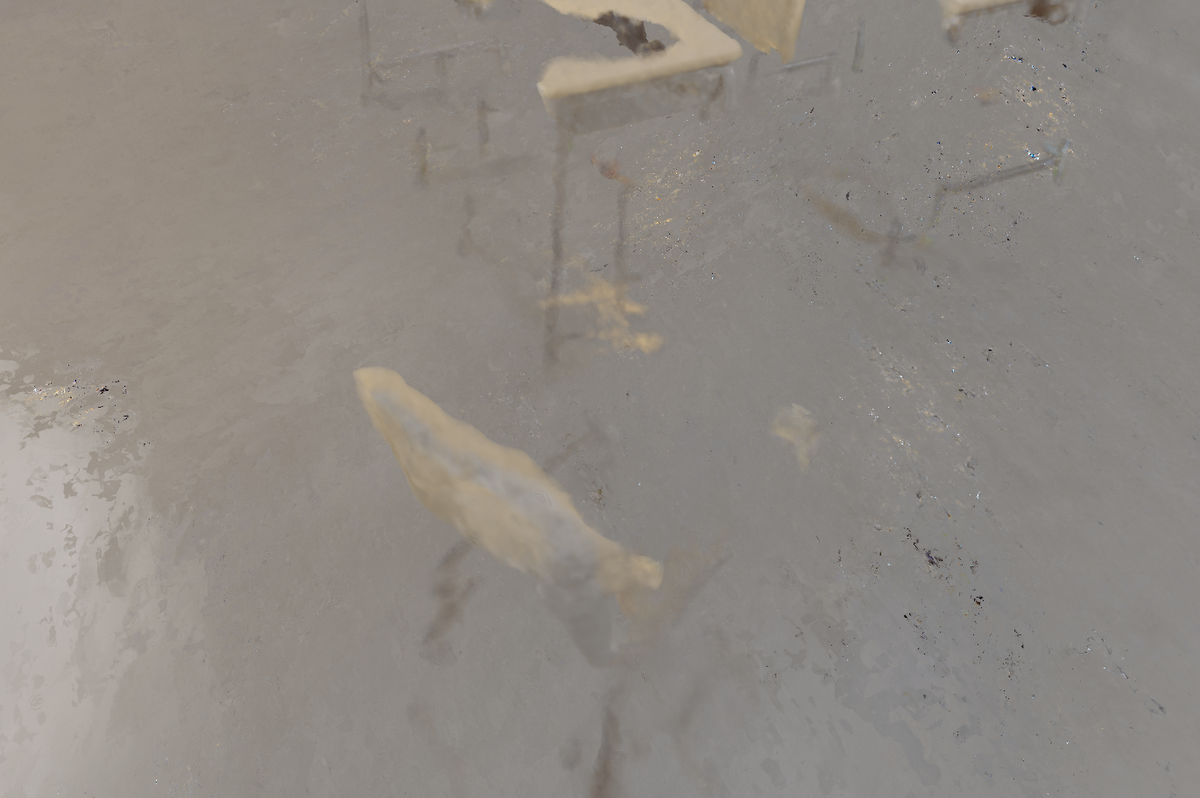}
        \end{minipage}%
        \hspace{0.01\linewidth}%
        \begin{minipage}[b]{0.23\linewidth}
            \centering
            \textsc{Chessboard}\\
                \vspace{0.15cm}
            \includegraphics[width=\linewidth]{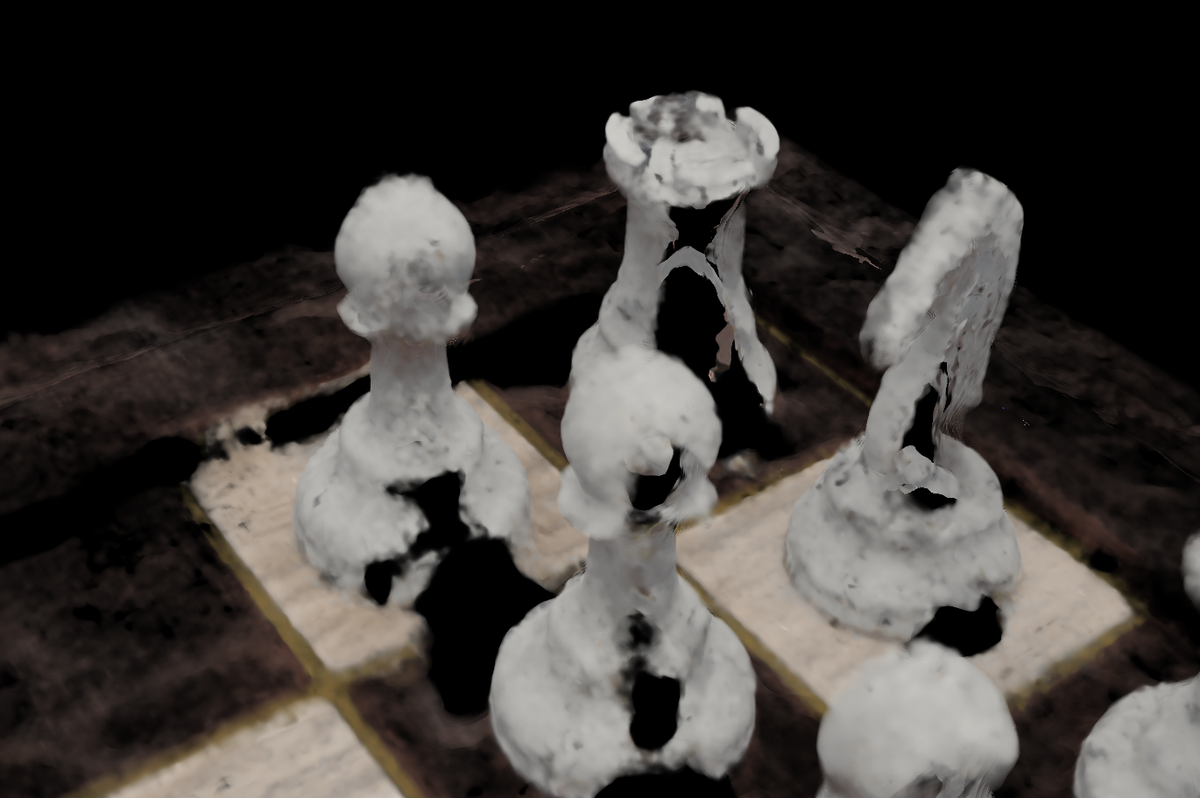}
        \end{minipage}%
        \hspace{0.01\linewidth}%
        \begin{minipage}[b]{0.23\linewidth}
            \centering
            \textsc{Office}\\
                \vspace{0.15cm}
            \includegraphics[width=\linewidth]{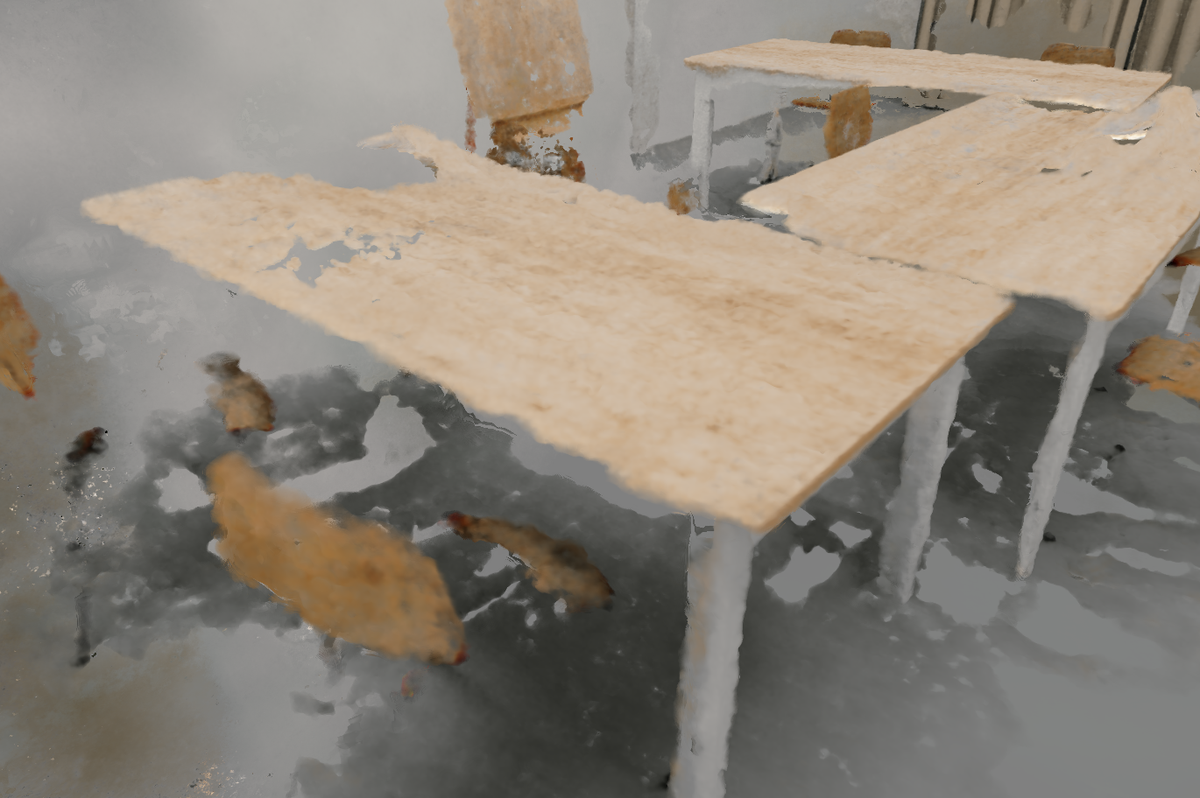}
        \end{minipage}%
    \end{minipage}\\
    \vspace{0.15cm}
    \begin{minipage}[b]{\linewidth}
        \rotatebox{90}{\hspace{0.5cm}Bayes' Rays}
        \begin{minipage}[b]{0.23\linewidth}
            \centering
            \includegraphics[width=\linewidth]{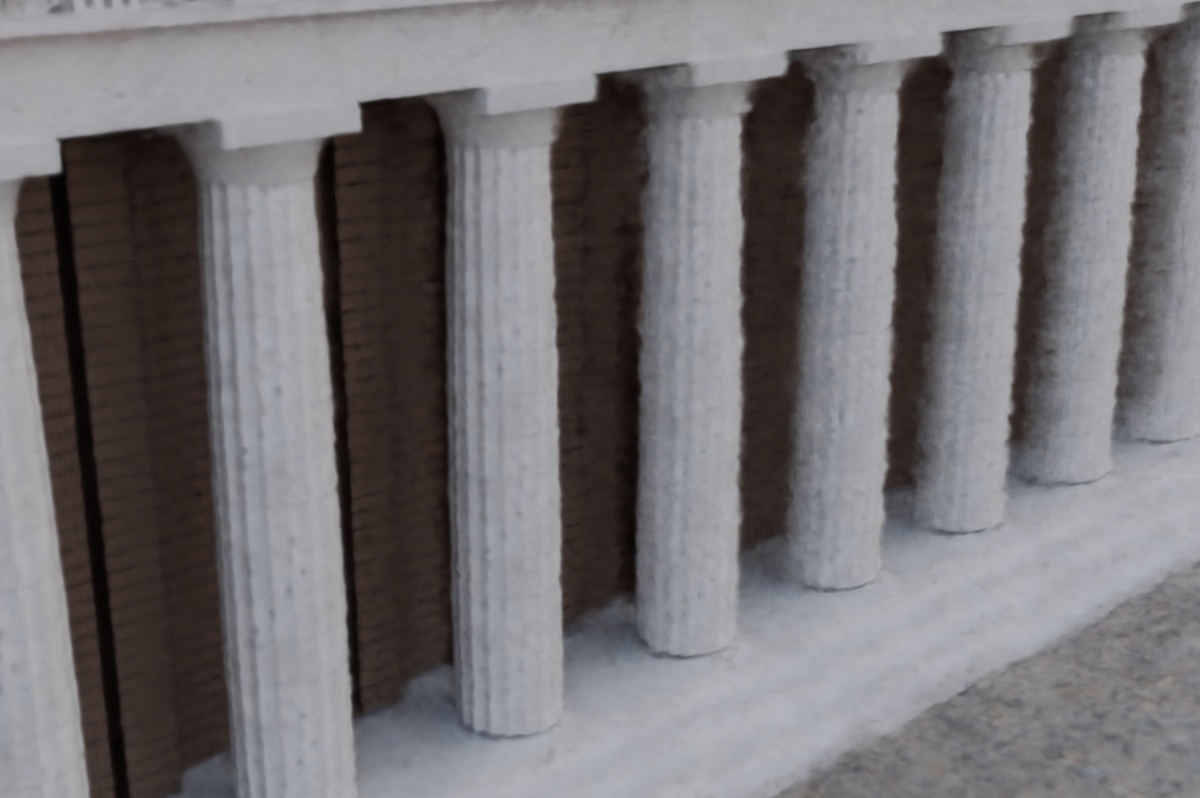}
        \end{minipage}%
        \hspace{0.01\linewidth}%
        \begin{minipage}[b]{0.23\linewidth}
            \centering
            \includegraphics[width=\linewidth]{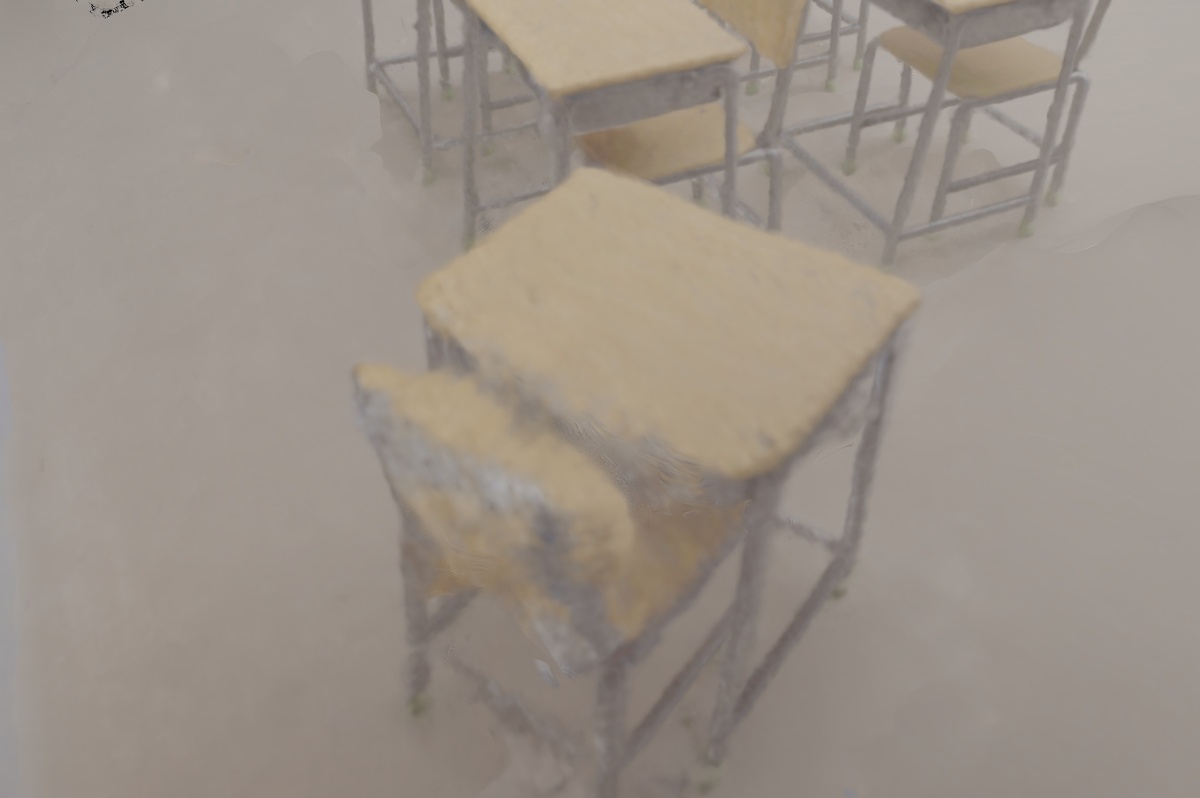}
        \end{minipage}%
        \hspace{0.01\linewidth}%
        \begin{minipage}[b]{0.23\linewidth}
            \centering
            \includegraphics[width=\linewidth]{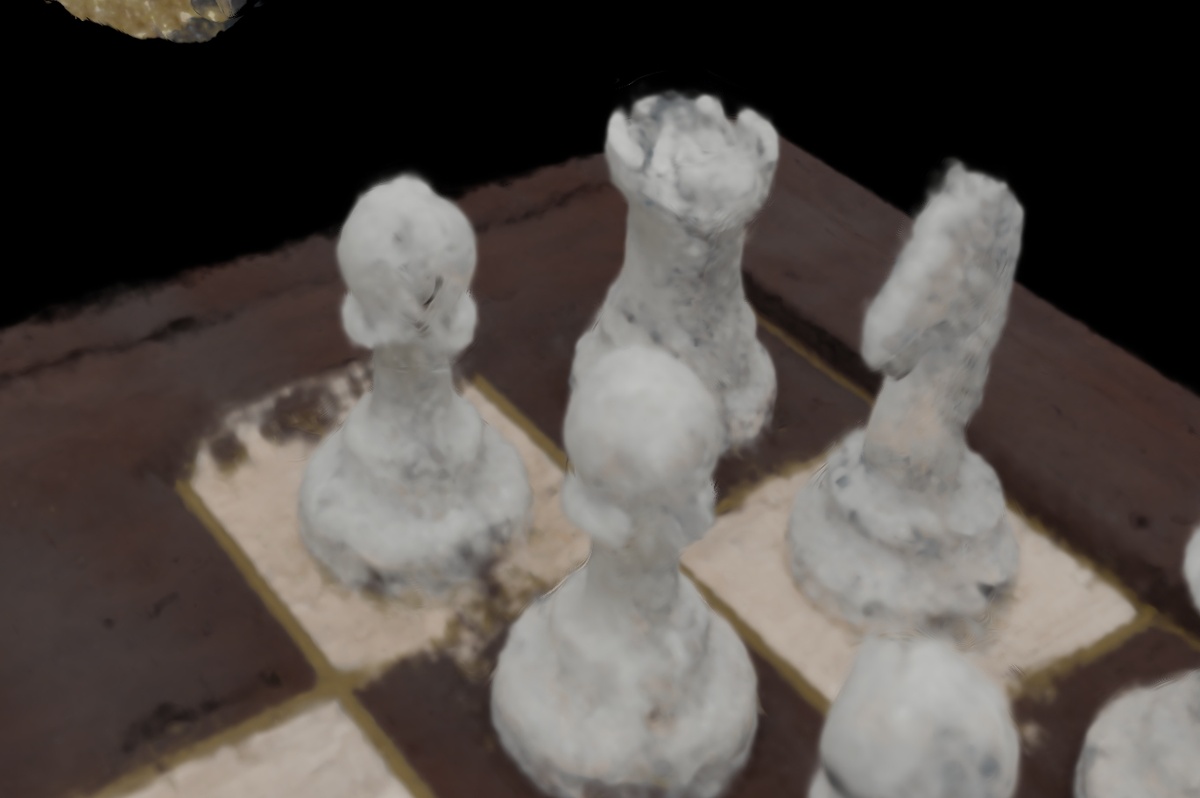}
        \end{minipage}%
        \hspace{0.01\linewidth}%
        \begin{minipage}[b]{0.23\linewidth}
            \centering
            \includegraphics[width=\linewidth]{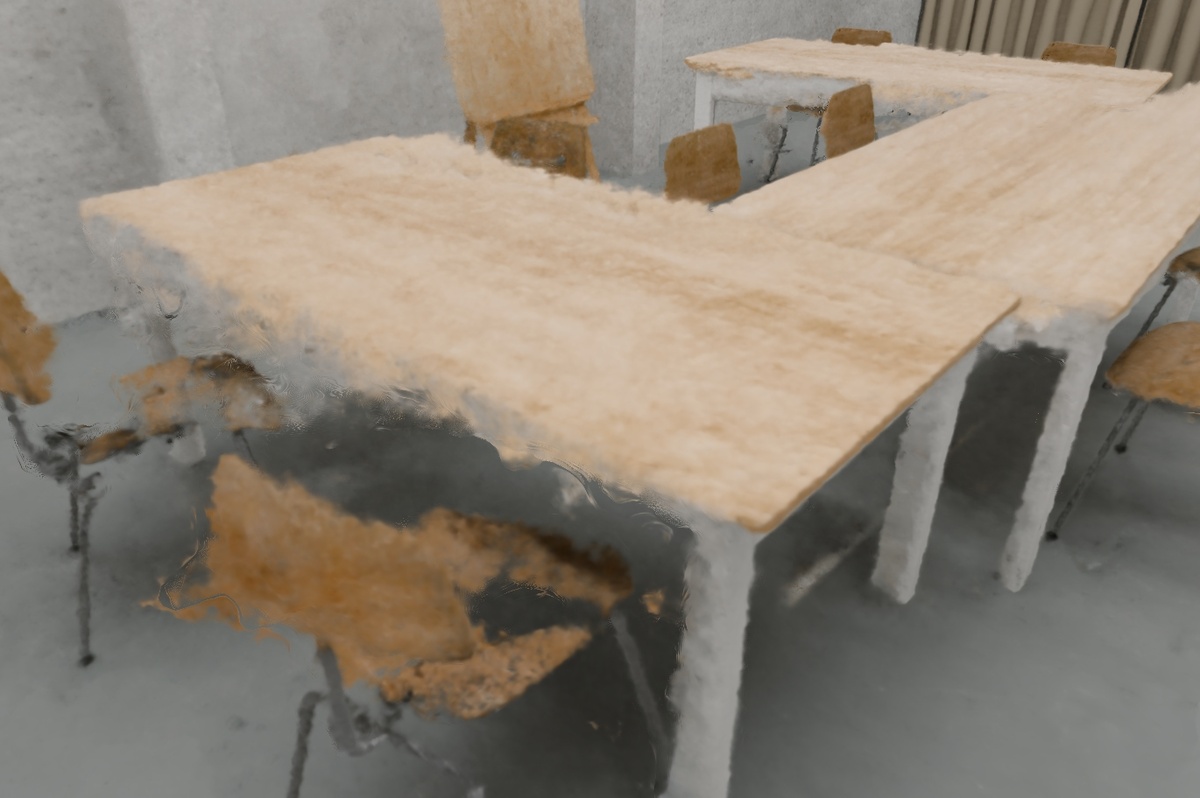}
        \end{minipage}%
    \end{minipage}\\
        \vspace{0.15cm}
    \begin{minipage}[b]{\linewidth}
        \rotatebox{90}{\hspace{0.8cm}3DGS*\phantom{y}}
        \begin{minipage}[b]{0.23\linewidth}
            \centering
            \includegraphics[width=\linewidth]{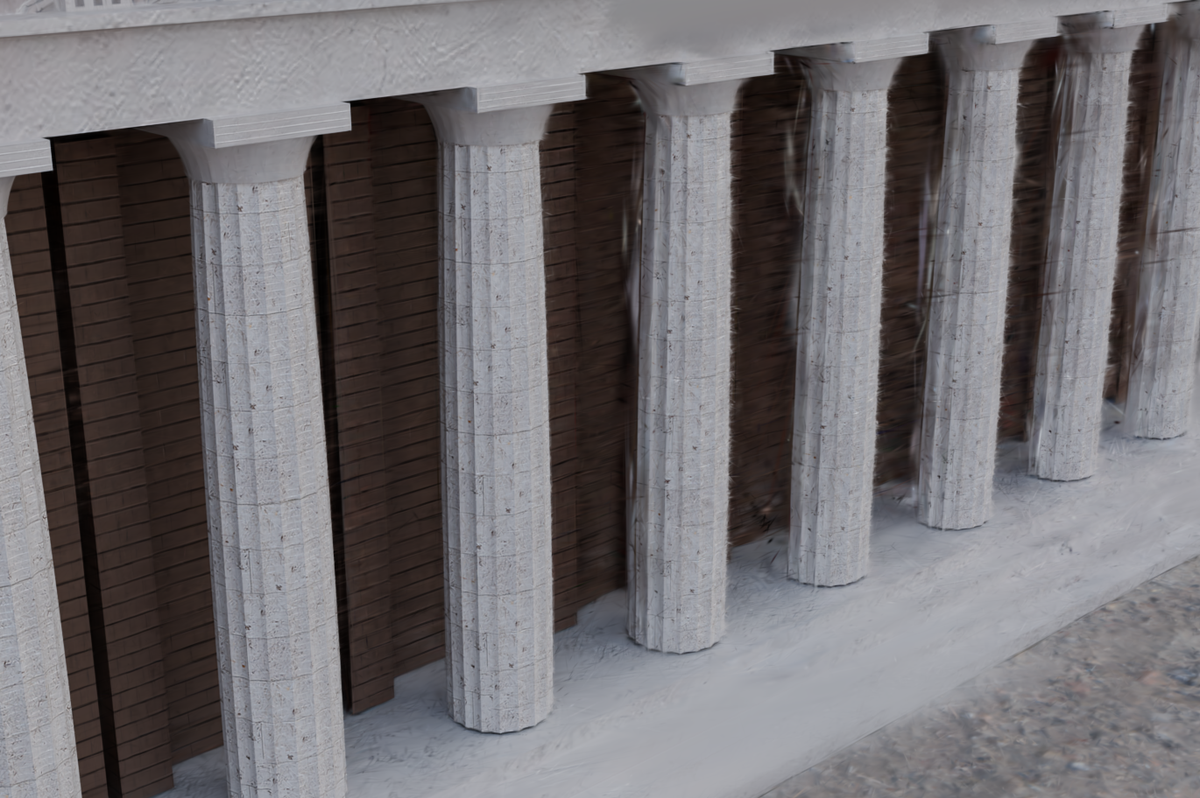}
        \end{minipage}%
        \hspace{0.01\linewidth}%
        \begin{minipage}[b]{0.23\linewidth}
            \centering
            \includegraphics[width=\linewidth]{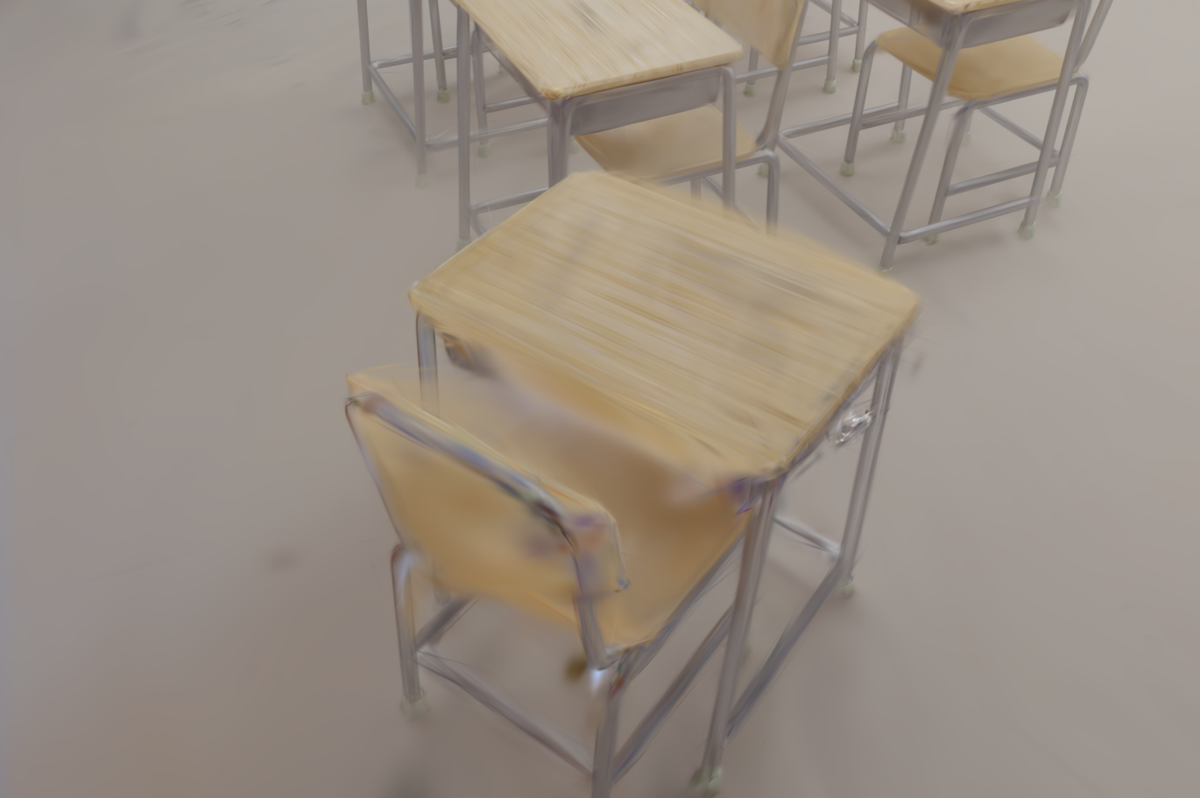}
        \end{minipage}%
        \hspace{0.01\linewidth}%
        \begin{minipage}[b]{0.23\linewidth}
            \centering
            \includegraphics[width=\linewidth]{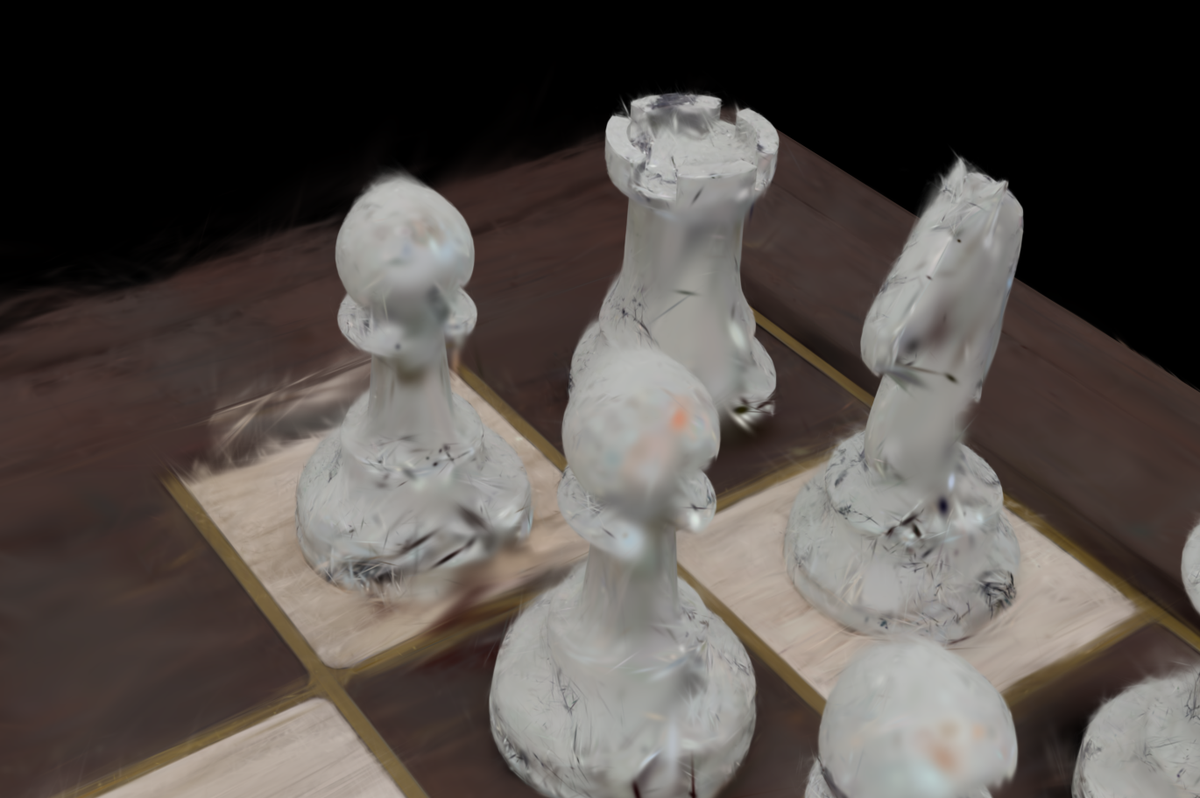}
        \end{minipage}%
        \hspace{0.01\linewidth}%
        \begin{minipage}[b]{0.23\linewidth}
            \centering
            \includegraphics[width=\linewidth]{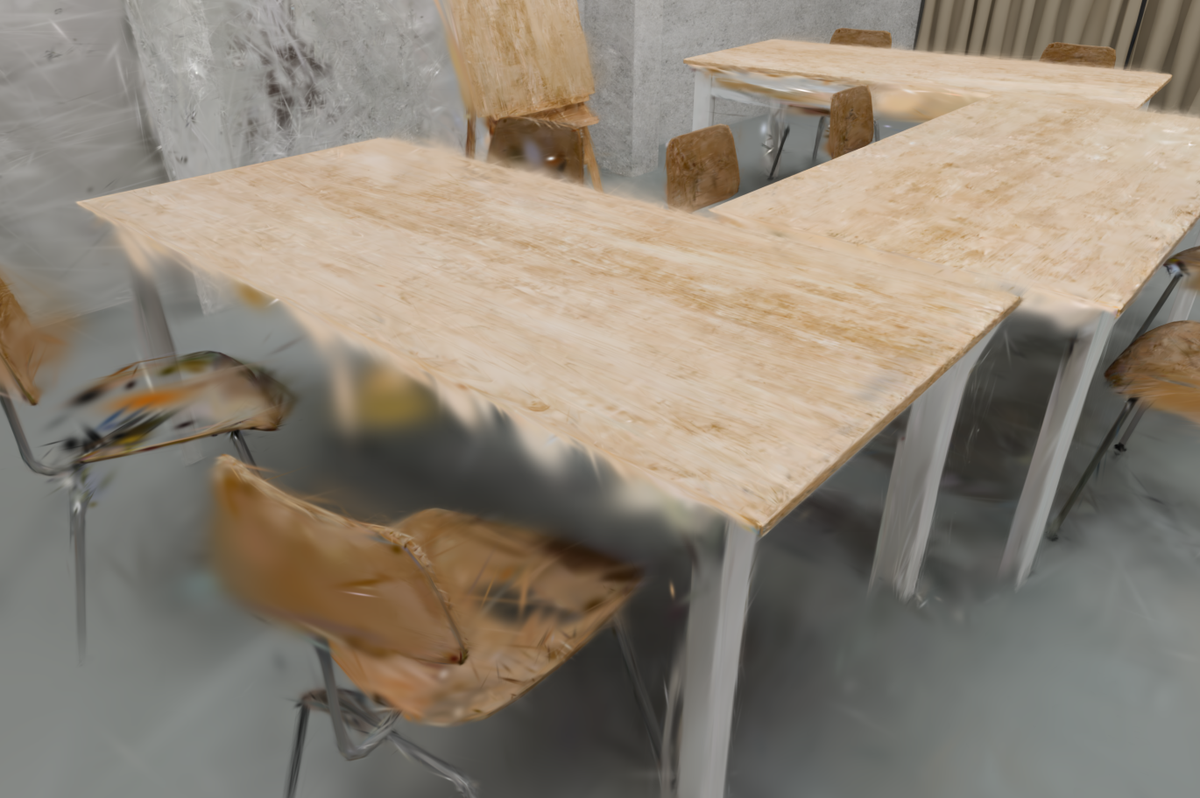}
        \end{minipage}%
    \end{minipage}\\%
        \vspace{0.15cm}
    \begin{minipage}[b]{\linewidth}
        \rotatebox{90}{\hspace{0.8cm}Ours\phantom{y}}
        \begin{minipage}[b]{0.23\linewidth}
            \centering
            \includegraphics[width=\linewidth]{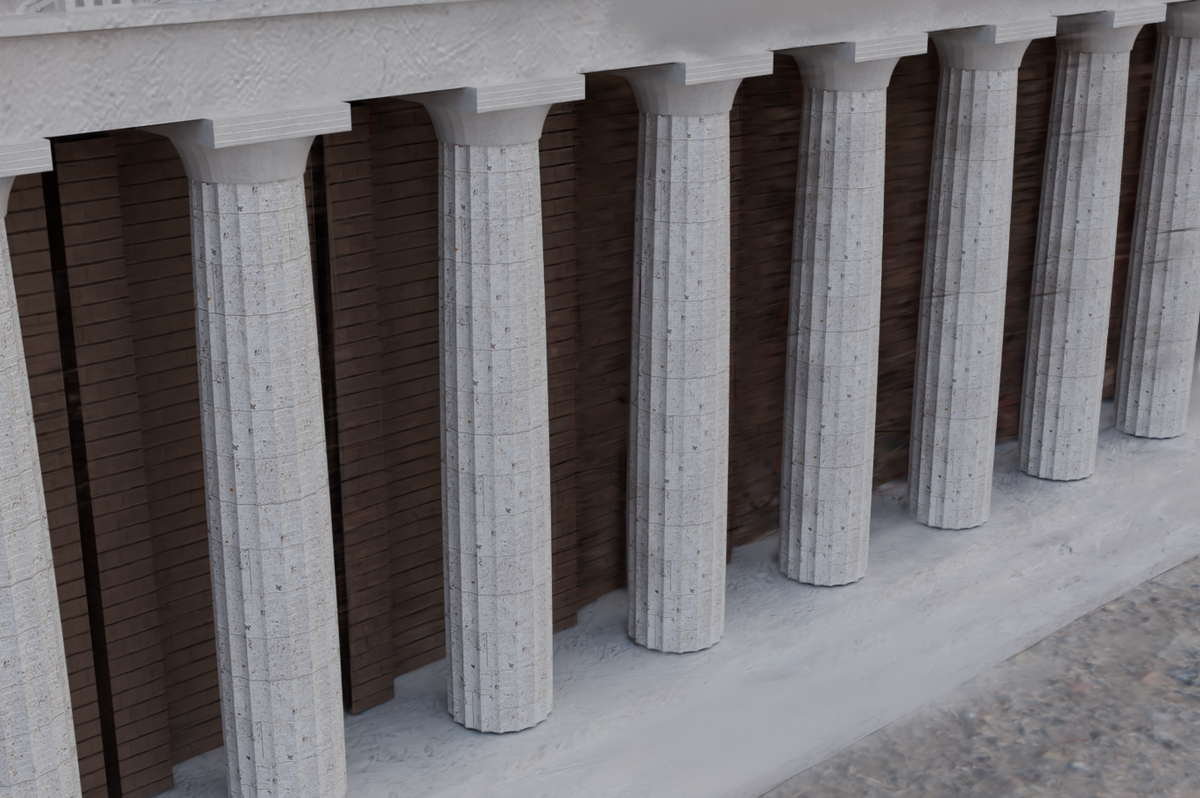}
        \end{minipage}%
        \hspace{0.01\linewidth}%
        \begin{minipage}[b]{0.23\linewidth}
            \centering
            \includegraphics[width=\linewidth]{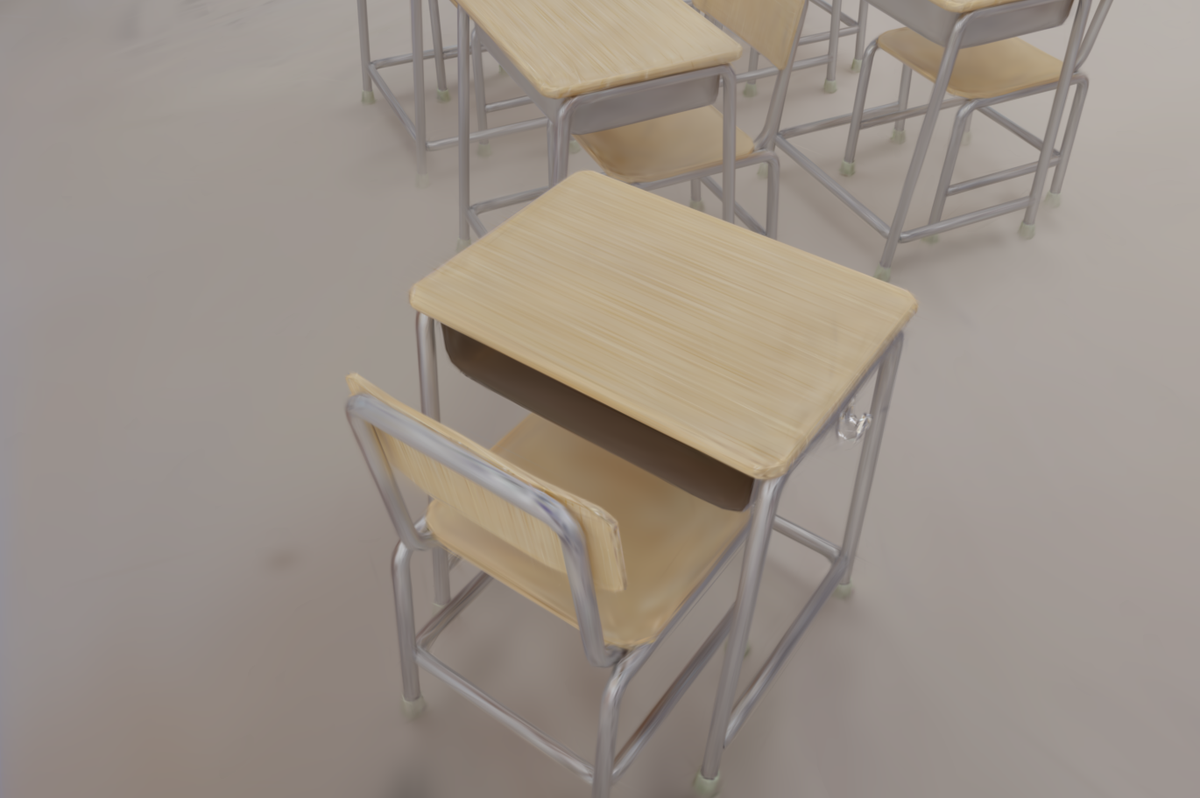}
        \end{minipage}%
        \hspace{0.01\linewidth}%
        \begin{minipage}[b]{0.23\linewidth}
            \centering
            \includegraphics[width=\linewidth]{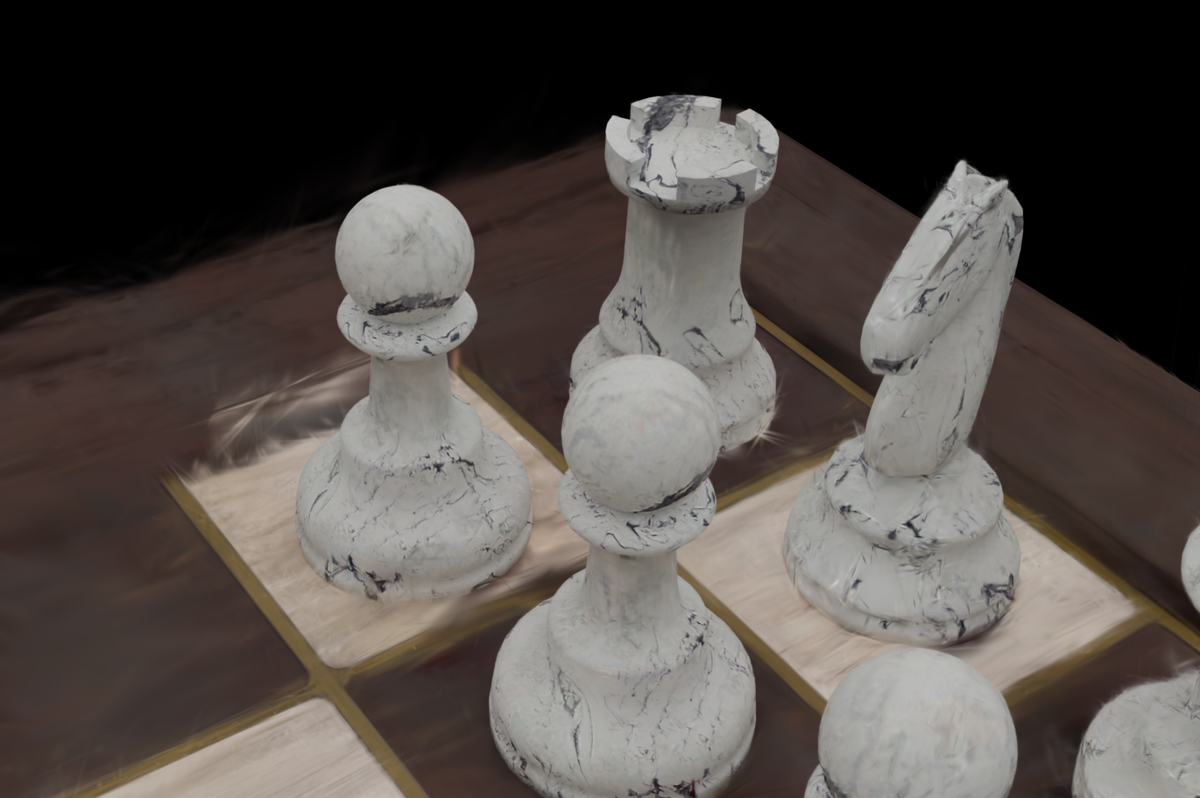}
        \end{minipage}%
        \hspace{0.01\linewidth}%
        \begin{minipage}[b]{0.23\linewidth}
            \centering
            \includegraphics[width=\linewidth]{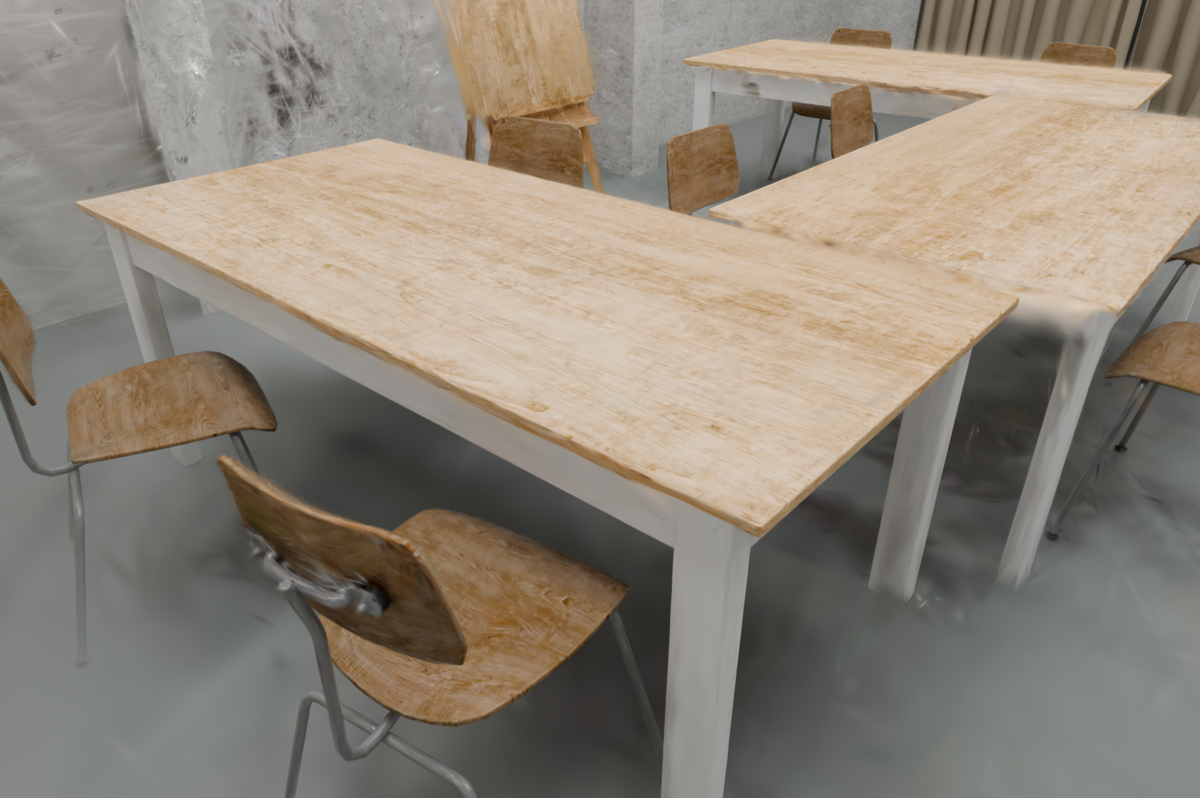}
        \end{minipage}%
    \end{minipage}\\
        \vspace{0.15cm}
    \begin{minipage}[b]{\linewidth}
        \rotatebox{90}{\hspace{0.5cm}Ground Truth\phantom{y}}
        \begin{minipage}[b]{0.23\linewidth}
            \centering
            \includegraphics[width=\linewidth]{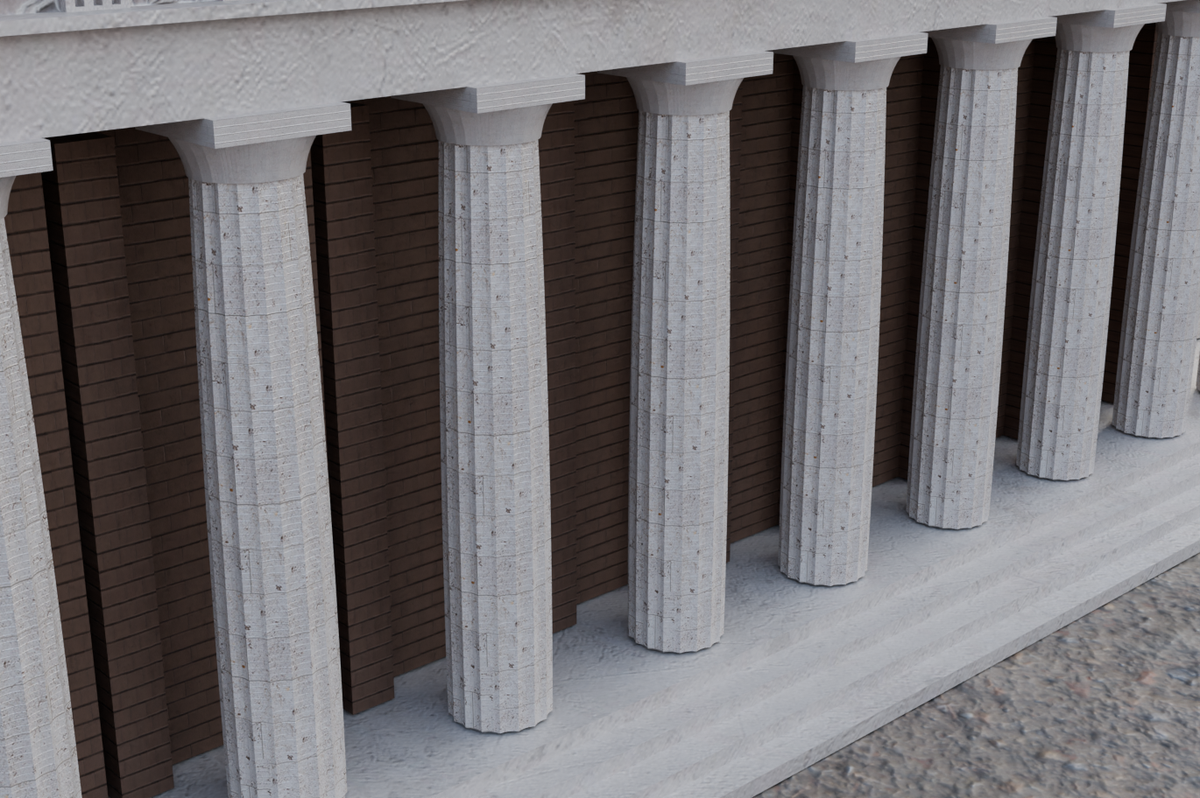}
        \end{minipage}%
        \hspace{0.01\linewidth}%
        \begin{minipage}[b]{0.23\linewidth}
            \centering
            \includegraphics[width=\linewidth]{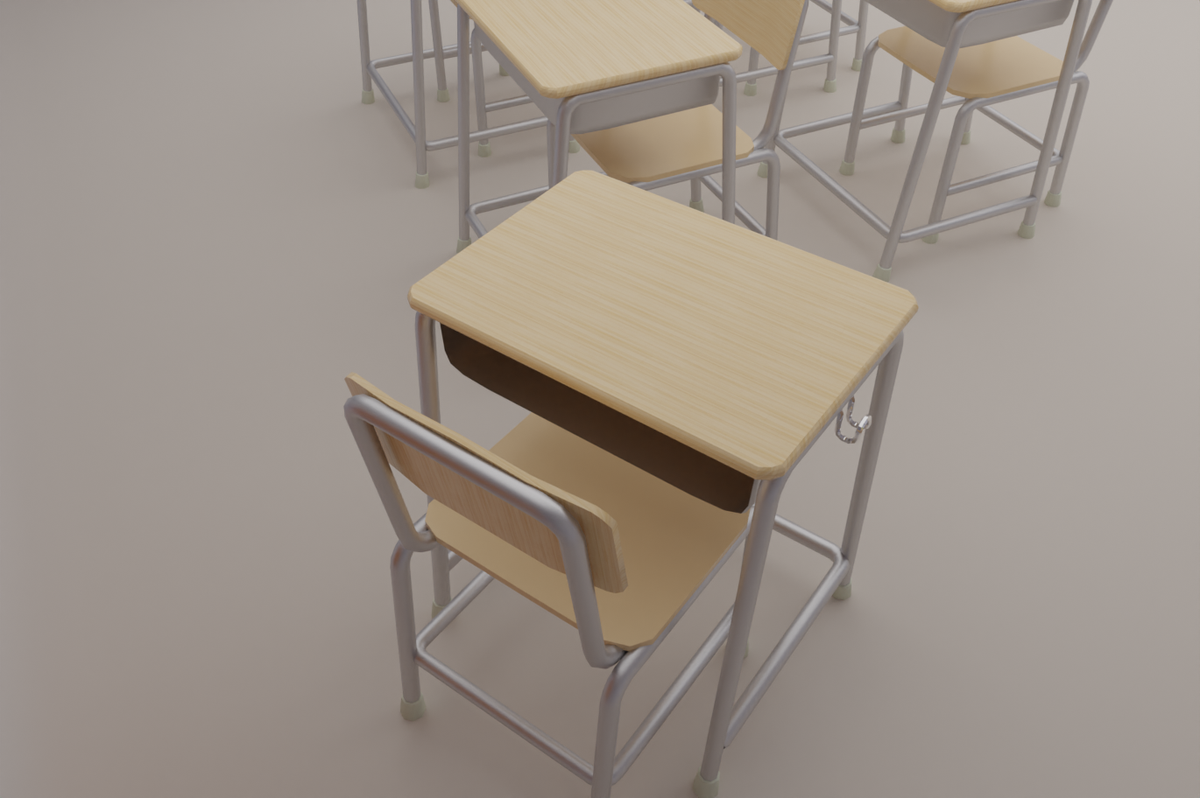}
        \end{minipage}%
        \hspace{0.01\linewidth}%
        \begin{minipage}[b]{0.23\linewidth}
            \centering
            \includegraphics[width=\linewidth]{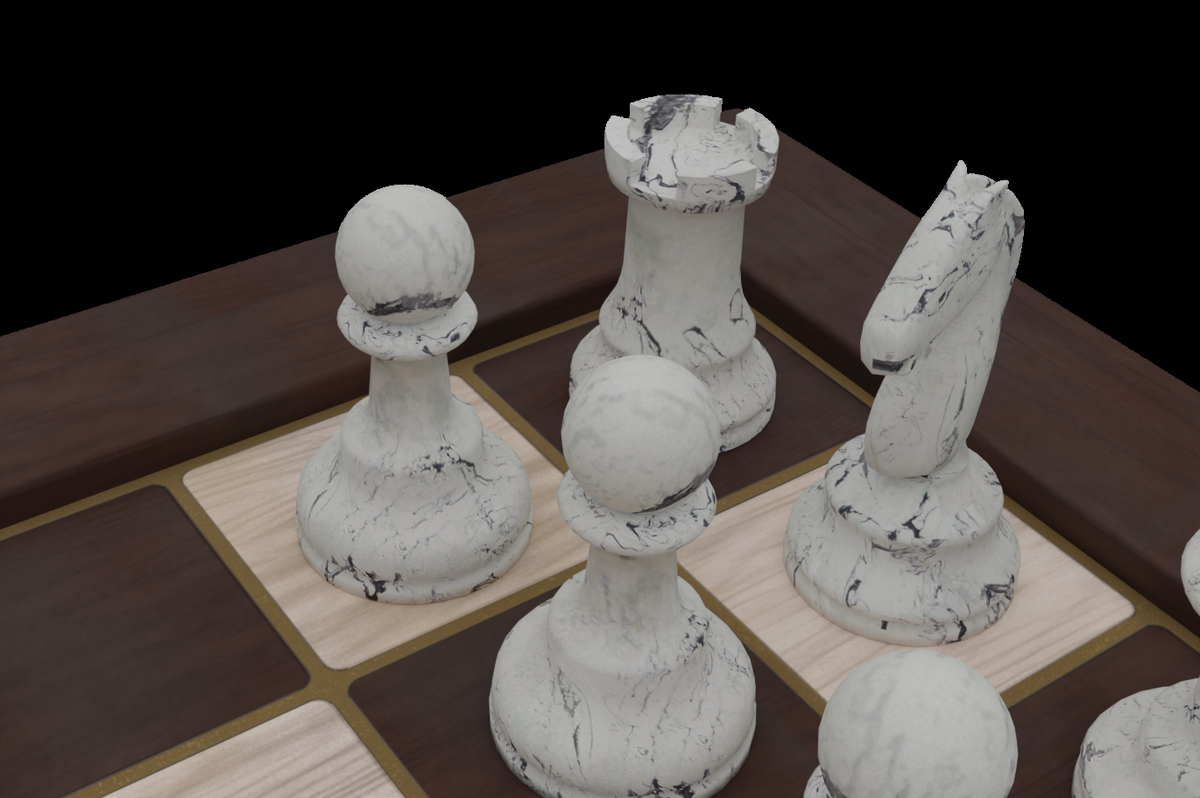}
        \end{minipage}%
        \hspace{0.01\linewidth}%
        \begin{minipage}[b]{0.23\linewidth}
            \centering
            \includegraphics[width=\linewidth]{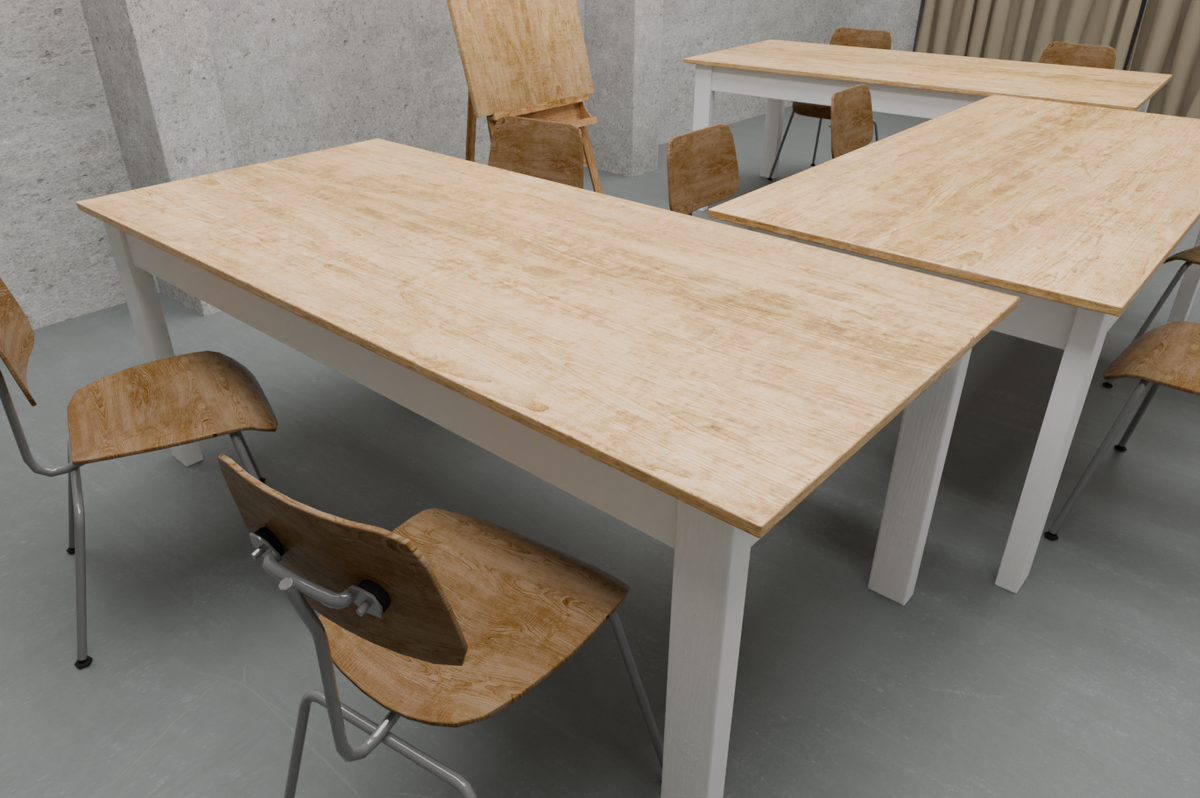}
        \end{minipage}%
    \end{minipage}%
    \caption{\textbf{Qualitative evaluation on synthetic scenes.} Each column corresponds to a different scene (\textsc{Temple}, \textsc{Classroom}, \textsc{Chessboard}, and \textsc{Office}), and each row shows results from different methods: Nerfbusters~\cite{warburg_nerfbusters_2023}, Bayes' Rays~\cite{goli_BayesRays_2023}, 3DGS*~\cite{kerbl_3Dgaussians_2023}, Ours, and the Ground Truth.}
    \label{fig:qualitative_synthetic}
\end{figure*}

\begin{figure*}[p]
    \centering
    \begin{minipage}[b]{\linewidth}
        \rotatebox{90}{\hspace{1.0cm}Nerfbusters\phantom{y}}
        \begin{minipage}[b]{0.32\linewidth}
            \centering
            \textsc{MeetingRoom}\\
                \vspace{0.15cm}
            \includegraphics[width=\linewidth]{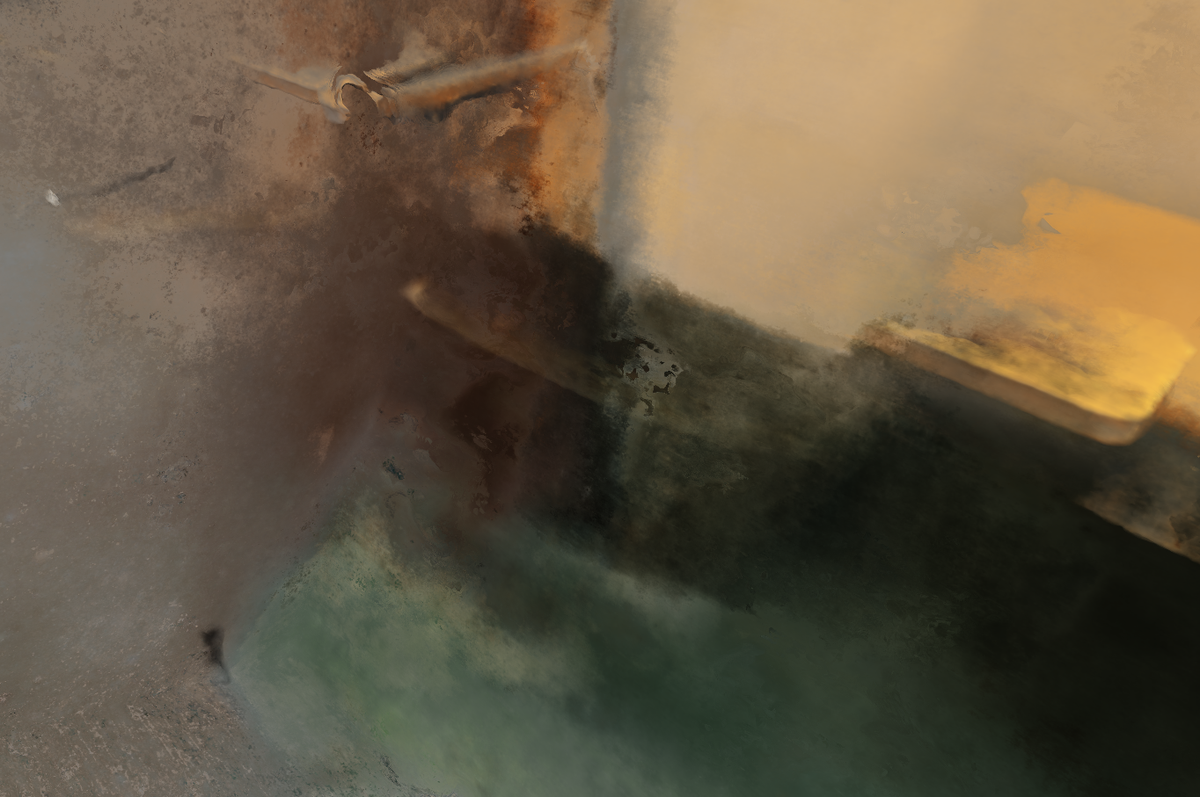}
        \end{minipage}%
        \hspace{0.01\linewidth}%
        \begin{minipage}[b]{0.32\linewidth}
            \centering
            \textsc{Pillars}\\
                \vspace{0.15cm}
            \includegraphics[width=\linewidth]{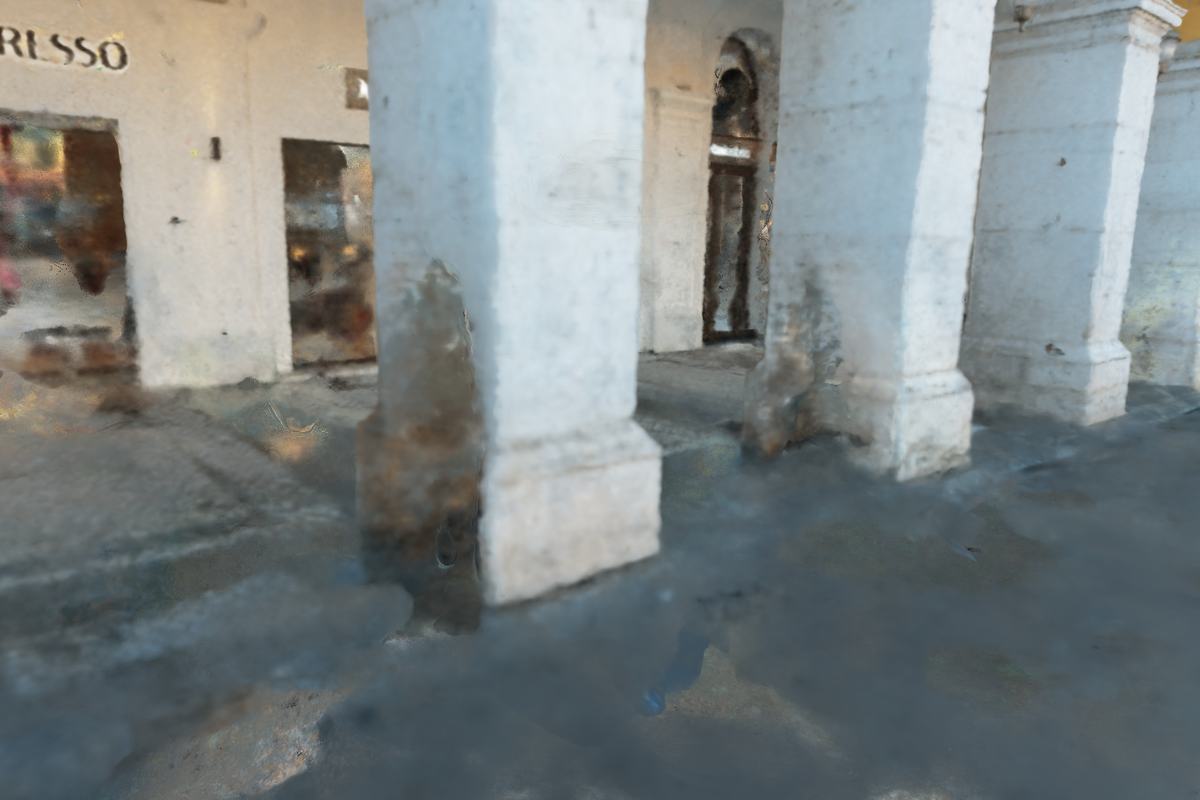}
        \end{minipage}%
        \hspace{0.01\linewidth}%
        \begin{minipage}[b]{0.32\linewidth}
            \centering
            \textsc{Facade}\\
                \vspace{0.15cm}
            \includegraphics[width=\linewidth]{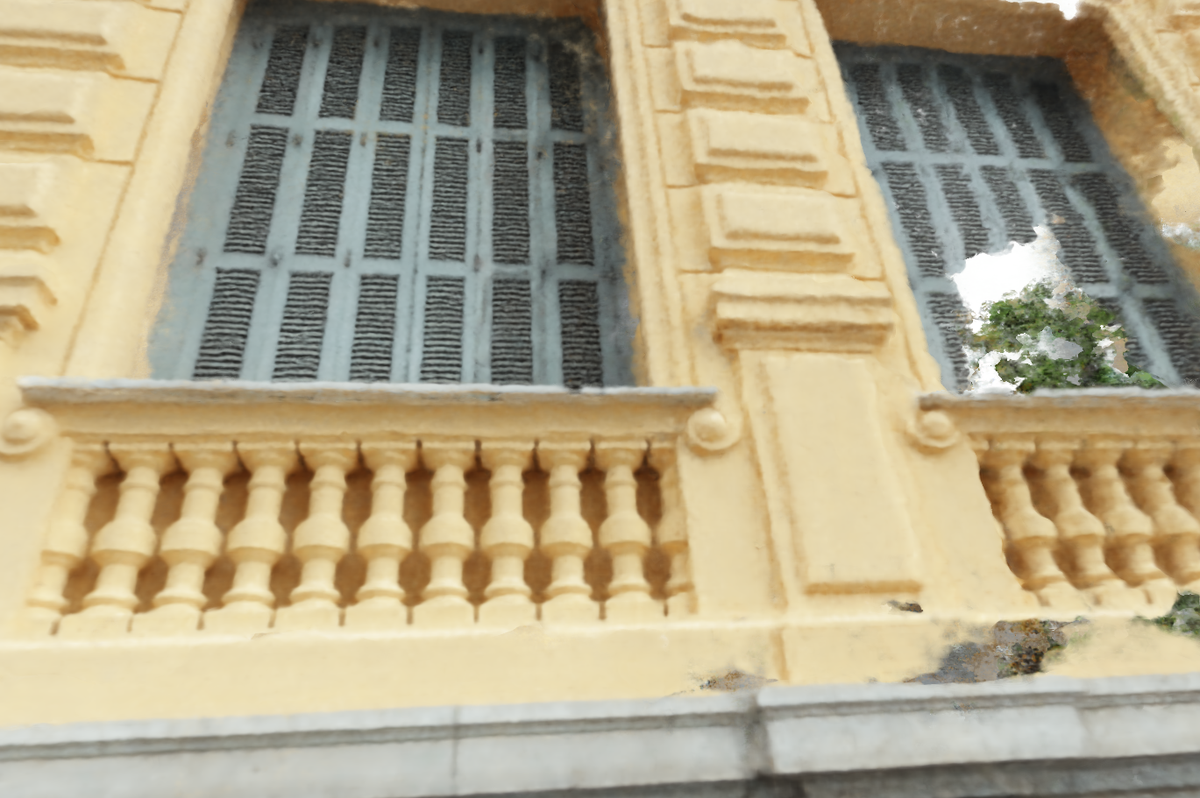}
        \end{minipage}%
        \hspace{0.01\linewidth}%
    \end{minipage}\\
    \vspace{0.15cm}
    \begin{minipage}[b]{\linewidth}
        \rotatebox{90}{\hspace{1.0cm}Bayes' Rays}
        \begin{minipage}[b]{0.32\linewidth}
            \centering
            \includegraphics[width=\linewidth]{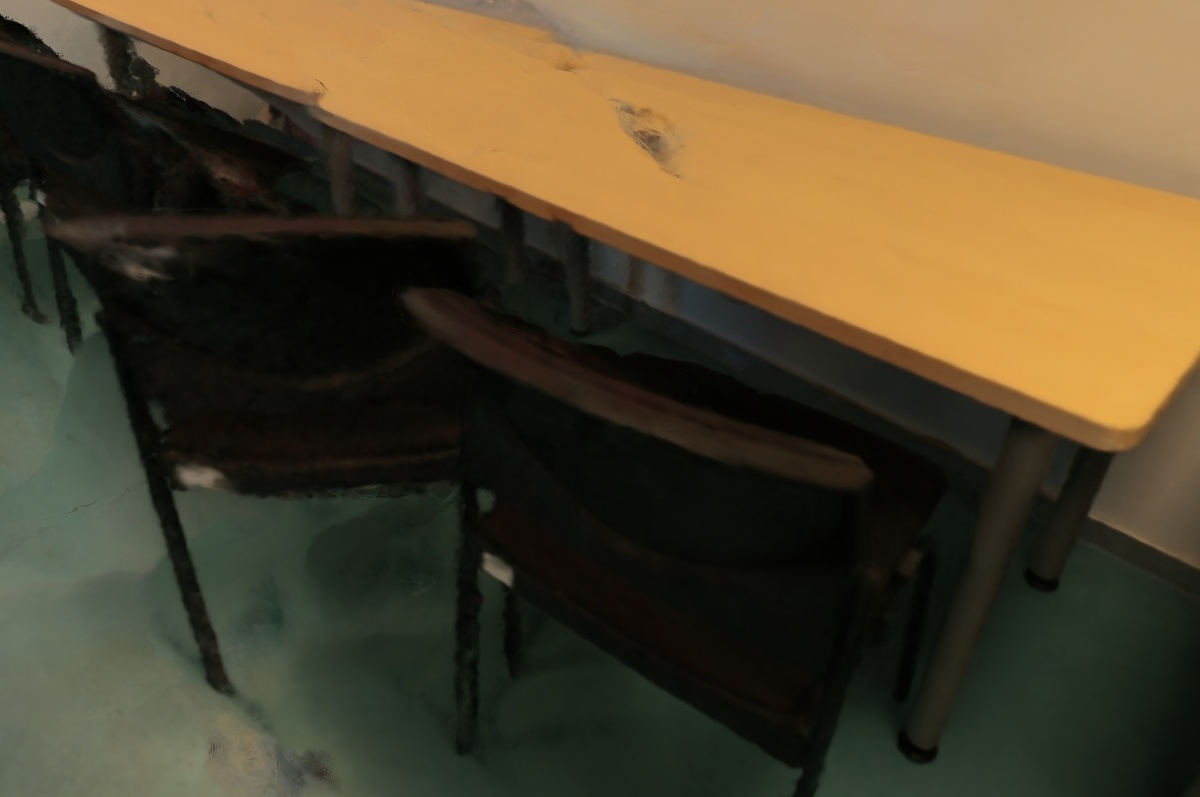}
        \end{minipage}%
        \hspace{0.01\linewidth}%
        \begin{minipage}[b]{0.32\linewidth}
            \centering
            \includegraphics[width=\linewidth]{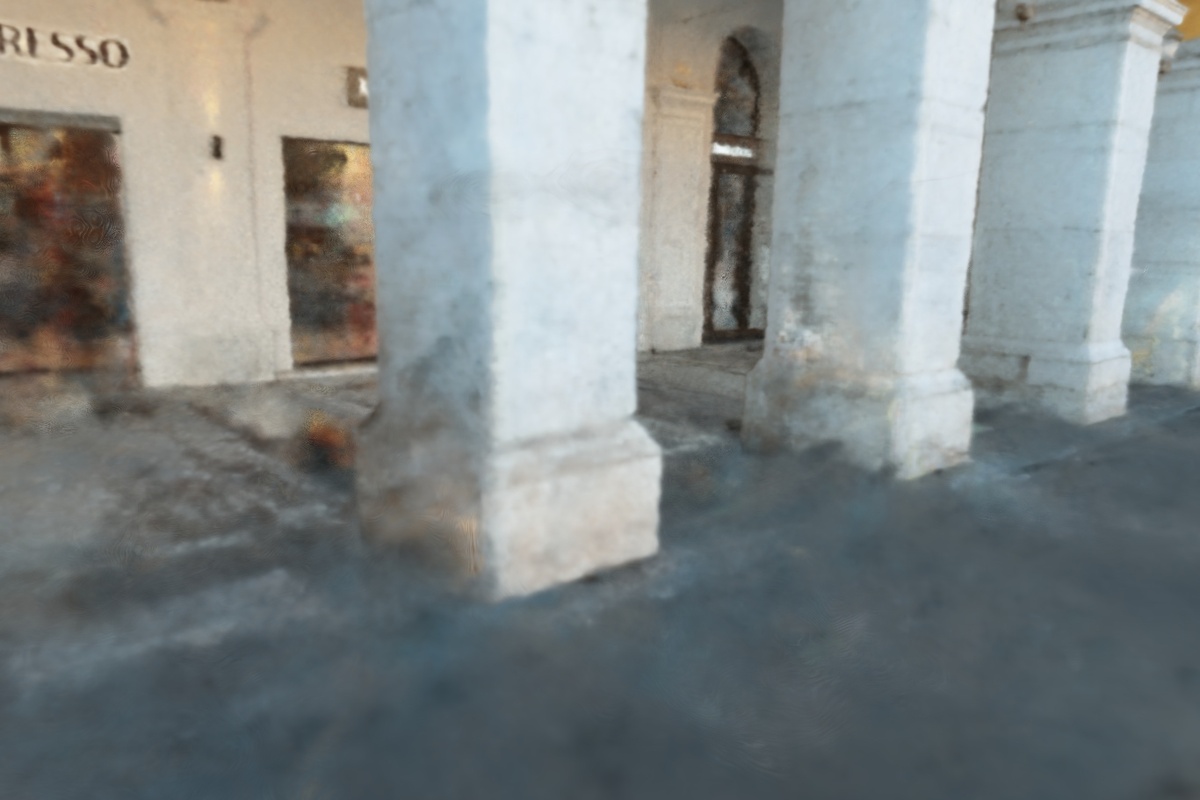}
        \end{minipage}%
        \hspace{0.01\linewidth}%
        \begin{minipage}[b]{0.32\linewidth}
            \centering
            \includegraphics[width=\linewidth]{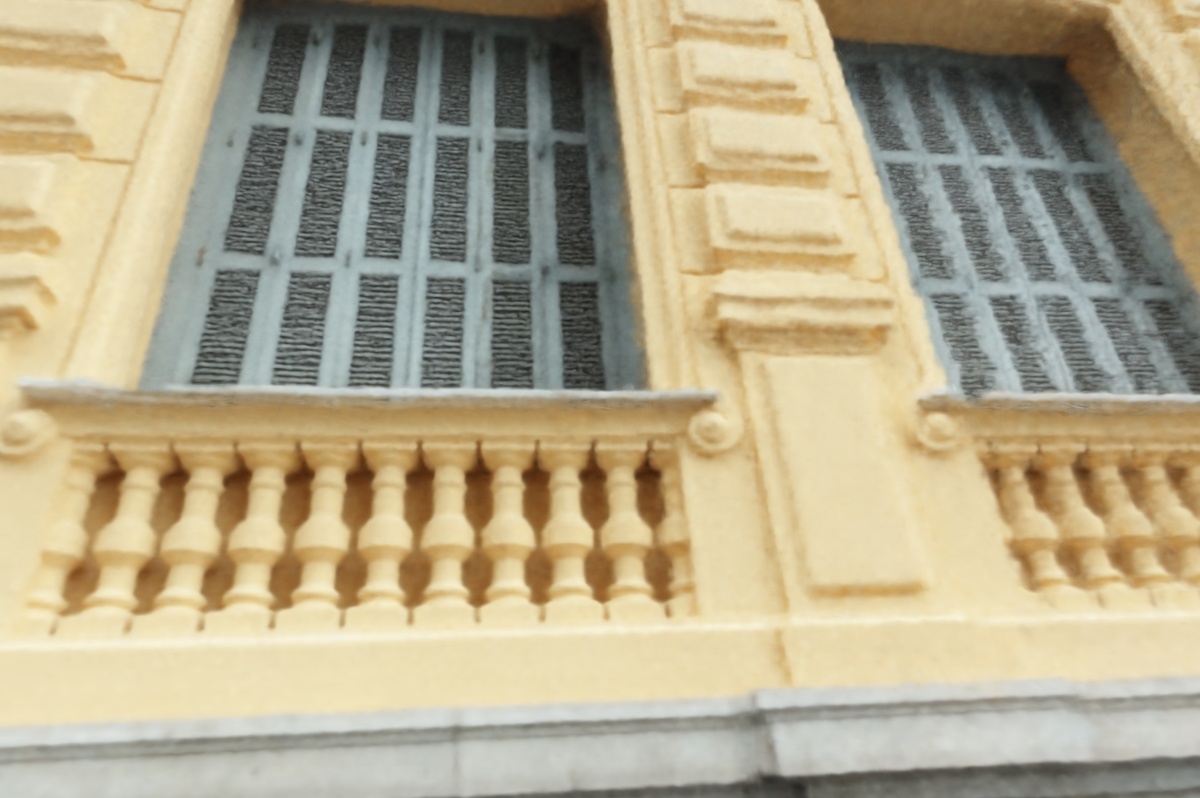}
        \end{minipage}%
        \hspace{0.01\linewidth}%
    \end{minipage}\\
        \vspace{0.15cm}
    \begin{minipage}[b]{\linewidth}
        \rotatebox{90}{\hspace{1.3cm}3DGS*\phantom{y}}
        \begin{minipage}[b]{0.32\linewidth}
            \centering
            \includegraphics[width=\linewidth]{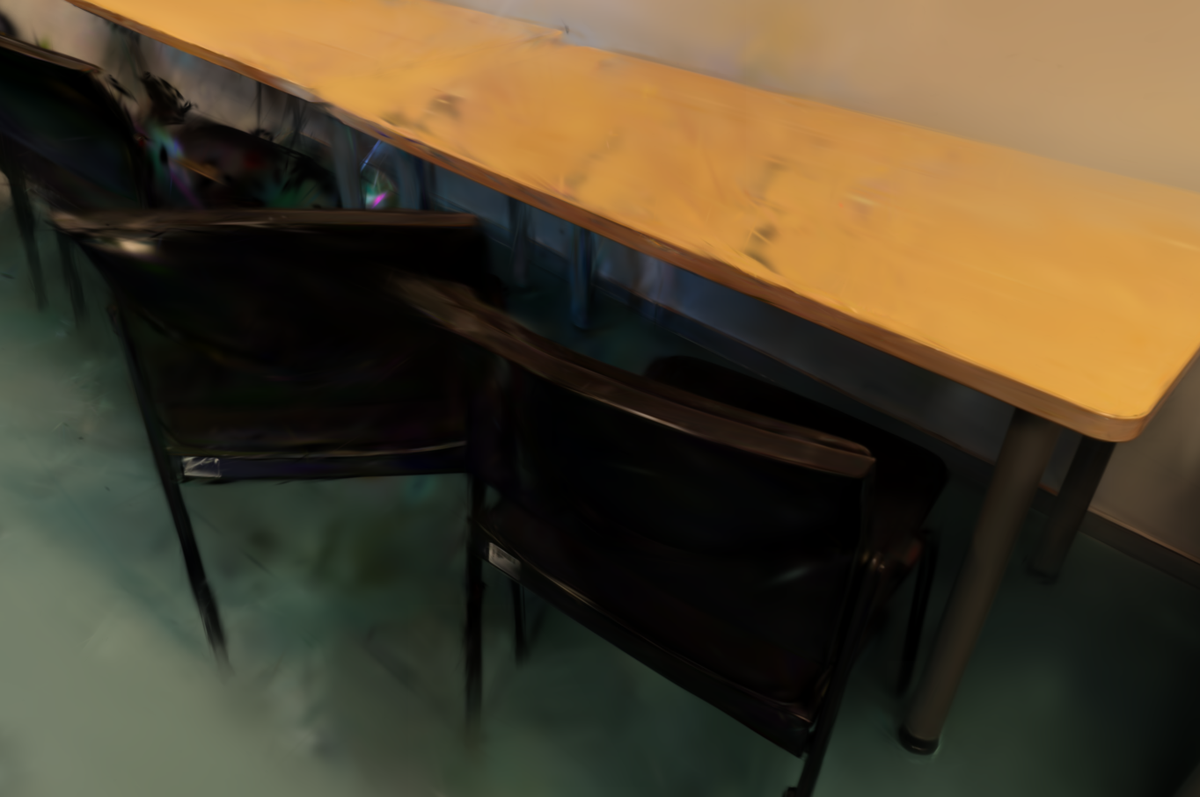}
        \end{minipage}%
        \hspace{0.01\linewidth}%
        \begin{minipage}[b]{0.32\linewidth}
            \centering
            \includegraphics[width=\linewidth]{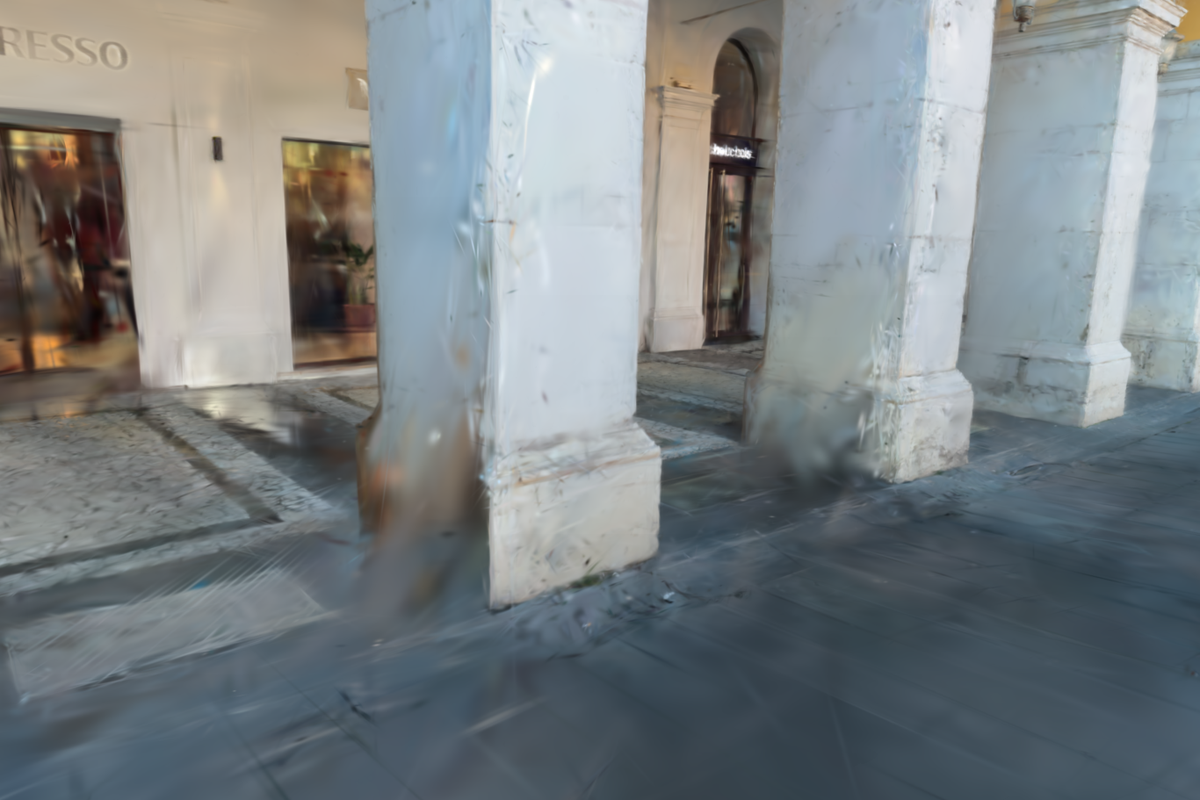}
        \end{minipage}%
        \hspace{0.01\linewidth}%
        \begin{minipage}[b]{0.32\linewidth}
            \centering
            \includegraphics[width=\linewidth]{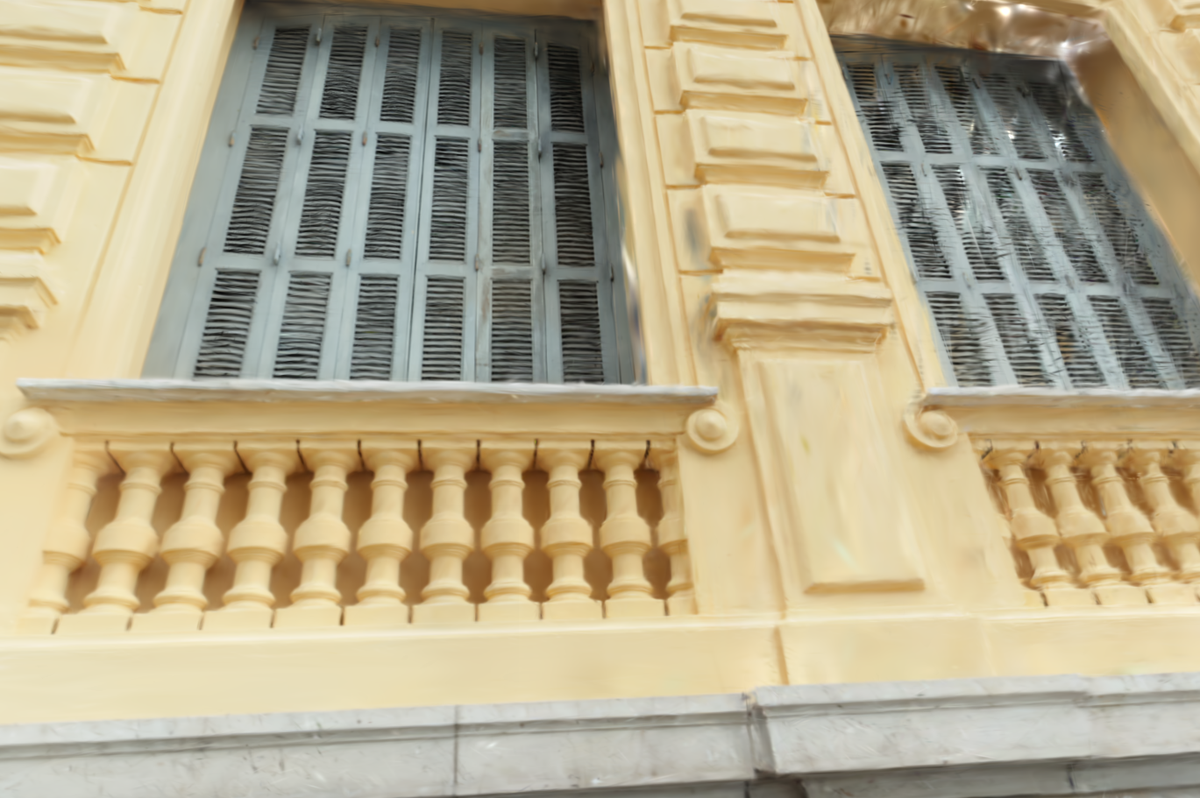}
        \end{minipage}%
        \hspace{0.01\linewidth}%
    \end{minipage}\\%
        \vspace{0.15cm}
    \begin{minipage}[b]{\linewidth}
        \rotatebox{90}{\hspace{1.3cm}Ours\phantom{y}}
        \begin{minipage}[b]{0.32\linewidth}
            \centering
            \includegraphics[width=\linewidth]{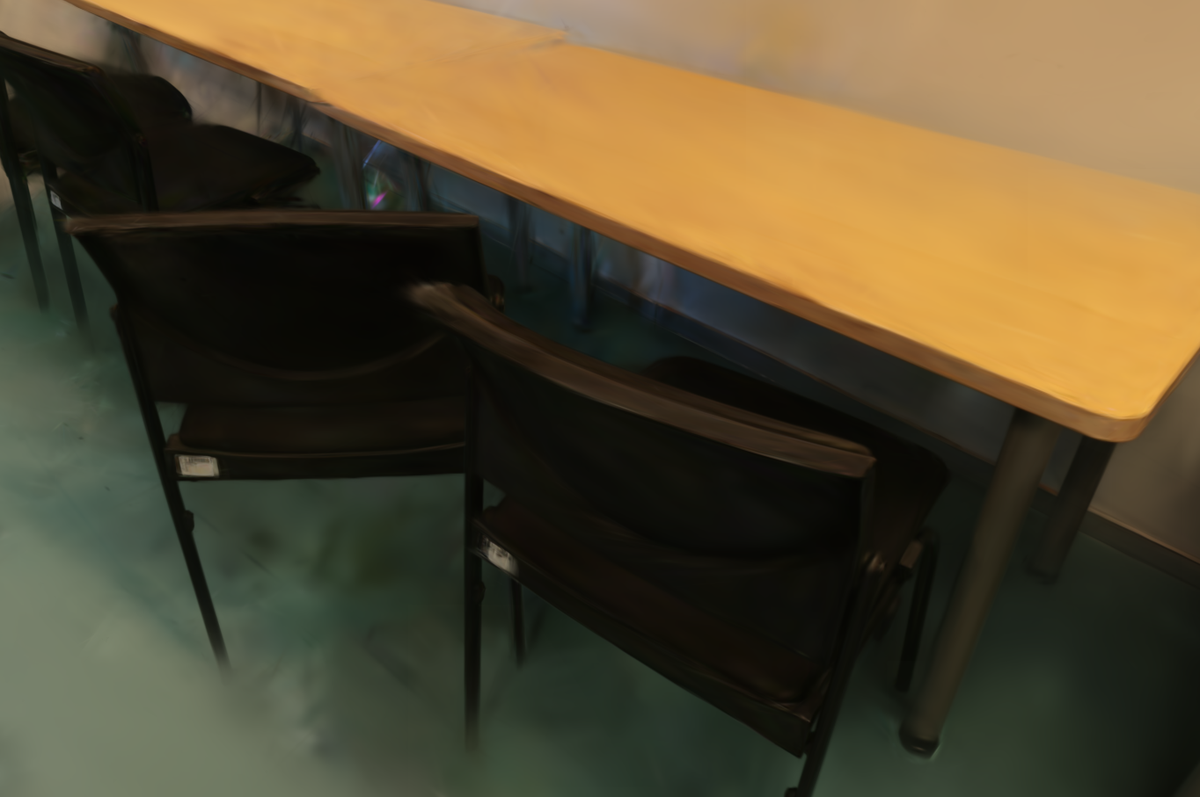}
        \end{minipage}%
        \hspace{0.01\linewidth}%
        \begin{minipage}[b]{0.32\linewidth}
            \centering
            \includegraphics[width=\linewidth]{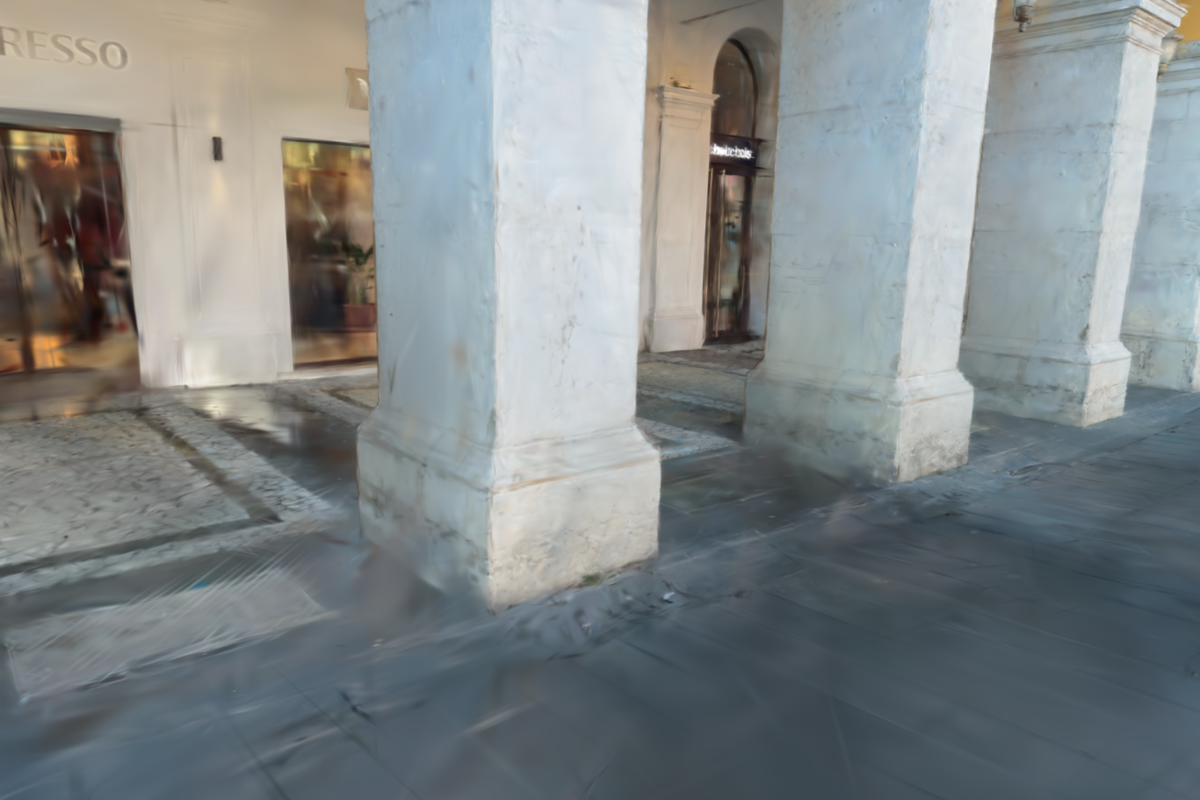}
        \end{minipage}%
        \hspace{0.01\linewidth}%
        \begin{minipage}[b]{0.32\linewidth}
            \centering
            \includegraphics[width=\linewidth]{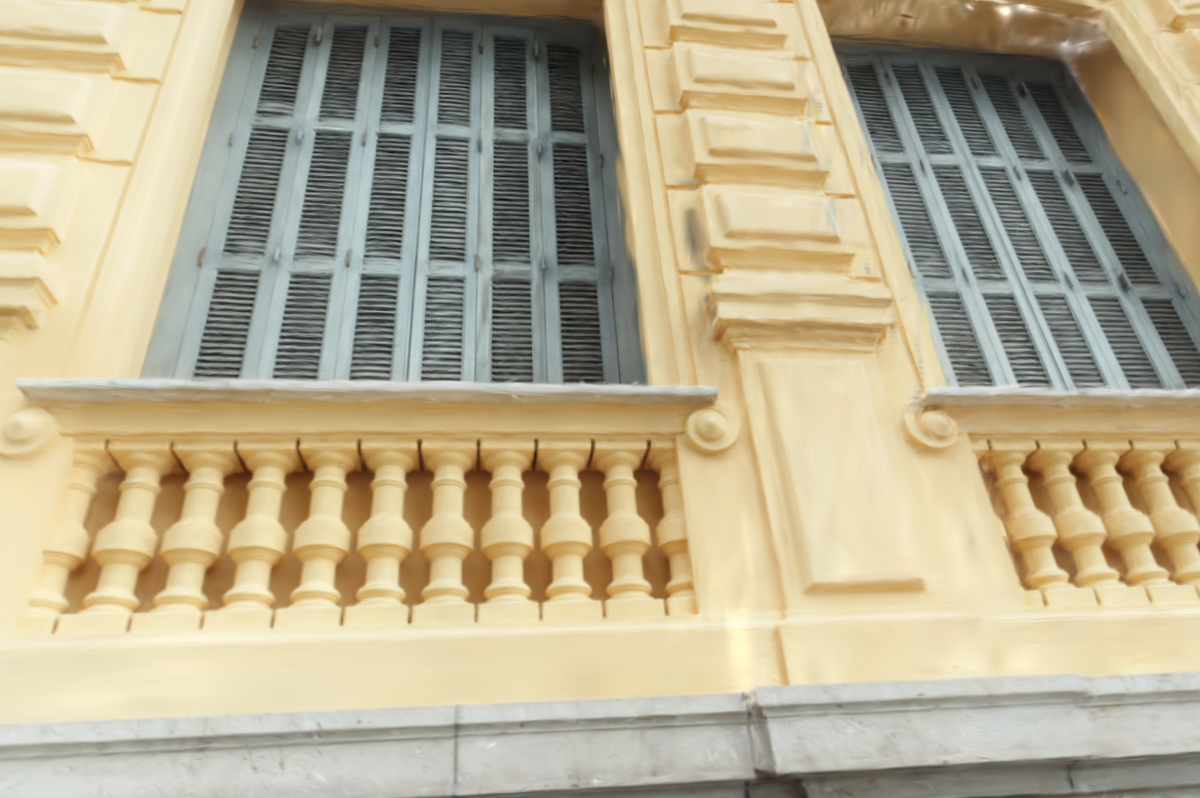}
        \end{minipage}%
        \hspace{0.01\linewidth}%
    \end{minipage}\\
        \vspace{0.15cm}
    \begin{minipage}[b]{\linewidth}
        \rotatebox{90}{\hspace{1.0cm}Ground Truth\phantom{y}}
        \begin{minipage}[b]{0.32\linewidth}
            \centering
            \includegraphics[width=\linewidth]{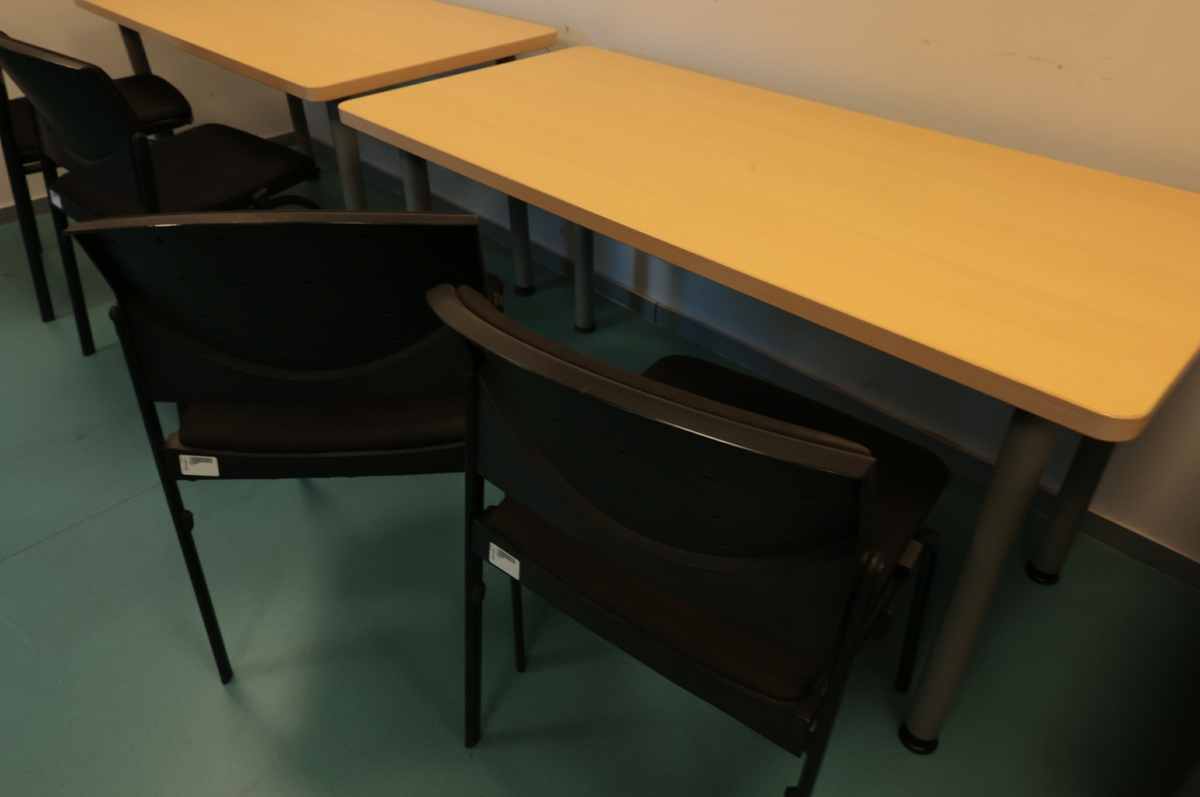}
        \end{minipage}%
        \hspace{0.01\linewidth}%
        \begin{minipage}[b]{0.32\linewidth}
            \centering
            \includegraphics[width=\linewidth]{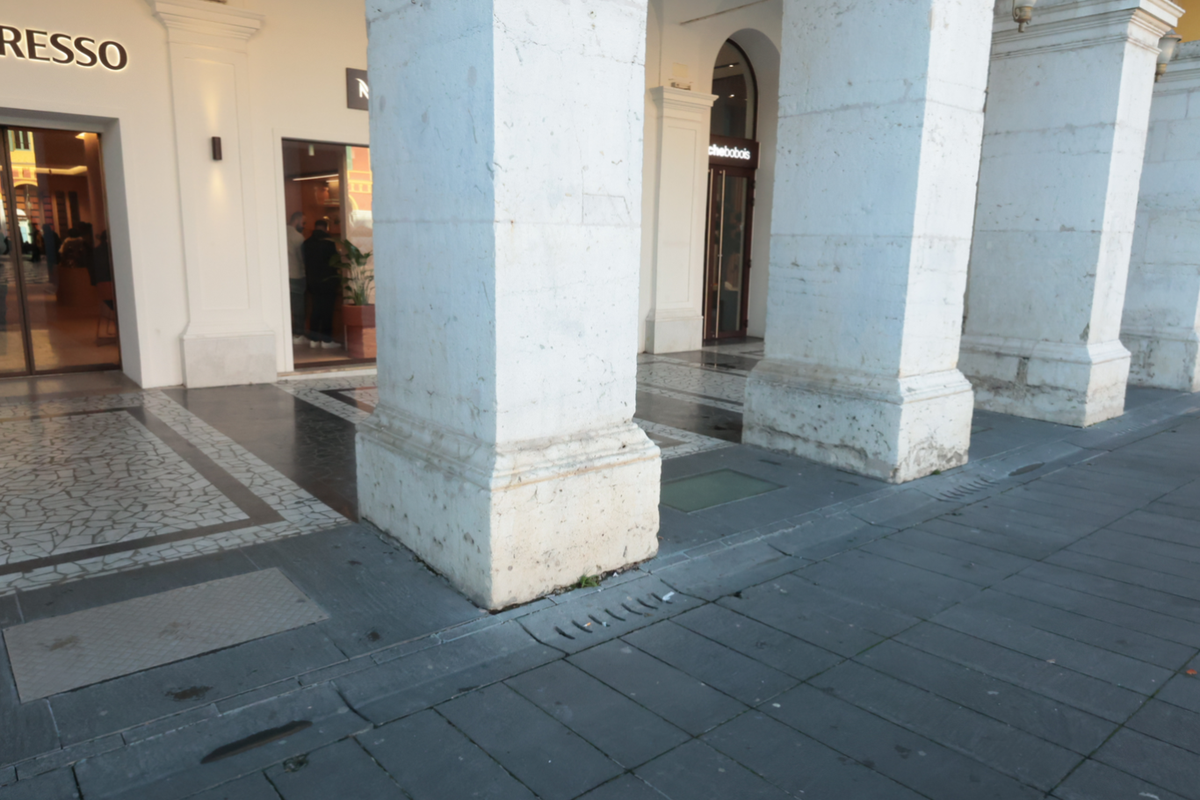}
        \end{minipage}%
        \hspace{0.01\linewidth}%
        \begin{minipage}[b]{0.32\linewidth}
            \centering
            \includegraphics[width=\linewidth]{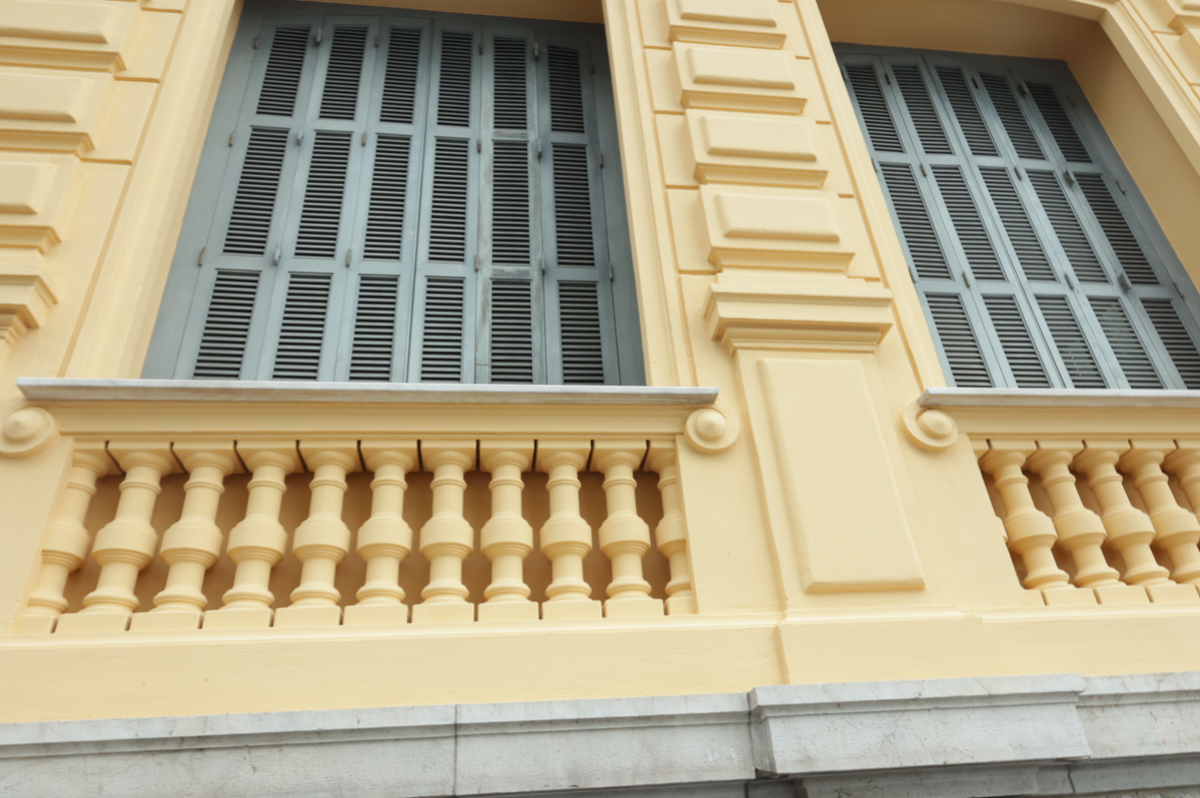}
        \end{minipage}%
        \hspace{0.01\linewidth}%
    \end{minipage}%
    \caption{\textbf{Qualitative evaluation on real scenes.} Each column corresponds to a different scene (\textsc{MeetingRoom}, \textsc{Pillars}, and \textsc{Facade}), and each row shows results from different methods: Nerfbusters~\cite{warburg_nerfbusters_2023}, Bayes' Rays~\cite{goli_BayesRays_2023}, 3DGS*~\cite{kerbl_3Dgaussians_2023}, Ours, and the Ground Truth.}
    \label{fig:qualitative_real}
\end{figure*}

\begin{figure*}
    \begin{minipage}[b]{\linewidth}
        \rotatebox{90}{\hspace{1.3cm}3DGS*\phantom{y}}
        \begin{minipage}[b]{0.32\linewidth}
            \centering
            \includegraphics[width=\linewidth]{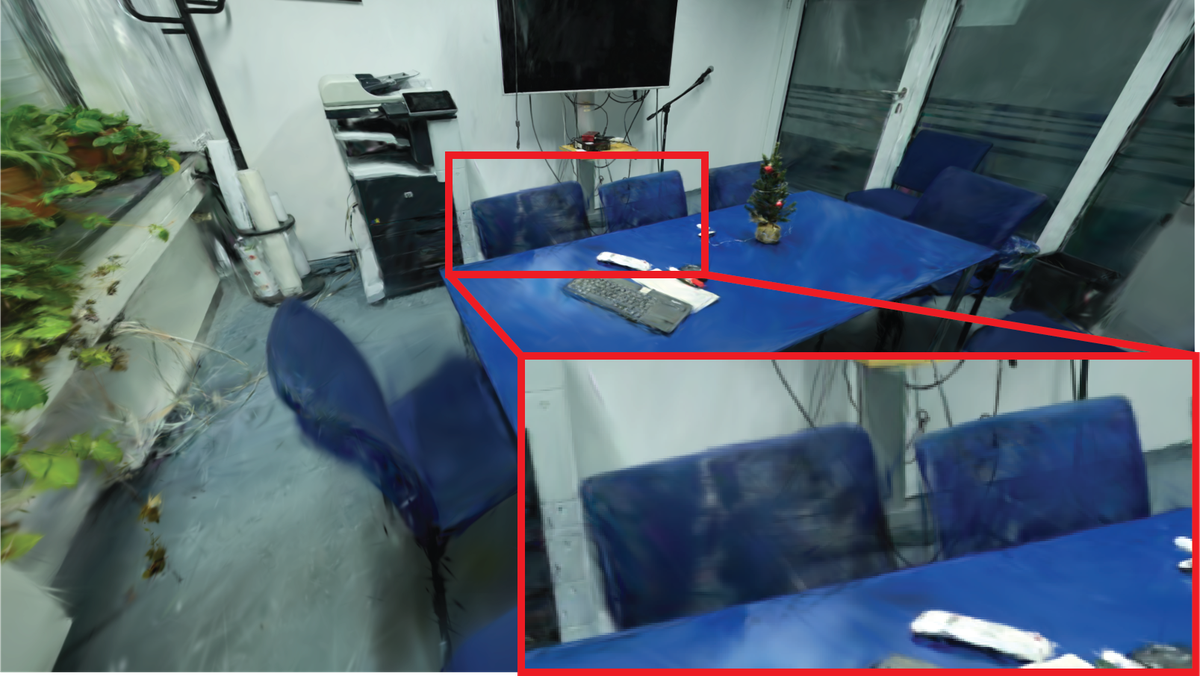}
        \end{minipage}%
        \hspace{0.01\linewidth}%
        \begin{minipage}[b]{0.32\linewidth}
            \centering
            \includegraphics[width=\linewidth]{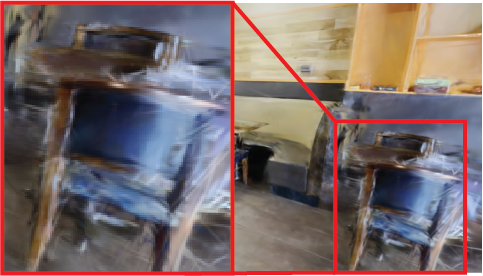}
        \end{minipage}%
        \hspace{0.01\linewidth}%
        \begin{minipage}[b]{0.32\linewidth}
            \centering
            \includegraphics[width=\linewidth]{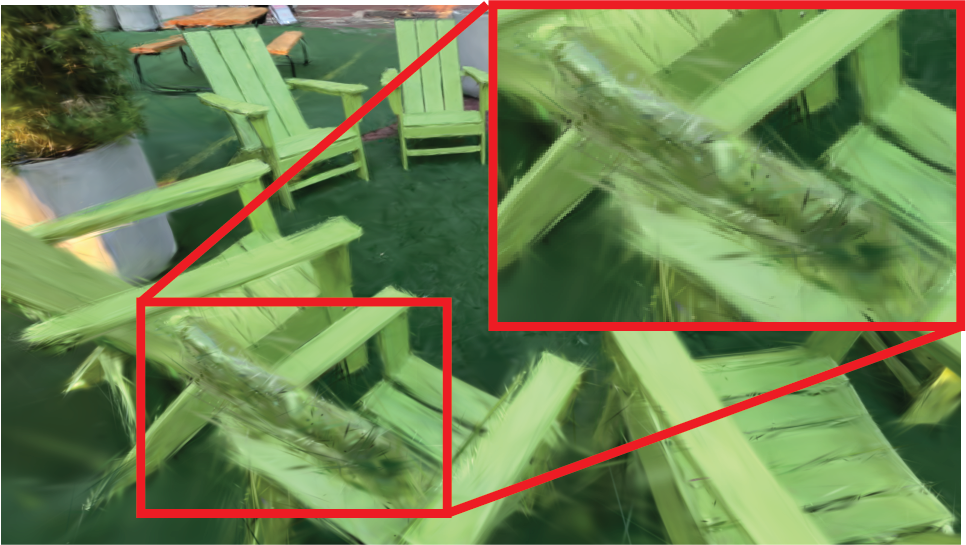}
        \end{minipage}%
        \hspace{0.01\linewidth}%
    \end{minipage}\\%
        \vspace{0.15cm}
    \begin{minipage}[b]{\linewidth}
        \rotatebox{90}{\hspace{1.3cm}Ours\phantom{y}}
        \begin{minipage}[b]{0.32\linewidth}
            \centering
            \includegraphics[width=\linewidth]{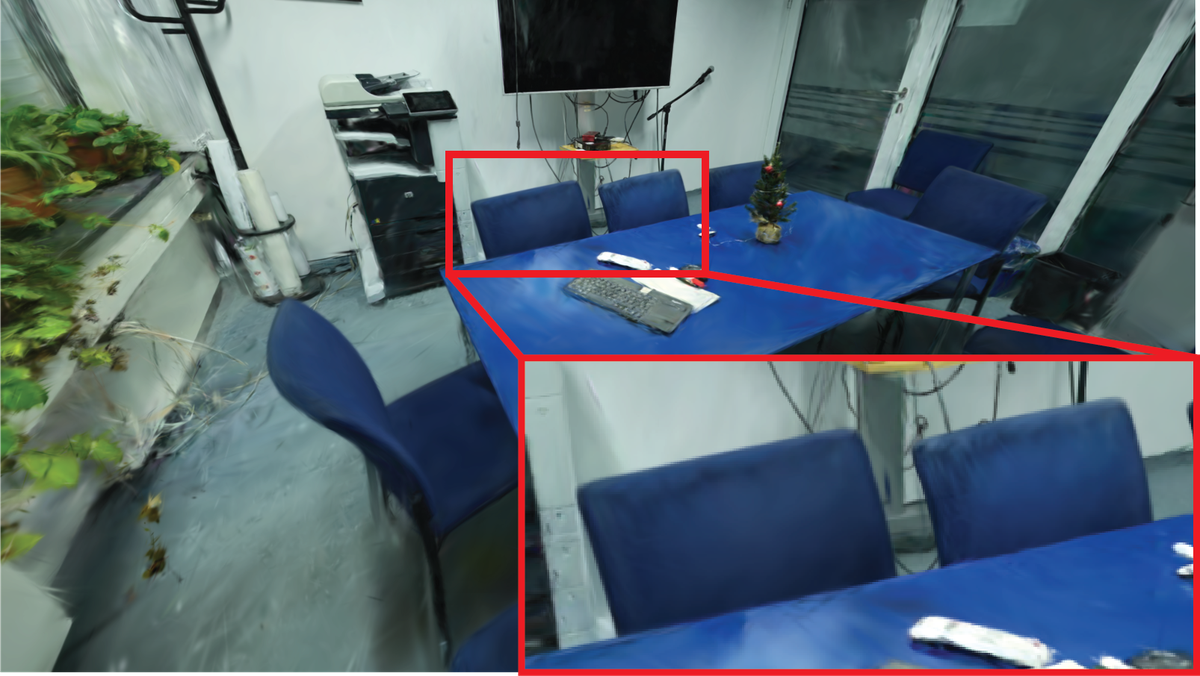}
        \end{minipage}%
        \hspace{0.01\linewidth}%
        \begin{minipage}[b]{0.32\linewidth}
            \centering
            \includegraphics[width=\linewidth]{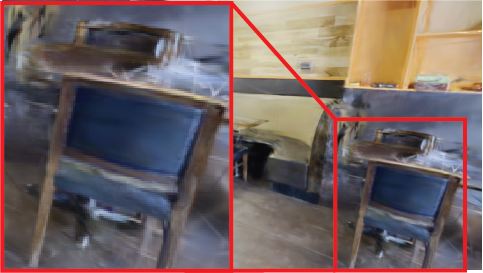}
        \end{minipage}%
        \hspace{0.01\linewidth}%
        \begin{minipage}[b]{0.32\linewidth}
            \centering
            \includegraphics[width=\linewidth]{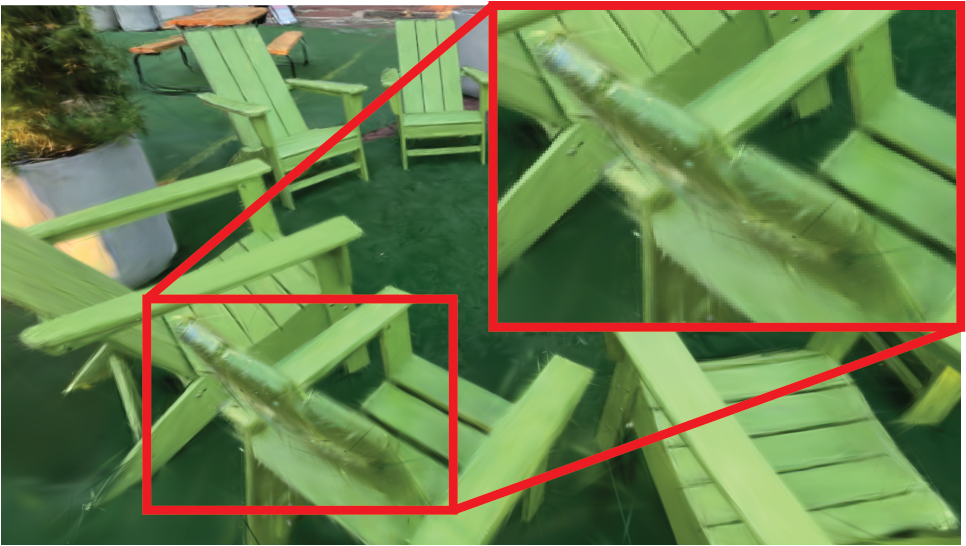}
        \end{minipage}%
        \hspace{0.01\linewidth}%
    \end{minipage}\\
        \vspace{0.15cm}
    \begin{minipage}[b]{\linewidth}
        \rotatebox{90}{\hspace{1.0cm}Ground Truth\phantom{y}}
        \begin{minipage}[b]{0.32\linewidth}
            \centering
            \includegraphics[width=\linewidth]{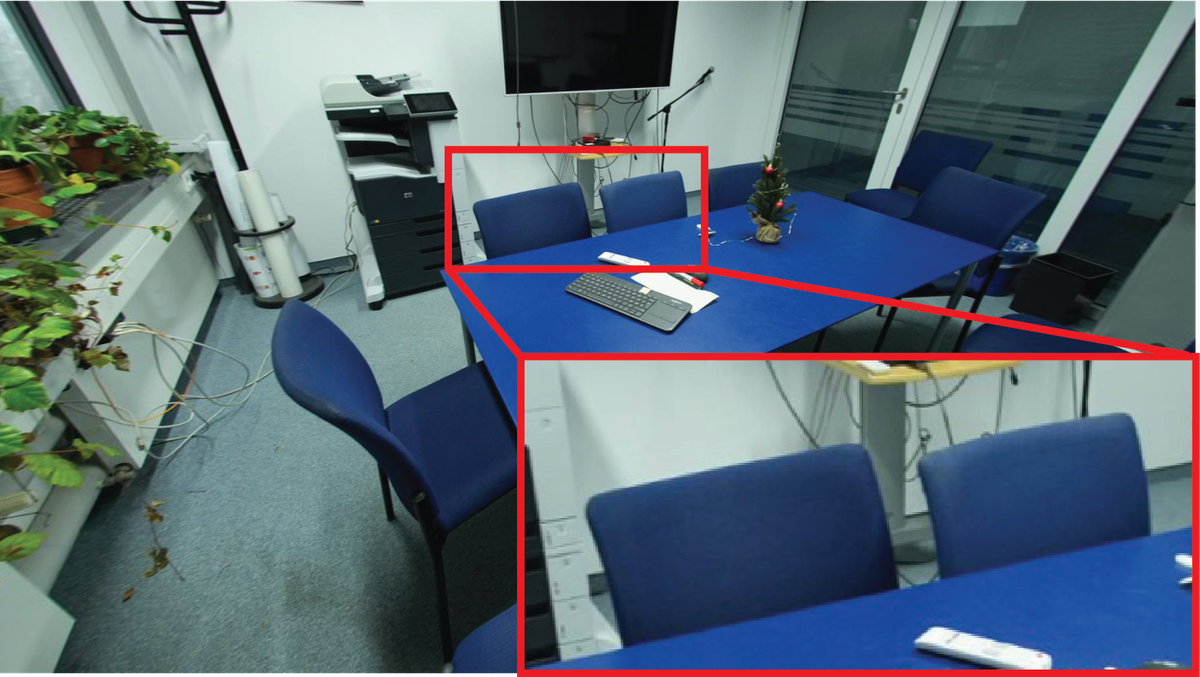}
        \end{minipage}%
        \hspace{0.01\linewidth}%
        \begin{minipage}[b]{0.32\linewidth}
            \centering
            \includegraphics[width=\linewidth]{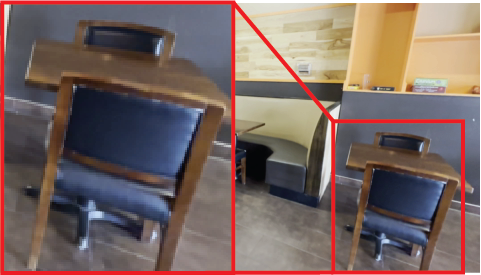}
        \end{minipage}%
        \hspace{0.01\linewidth}%
        \begin{minipage}[b]{0.32\linewidth}
            \centering
            \includegraphics[width=\linewidth]{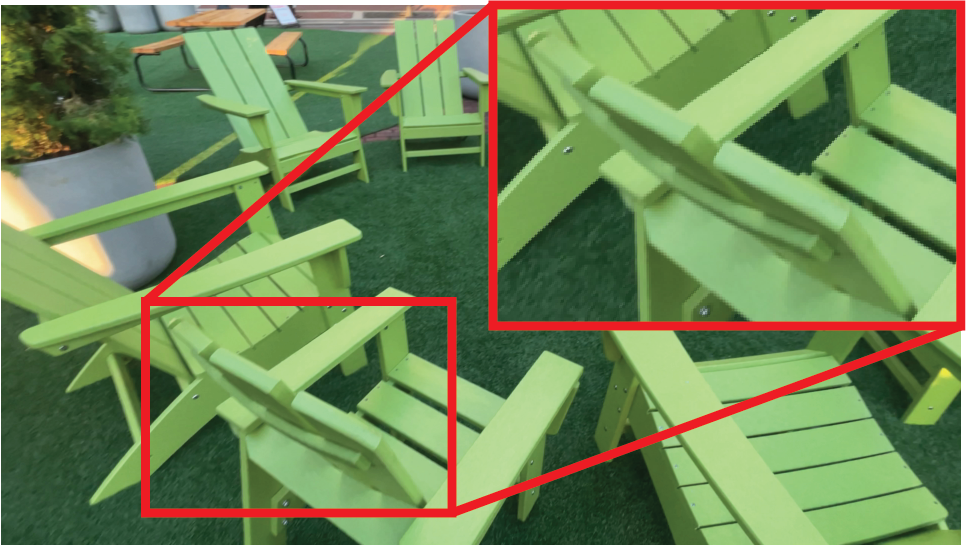}
        \end{minipage}%
        \hspace{0.01\linewidth}%
    \end{minipage}%
    \caption{\textbf{Qualitative evaluation on real scenes from ScanNet++ and DL3DV} Each column corresponds to a different scene (fist column from ScanNet++~\cite{yeshwanth2023scannet++}, second and third from DL3DV~\cite{ling2024dl3dv}), and each row shows results from different methods: 
    3DGS*~\cite{kerbl_3Dgaussians_2023}, Ours, and the Ground Truth.}
    \label{fig:qualitative_extra}
\end{figure*}

\textit{Qualitative Evaluation.} We show different unseen test views in  Fig.~\ref{fig:qualitative_real} for real scenes, Fig.~\ref{fig:qualitative_synthetic} for synthetic scenes, and Fig~\ref{fig:qualitative_extra} for ScanNet++/DL3DV. Our reconstruction, based on a shared representation for repetitive elements, shows better overall appearance for the repetitive objects. For instance, in \textsc{Office} scene of Fig.~\ref{fig:qualitative_synthetic}, the chairs and table have severe occlusion in the training views, leading to poor reconstruction by the baselines. In contrast, our shared representation incorporates the information from visible instances to improve the reconstruction. \\

\begin{figure}
    \centering
    \includegraphics[width=\columnwidth]{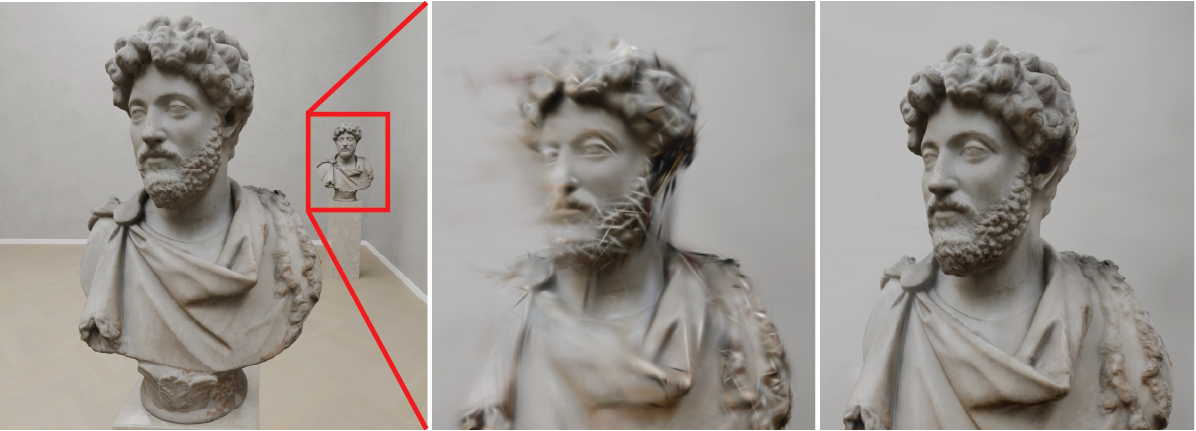}
    \begin{minipage}[b]{0.33\linewidth} 
        \centering
        Train view
    \end{minipage}%
    \begin{minipage}[b]{0.33\linewidth} 
        \centering
        \hspace{0.2cm}3DGS*
    \end{minipage}%
    \begin{minipage}[b]{0.33\linewidth} 
        \centering
        \hspace{0.2cm}Ours
    \end{minipage}
    \caption{\textbf{Zoom-in.} Our method improves the reconstruction of the bust in the background, allowing for a successful close-up shot (\textit{middle and right}). This happens because a repetition of the bust is seen up-close in the training views, benefiting the shared representation (\textit{left}).}
\label{fig:zoom_in}
\end{figure}

\textit{Zooming In.} Our method can also render detailed zoomed-in novel view of objects that are seen from afar in the training sequence, if one of its repetition is seen from up close.  We show this in Fig.~\ref{fig:zoom_in}, where the reconstruction of the bust in the background is improved using the bust in the front. Our method removes artifacts and adds the fine-grain details that were missing in the 3DGS* reconstruction, due to this instance being captured from far away.\\

\textit{Quantitative Evaluation.} Our quantitative results are summarized in Table~\ref{tab:quantitative}, on the synthetic and real scenes. We report the standard reconstruction error metrics PSNR and SSIM~\cite{wang_ssim_2003}, and the perceptual metric LPIPS~\cite{zhang_lpips_2018}. In addition, we compute KID~\cite{binkowski_demystifying_2021} to compare the distribution of rendered and ground truth views, which is typically used to evaluate the overall realism of rendered images. We compute metrics on the full rendered test views, as well as on the masked regions corresponding to repetitions to isolate the impact of our method on the repetitive elements. Our visual improvement translates directly to the quantitative results, where we improve on all metrics by a large margin. Most importantly, on the real scenes, our approach improves by 1.59 dB the PSNR compared to the second-best performing method and 2.72 dB if only considering the masked region with repetitions. For the ScanNet++/DL3DV scenes we obtain an average improvement of 1.28 dB in PSNR in the masked regions.

\subsection{Analysis}\label{subsec:analysis}

\begin{figure}
    \centering
        \includegraphics[width=\columnwidth]{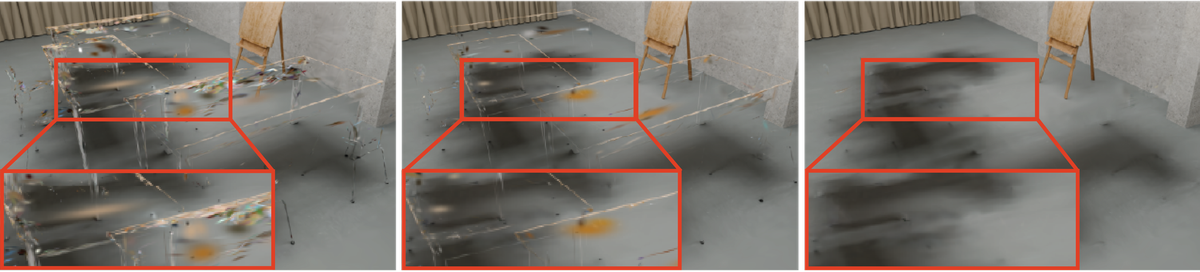}
    \caption{\textbf{Segmentation ablation.} Without opacity and scale regularization or post-processing (\textit{left}) more gaussians are left over. Using the regularization (without post-processing) shows a reduced number of remaining gaussians (\textit{middle}). Our complete solution with both regularization and post-processing produces the cleanest results (\textit{right) .}
	}
\label{fig:ablation_seg}
\end{figure}

\textit{Segmentation.} In Fig.~\ref{fig:ablation_seg}, we qualitatively ablate our opacity and scale regularization (see Eq.~\ref{eq:l1_reg_opacity_scale}) and our space carving post-processing. As shown on the left, without the opacity and scale regularization, many Gaussian primitives fail to meet the contrastive features threshold, and remain where the object was. Our regularization reduces the number of trailing Gaussians (middle) but some remain at the border of the object. We hypothesize that they do not pass the contrastive features threshold because they can contribute both to the appearance of the object and to the background, being at the border. Our space carving approach successfully removes them (right), achieving a clean segmentation. 
\\

\begin{figure}
    \centering
    \includegraphics[width=\columnwidth]{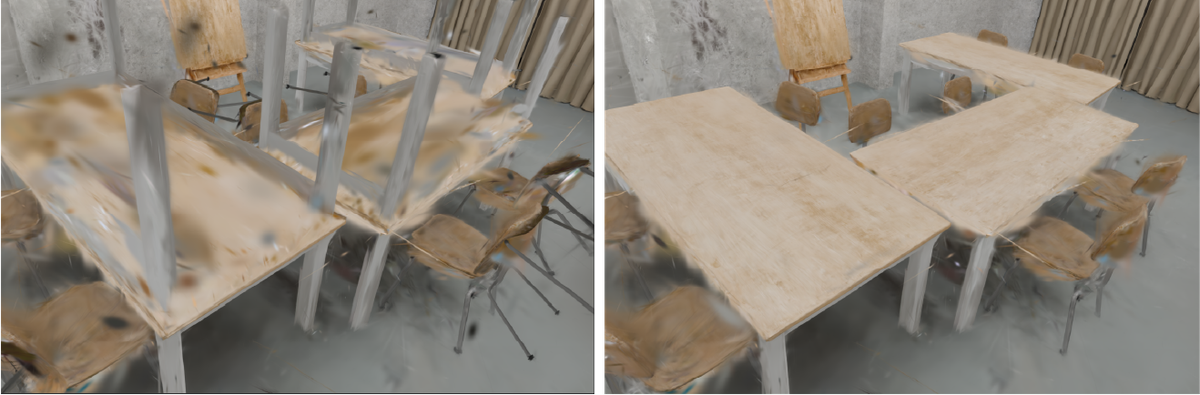}
    \caption{\textbf{Registration comparison.} We compare our registration scheme (\textit{right}) with FPFH~\cite{Rusu_histogram3DFPFH_2009} on the Gaussian primitives (\textit{left}). We replace each instance by the union of all its registered repetitions. 
    Note that the table and chair both have two modes for their pose for FPFH, while with our registration they are aligned. This is an intermediary visualization of our method, after which the shared representation is optimized with gradients flowing from all the repetitions.
    }\label{fig:ablation_register_3d}
\end{figure}

\textit{Registration.} 
We compare our registration procedure with a  global registration approach based on 3D features, FPFH~\cite{Rusu_histogram3DFPFH_2009} and a RANSAC procedure to register the two point clouds based on matching the features. Fig.~\ref{fig:ablation_register_3d} showcases the limitations of this approach in the context of 3DGS, where the point cloud data is very noisy and not concentrated only on surfaces. By replacing each instance with the union of its registered repetitions, we observe that the registered instances for the chair and tables are not aligned. This is confirmed by our quantitative analysis in Table~\ref{tab:registration}, where we measure the angular error of the alignment on the synthetic scenes. Our method averages a half-degree error, while~\cite{Rusu_histogram3DFPFH_2009} fails with 50 degrees of error.
We also ablate the main components of our registration approach. Using the train views instead of sampling new views in our virtual sphere degrades the performance by 4 degrees. Indeed, the train views cannot ensure finding a pair of views from a similar perspective and roughly from the same distance, which makes 2D matching harder for MASt3R. We also ablate the refinement step showing that performances degrades by 5 degrees without ICP. We also ablate the dense matching with MASt3R using SIFT~\cite{lowe_distinctive_2004} and SuperPoint~\cite{detone_superpoint_2018} as alternatives. SuperPoint failed in the \textsc{Temple} scene, we report the average on the other synthetic scenes. On real scenes, SuperPoint also failed in \textsc{Pillars}, while SIFT failed in \textsc{Facade} and \textsc{MeetingRoom}. These failures emphasize the need for a robust matcher like MASt3R.
\\

\textit{Shared Representation.}
A key element of the shared representation is its ability to adapt to the specific appearance of each instance via the SH offsets while aggregating information with a shared SH component. We report our ablations in Table~\ref{tab:ablations} on synthetic scenes. If no SH offsets are used, the shared representation cannot adapt to each instance and learns a suboptimal appearance.
On the other hand, when no shared SHs are used, the representation does not aggregate the information from all instances effectively, resulting in degrading quality, and lower PSNR.  

We also experimented with a coordinate-based MLP shared across instances, that takes as input a Gaussian primitive position in the shared template, the rigid transformation of the instance, the viewing direction and the SHs in the base 3DGS reconstruction for this primitive, and outputs the RGB color of the gaussian primitive. We observed similar quality but found it more than one and a half times slower.  Additionally, in this representation, the reflectance's shared component and the instance-specific components are not disentangled.
\\

\textit{Computational Cost.}
Our method extends a base 3DGS representation, requiring additional training time for contrastive features and shared representation finetuning, similar to prior 3DGS extensions. We provide a time breakdown for each step with our default setup: (i) 3DGS: 22min, (ii) Contrastive features: 21min, (iii) 3D Segmentation: 19sec, (iv) Registration: 123sec, (v) Finetuning: 51min. \\

\hspace{-3.5mm}%
\begin{minipage}[t]{0.39\columnwidth}
    \vspace{-0.67cm}
    \begin{table}[H]
    \caption{\textbf{Shared representation ablation.} We report average PSNR across test views on two synthetic scenes. We also include average training iterations per second.}
    \vspace{-2mm}
    \label{tab:ablations}
    \resizebox{\linewidth}{!}{
    \begin{tabular}{llll}
    \toprule
     &  PSNR $\uparrow$ & it/s $\uparrow$ \\
    \midrule
    SHs  &  27.33  &  \textbf{2.6} \\
    \hspace{0.2cm} w/o offset  
    & 26.63  & \textbf{2.6}\\
    \hspace{0.2cm} w/o shared 
    & 25.19 & \textbf{2.6}\\
    MLP  & \textbf{27.34} & 1.5  \\
    \bottomrule
    \end{tabular}
    }
    \end{table}
\end{minipage}
\hspace{0.02\columnwidth}
\begin{minipage}[t]{0.59\columnwidth}
    \vspace{-0.67cm}
    \begin{table}[H]
    \caption{\textbf{Registration comparison and ablation.} On the synthetic dataset, we report mean absolute error (MAE) for the predicted rotation \textbf{R} (geodesic distance in degrees) and the translation vector \textbf{t}. $^{\dagger}$ Both ablations fail in some real scenes. SIFT also fails in \textsc{Temple}, which we exclude to compute the metrics.}
    \vspace{-2mm}
    \resizebox{\linewidth}{!}{
    \begin{tabular}{llll}
    \toprule
     &  \text{MAE}(\textbf{R}) $\downarrow$    &  \text{MAE}(\textbf{t})  $\downarrow$ \\
    \midrule
    Ours                                & 0.490 & 0.065 \\
    \hspace{0.2cm} w/o Virtual view     & 4.367 &  0.204 \\
    \hspace{0.2cm} w/o ICP              & 5.455 &  0.114  \\
    \hspace{0.2cm} w/ SIFT$^{\dagger}$         & 0.379 &  0.059  \\
    \hspace{0.2cm} w/ SuperPoint$^{\dagger}$        & 0.458 &  0.075  \\
    FPFH~\cite{Rusu_histogram3DFPFH_2009}                     & 50.58 &  2.547 \\
    \hspace{0.2cm} w/o ICP              & 66.94 &  2.606 \\
    \bottomrule
    \end{tabular}
    }
    \label{tab:registration}
    \end{table}
    \vspace{-3.2mm}
\end{minipage}%

\subsection{Limitations \& Future Work}
A natural limitation of our method is given by the number of repetitions in the scene and the variability they show. Our method  benefits the most when several objects are seen from different points of view: this provides a stronger signal for our shared representation. However, our method cannot improve the background, and some artifacts remain in the test views as a result. Also, the identification of the repetitive elements in a scene require user interaction.

Our method is limited by strong changes in illumination in the scene (for example, strong highlights in specular surfaces), which produce pronounced differences in appearance among instances of the same object. Scenes with uniform illumination are easier to handle with our method. Strong differences in appearance are still challenging to model and we leave their specialized treatment for future work.

As discussed in Sec ~\ref{subsec:analysis}, our method requires additional training for the contrastive features and the shared representation. Contrastive features can be trained in 4 min (1k iterations) instead of 21 min (5k), with minimal performance difference (0.969 vs. 0.966 mAcc, 0.965 vs. 0.963 mIoU). Furthermore, we can reduce finetuning iterations from 7k to 4k (29min) and keep the performance (PSNR=27.62@7k vs PSNR=27.63@4k in synthetic scenes). Recent advances like Taming 3DGS~\cite{mallick2024taming} should reduce the cost of contrastive training and finetuning.

An interesting direction for future work is to leverage the shared representation to reduce the memory requirements of 3DGS: if $N$ instances can be represented with a single and compact shared representation, the total number of Gaussians required for all instances is reduced. Here, the major challenge is how to encode the  appearance of each instance in a compact representation.

Another promising direction is improving inverse rendering with our shared representation. A direct extension would be to also share the same material parameters of a BRDF. The shared representation would receive multi-illumination information for all instances, helping material and illumination disentanglement.

\section{Conclusions}
 We show that repetitions in 3D scenes can be leveraged to improve reconstruction and novel view synthesis. After an initial 3DGS reconstruction, our method detects and fuses the multiple occurrences of a given object into a shared representation with common geometry and base appearance, while individual appearance is modeled as an offset for each instance. Then, each instance is replaced with this representation. Our key insight is that this shared representation and the offsets can be jointly optimized using all information available in the scene for that object, which improves the geometry and appearance for all instances.

\begin{acks}
This work was funded by the European Research Council (ERC) Advanced Grant NERPHYS, number 101141721 \url{https://project.inria.fr/nerphys/}. The authors are grateful to the OPAL infrastructure of the Universit\'e C\^ote d'Azur for providing resources and support, as well as Adobe and NVIDIA for software and hardware donations. 
The authors thank the anonymous reviewers for their valuable feedback.
\end{acks}

\renewcommand\UrlFont{\ttfamily\color{black}}

\bibliographystyle{ACM-Reference-Format}
\bibliography{refs}


\begin{thebibliography}{72}


\ifx \showCODEN    \undefined \def \showCODEN     #1{\unskip}     \fi
\ifx \showISBNx    \undefined \def \showISBNx     #1{\unskip}     \fi
\ifx \showISBNxiii \undefined \def \showISBNxiii  #1{\unskip}     \fi
\ifx \showISSN     \undefined \def \showISSN      #1{\unskip}     \fi
\ifx \showLCCN     \undefined \def \showLCCN      #1{\unskip}     \fi
\ifx \shownote     \undefined \def \shownote      #1{#1}          \fi
\ifx \showarticletitle \undefined \def \showarticletitle #1{#1}   \fi
\ifx \showURL      \undefined \def \showURL       {\relax}        \fi
\providecommand\bibfield[2]{#2}
\providecommand\bibinfo[2]{#2}
\providecommand\natexlab[1]{#1}
\providecommand\showeprint[2][]{arXiv:#2}

\bibitem[Barron et~al\mbox{.}(2022)]%
        {barron2022mipnerf360}
\bibfield{author}{\bibinfo{person}{Jonathan~T. Barron}, \bibinfo{person}{Ben
  Mildenhall}, \bibinfo{person}{Dor Verbin}, \bibinfo{person}{Pratul~P.
  Srinivasan}, {and} \bibinfo{person}{Peter Hedman}.}
  \bibinfo{year}{2022}\natexlab{}.
\newblock \showarticletitle{Mip-NeRF 360: Unbounded Anti-Aliased Neural
  Radiance Fields}.
\newblock \bibinfo{journal}{\emph{CVPR}} (\bibinfo{year}{2022}).
\newblock


\bibitem[Bhalgat et~al\mbox{.}(2023)]%
        {bhalgat_contrastive_2023}
\bibfield{author}{\bibinfo{person}{Yash Bhalgat}, \bibinfo{person}{Iro Laina},
  \bibinfo{person}{Jo{\~a}o~F Henriques}, \bibinfo{person}{Andrew Zisserman},
  {and} \bibinfo{person}{Andrea Vedaldi}.} \bibinfo{year}{2023}\natexlab{}.
\newblock \showarticletitle{Contrastive Lift: 3D Object Instance Segmentation
  by Slow-Fast Contrastive Fusion}. In \bibinfo{booktitle}{\emph{Thirty-seventh
  Conference on Neural Information Processing Systems}}.
\newblock
\urldef\tempurl%
\url{https://openreview.net/forum?id=bbbbbov4Xu}
\showURL{%
\tempurl}


\bibitem[Bi et~al\mbox{.}(2020)]%
        {bi2020deep}
\bibfield{author}{\bibinfo{person}{Sai Bi}, \bibinfo{person}{Zexiang Xu},
  \bibinfo{person}{Kalyan Sunkavalli}, \bibinfo{person}{Milo{\v{s}}
  Ha{\v{s}}an}, \bibinfo{person}{Yannick Hold-Geoffroy}, \bibinfo{person}{David
  Kriegman}, {and} \bibinfo{person}{Ravi Ramamoorthi}.}
  \bibinfo{year}{2020}\natexlab{}.
\newblock \showarticletitle{Deep reflectance volumes: Relightable
  reconstructions from multi-view photometric images}. In
  \bibinfo{booktitle}{\emph{Computer Vision--ECCV 2020: 16th European
  Conference, Glasgow, UK, August 23--28, 2020, Proceedings, Part III 16}}.
  Springer, \bibinfo{pages}{294--311}.
\newblock


\bibitem[Bińkowski et~al\mbox{.}(2021)]%
        {binkowski_demystifying_2021}
\bibfield{author}{\bibinfo{person}{Mikołaj Bińkowski},
  \bibinfo{person}{Danica~J. Sutherland}, \bibinfo{person}{Michael Arbel},
  {and} \bibinfo{person}{Arthur Gretton}.} \bibinfo{year}{2021}\natexlab{}.
\newblock \bibinfo{title}{Demystifying {MMD} {GANs}}.
\newblock
\urldef\tempurl%
\url{http://arxiv.org/abs/1801.01401}
\showURL{%
\tempurl}
\newblock
\shownote{arXiv:1801.01401 [cs, stat]}.


\bibitem[Caron et~al\mbox{.}(2021)]%
        {caron_dino_2021}
\bibfield{author}{\bibinfo{person}{Mathilde Caron}, \bibinfo{person}{Hugo
  Touvron}, \bibinfo{person}{Ishan Misra}, \bibinfo{person}{Herv\'e J\'egou},
  \bibinfo{person}{Julien Mairal}, \bibinfo{person}{Piotr Bojanowski}, {and}
  \bibinfo{person}{Armand Joulin}.} \bibinfo{year}{2021}\natexlab{}.
\newblock \showarticletitle{Emerging Properties in Self-Supervised Vision
  Transformers}. In \bibinfo{booktitle}{\emph{Proceedings of the International
  Conference on Computer Vision (ICCV)}}.
\newblock


\bibitem[Cen et~al\mbox{.}(2024)]%
        {cen_segment_2024}
\bibfield{author}{\bibinfo{person}{Jiazhong Cen}, \bibinfo{person}{Jiemin
  Fang}, \bibinfo{person}{Chen Yang}, \bibinfo{person}{Lingxi Xie},
  \bibinfo{person}{Xiaopeng Zhang}, \bibinfo{person}{Wei Shen}, {and}
  \bibinfo{person}{Qi Tian}.} \bibinfo{year}{2024}\natexlab{}.
\newblock \bibinfo{title}{Segment {Any} {3D} {Gaussians}}.
\newblock
\href{https://doi.org/10.48550/arXiv.2312.00860}{doi:\nolinkurl{10.48550/arXiv.2312.00860}}
\newblock
\shownote{arXiv:2312.00860}.


\bibitem[Chen and Wang(2024)]%
        {chen2024survey3dgs}
\bibfield{author}{\bibinfo{person}{Guikun Chen} {and} \bibinfo{person}{Wenguan
  Wang}.} \bibinfo{year}{2024}\natexlab{}.
\newblock \bibinfo{title}{A Survey on 3D Gaussian Splatting}.
\newblock
\showeprint[arxiv]{2401.03890}~[cs.CV]
\urldef\tempurl%
\url{https://arxiv.org/abs/2401.03890}
\showURL{%
\tempurl}


\bibitem[Cheng et~al\mbox{.}(2023)]%
        {cheng_structure_2023}
\bibfield{author}{\bibinfo{person}{Tianhang Cheng}, \bibinfo{person}{Wei-Chiu
  Ma}, \bibinfo{person}{Kaiyu Guan}, \bibinfo{person}{Antonio Torralba}, {and}
  \bibinfo{person}{Shenlong Wang}.} \bibinfo{year}{2023}\natexlab{}.
\newblock \showarticletitle{Structure from {Duplicates}: {Neural} {Inverse}
  {Graphics} from a {Pile} of {Objects}}.
\newblock
\urldef\tempurl%
\url{https://openreview.net/forum?id=7irm2VJARb}
\showURL{%
\tempurl}


\bibitem[Choi et~al\mbox{.}(2024)]%
        {choi_clickGaussian_2024}
\bibfield{author}{\bibinfo{person}{Seokhun Choi}, \bibinfo{person}{Hyeonseop
  Song}, \bibinfo{person}{Jaechul Kim}, \bibinfo{person}{Taehyeong Kim}, {and}
  \bibinfo{person}{Hoseok Do}.} \bibinfo{year}{2024}\natexlab{}.
\newblock \showarticletitle{Click-Gaussian: Interactive Segmentation to Any 3D
  Gaussians}. In \bibinfo{booktitle}{\emph{ECCV}}.
\newblock


\bibitem[Choi et~al\mbox{.}(2025)]%
        {choi_clickgaussian_2025}
\bibfield{author}{\bibinfo{person}{Seokhun Choi}, \bibinfo{person}{Hyeonseop
  Song}, \bibinfo{person}{Jaechul Kim}, \bibinfo{person}{Taehyeong Kim}, {and}
  \bibinfo{person}{Hoseok Do}.} \bibinfo{year}{2025}\natexlab{}.
\newblock \showarticletitle{Click-gaussian: Interactive segmentation to any 3d
  gaussians}. In \bibinfo{booktitle}{\emph{European Conference on Computer
  Vision}}. Springer, \bibinfo{pages}{289--305}.
\newblock


\bibitem[Community(2018)]%
        {blender}
\bibfield{author}{\bibinfo{person}{Blender~Online Community}.}
  \bibinfo{year}{2018}\natexlab{}.
\newblock \bibinfo{booktitle}{\emph{Blender - a 3D modelling and rendering
  package}}.
\newblock Blender Foundation, Stichting Blender Foundation, Amsterdam.
\newblock
\urldef\tempurl%
\url{http://www.blender.org}
\showURL{%
\tempurl}


\bibitem[Debevec et~al\mbox{.}(2000)]%
        {debevec2000acquiring}
\bibfield{author}{\bibinfo{person}{Paul Debevec}, \bibinfo{person}{Tim
  Hawkins}, \bibinfo{person}{Chris Tchou}, \bibinfo{person}{Haarm-Pieter
  Duiker}, \bibinfo{person}{Westley Sarokin}, {and} \bibinfo{person}{Mark
  Sagar}.} \bibinfo{year}{2000}\natexlab{}.
\newblock \showarticletitle{Acquiring the reflectance field of a human face}.
  In \bibinfo{booktitle}{\emph{Proceedings of the 27th annual conference on
  Computer graphics and interactive techniques}}. \bibinfo{pages}{145--156}.
\newblock


\bibitem[DeTone et~al\mbox{.}(2018)]%
        {detone_superpoint_2018}
\bibfield{author}{\bibinfo{person}{Daniel DeTone}, \bibinfo{person}{Tomasz
  Malisiewicz}, {and} \bibinfo{person}{Andrew Rabinovich}.}
  \bibinfo{year}{2018}\natexlab{}.
\newblock \showarticletitle{{SuperPoint}: {Self}-{Supervised} {Interest}
  {Point} {Detection} and {Description}}. In \bibinfo{booktitle}{\emph{2018
  {IEEE}/{CVF} {Conference} on {Computer} {Vision} and {Pattern} {Recognition}
  {Workshops} ({CVPRW})}}. \bibinfo{pages}{337--33712}.
\newblock
\href{https://doi.org/10.1109/CVPRW.2018.00060}{doi:\nolinkurl{10.1109/CVPRW.2018.00060}}
\newblock
\shownote{ISSN: 2160-7516}.


\bibitem[Fan et~al\mbox{.}(2023)]%
        {fan_nerfsos_2023}
\bibfield{author}{\bibinfo{person}{Zhiwen Fan}, \bibinfo{person}{Peihao Wang},
  \bibinfo{person}{Yifan Jiang}, \bibinfo{person}{Xinyu Gong},
  \bibinfo{person}{Dejia Xu}, {and} \bibinfo{person}{Zhangyang Wang}.}
  \bibinfo{year}{2023}\natexlab{}.
\newblock \showarticletitle{NeRF-SOS: Any-View Self-supervised Object
  Segmentation on Complex Scenes}. In \bibinfo{booktitle}{\emph{The Eleventh
  International Conference on Learning Representations}}.
\newblock


\bibitem[Fei et~al\mbox{.}(2024)]%
        {fei20243d}
\bibfield{author}{\bibinfo{person}{Ben Fei}, \bibinfo{person}{Jingyi Xu},
  \bibinfo{person}{Rui Zhang}, \bibinfo{person}{Qingyuan Zhou},
  \bibinfo{person}{Weidong Yang}, {and} \bibinfo{person}{Ying He}.}
  \bibinfo{year}{2024}\natexlab{}.
\newblock \showarticletitle{3d gaussian splatting as new era: A survey}.
\newblock \bibinfo{journal}{\emph{IEEE Transactions on Visualization and
  Computer Graphics}} (\bibinfo{year}{2024}).
\newblock


\bibitem[Fischler and Bolles(1981)]%
        {fischler_ransac_1981}
\bibfield{author}{\bibinfo{person}{Martin~A. Fischler} {and}
  \bibinfo{person}{Robert~C. Bolles}.} \bibinfo{year}{1981}\natexlab{}.
\newblock \showarticletitle{Random sample consensus: a paradigm for model
  fitting with applications to image analysis and automated cartography}. In
  \bibinfo{booktitle}{\emph{Commun. ACM}}, Vol.~\bibinfo{volume}{24}.
  \bibinfo{publisher}{Association for Computing Machinery},
  \bibinfo{address}{New York, NY, USA}, \bibinfo{pages}{381–395}.
\newblock
\showISSN{0001-0782}
\href{https://doi.org/10.1145/358669.358692}{doi:\nolinkurl{10.1145/358669.358692}}


\bibitem[Gao* et~al\mbox{.}(2024)]%
        {gao_cat3d_2024}
\bibfield{author}{\bibinfo{person}{Ruiqi Gao*}, \bibinfo{person}{Aleksander
  Holynski*}, \bibinfo{person}{Philipp Henzler}, \bibinfo{person}{Arthur
  Brussee}, \bibinfo{person}{Ricardo Martin-Brualla},
  \bibinfo{person}{Pratul~P. Srinivasan}, \bibinfo{person}{Jonathan~T. Barron},
  {and} \bibinfo{person}{Ben Poole*}.} \bibinfo{year}{2024}\natexlab{}.
\newblock \showarticletitle{CAT3D: Create Anything in 3D with Multi-View
  Diffusion Models}.
\newblock \bibinfo{journal}{\emph{Advances in Neural Information Processing
  Systems}} (\bibinfo{year}{2024}).
\newblock


\bibitem[Goel et~al\mbox{.}(2023)]%
        {goel_isrf_2023}
\bibfield{author}{\bibinfo{person}{Rahul Goel}, \bibinfo{person}{Dhawal
  Sirikonda}, \bibinfo{person}{Saurabh Saini}, {and} \bibinfo{person}{P.J.
  Narayanan}.} \bibinfo{year}{2023}\natexlab{}.
\newblock \showarticletitle{{Interactive Segmentation of Radiance Fields}}. In
  \bibinfo{booktitle}{\emph{{Proceedings of the IEEE/CVF Conference on Computer
  Vision and Pattern Recognition (CVPR)}}}.
\newblock


\bibitem[Goli et~al\mbox{.}(2024)]%
        {goli_BayesRays_2023}
\bibfield{author}{\bibinfo{person}{Lily Goli}, \bibinfo{person}{Cody Reading},
  \bibinfo{person}{Silvia Sellán}, \bibinfo{person}{Alec Jacobson}, {and}
  \bibinfo{person}{Andrea Tagliasacchi}.} \bibinfo{year}{2024}\natexlab{}.
\newblock \showarticletitle{{Bayes' Rays}: Uncertainty Quantification in Neural
  Radiance Fields}.
\newblock \bibinfo{journal}{\emph{CVPR}} (\bibinfo{year}{2024}).
\newblock


\bibitem[Gu et~al\mbox{.}(2024)]%
        {gu_egolifter_2024}
\bibfield{author}{\bibinfo{person}{Qiao Gu}, \bibinfo{person}{Zhaoyang Lv},
  \bibinfo{person}{Duncan Frost}, \bibinfo{person}{Simon Green},
  \bibinfo{person}{Julian Straub}, {and} \bibinfo{person}{Chris Sweeney}.}
  \bibinfo{year}{2024}\natexlab{}.
\newblock \bibinfo{title}{{EgoLifter}: {Open}-world {3D} {Segmentation} for
  {Egocentric} {Perception}}.
\newblock
\urldef\tempurl%
\url{http://arxiv.org/abs/2403.18118}
\showURL{%
\tempurl}
\newblock
\shownote{arXiv:2403.18118 [cs]}.


\bibitem[Huang et~al\mbox{.}(2021)]%
        {huang_surveyregistration_2021}
\bibfield{author}{\bibinfo{person}{Xiaoshui Huang}, \bibinfo{person}{Guofeng
  Mei}, \bibinfo{person}{Jian Zhang}, {and} \bibinfo{person}{Rana Abbas}.}
  \bibinfo{year}{2021}\natexlab{}.
\newblock \showarticletitle{A comprehensive survey on point cloud
  registration}.
\newblock \bibinfo{journal}{\emph{arXiv preprint arXiv:2103.02690}}
  (\bibinfo{year}{2021}).
\newblock


\bibitem[Je et~al\mbox{.}(2024)]%
        {je_robust_2024}
\bibfield{author}{\bibinfo{person}{Jihyeon Je}, \bibinfo{person}{Jiayi Liu},
  \bibinfo{person}{Guandao Yang}, \bibinfo{person}{Boyang Deng},
  \bibinfo{person}{Shengqu Cai}, \bibinfo{person}{Gordon Wetzstein},
  \bibinfo{person}{Or Litany}, {and} \bibinfo{person}{Leonidas Guibas}.}
  \bibinfo{year}{2024}\natexlab{}.
\newblock \showarticletitle{Robust {Symmetry} {Detection} via {Riemannian}
  {Langevin} {Dynamics}}.
\newblock
\href{https://doi.org/10.1145/3680528.3687682}{doi:\nolinkurl{10.1145/3680528.3687682}}
\newblock
\shownote{arXiv:2410.02786 [cs]}.


\bibitem[Johnson and Hebert(1999)]%
        {johnson_spinimage_1999}
\bibfield{author}{\bibinfo{person}{Andrew~E Johnson} {and}
  \bibinfo{person}{Martial Hebert}.} \bibinfo{year}{1999}\natexlab{}.
\newblock \showarticletitle{Using spin images for efficient object recognition
  in cluttered 3D scenes}.
\newblock \bibinfo{journal}{\emph{IEEE Transactions on pattern analysis and
  machine intelligence}} \bibinfo{volume}{21}, \bibinfo{number}{5},
  \bibinfo{pages}{433--449}.
\newblock


\bibitem[Ke et~al\mbox{.}(2023)]%
        {ke_samhq_2023}
\bibfield{author}{\bibinfo{person}{Lei Ke}, \bibinfo{person}{Mingqiao Ye},
  \bibinfo{person}{Martin Danelljan}, \bibinfo{person}{Yifan Liu},
  \bibinfo{person}{Yu-Wing Tai}, \bibinfo{person}{Chi-Keung Tang}, {and}
  \bibinfo{person}{Fisher Yu}.} \bibinfo{year}{2023}\natexlab{}.
\newblock \showarticletitle{Segment Anything in High Quality}. In
  \bibinfo{booktitle}{\emph{NeurIPS}}.
\newblock


\bibitem[Kerbl et~al\mbox{.}(2023)]%
        {kerbl_3Dgaussians_2023}
\bibfield{author}{\bibinfo{person}{Bernhard Kerbl}, \bibinfo{person}{Georgios
  Kopanas}, \bibinfo{person}{Thomas Leimk{\"u}hler}, {and}
  \bibinfo{person}{George Drettakis}.} \bibinfo{year}{2023}\natexlab{}.
\newblock \showarticletitle{3D Gaussian Splatting for Real-Time Radiance Field
  Rendering}.
\newblock \bibinfo{journal}{\emph{ACM Transactions on Graphics}}
  \bibinfo{volume}{42}, \bibinfo{number}{4}.
\newblock
\urldef\tempurl%
\url{https://repo-sam.inria.fr/fungraph/3d-gaussian-splatting/}
\showURL{%
\tempurl}


\bibitem[Kerbl et~al\mbox{.}(2024)]%
        {Kerbl_hierarchicalgaussians_2024}
\bibfield{author}{\bibinfo{person}{Bernhard Kerbl}, \bibinfo{person}{Andreas
  Meuleman}, \bibinfo{person}{Georgios Kopanas}, \bibinfo{person}{Michael
  Wimmer}, \bibinfo{person}{Alexandre Lanvin}, {and} \bibinfo{person}{George
  Drettakis}.} \bibinfo{year}{2024}\natexlab{}.
\newblock \showarticletitle{A Hierarchical 3D Gaussian Representation for
  Real-Time Rendering of Very Large Datasets}. In \bibinfo{booktitle}{\emph{ACM
  Transactions on Graphics}}, Vol.~\bibinfo{volume}{43}.
\newblock
\urldef\tempurl%
\url{https://repo-sam.inria.fr/fungraph/hierarchical-3d-gaussians/}
\showURL{%
\tempurl}


\bibitem[Kim et~al\mbox{.}(2024)]%
        {kim_garfield_2024}
\bibfield{author}{\bibinfo{person}{Chung~Min Kim}, \bibinfo{person}{Mingxuan
  Wu}, \bibinfo{person}{Justin Kerr}, \bibinfo{person}{Ken Goldberg},
  \bibinfo{person}{Matthew Tancik}, {and} \bibinfo{person}{Angjoo Kanazawa}.}
  \bibinfo{year}{2024}\natexlab{}.
\newblock \bibinfo{title}{{GARField}: {Group} {Anything} with {Radiance}
  {Fields}}.
\newblock
\href{https://doi.org/10.48550/arXiv.2401.09419}{doi:\nolinkurl{10.48550/arXiv.2401.09419}}
\newblock
\shownote{arXiv:2401.09419 [cs]}.


\bibitem[Kobayashi et~al\mbox{.}(2022)]%
        {kobayashi_FFD_2022}
\bibfield{author}{\bibinfo{person}{Sosuke Kobayashi}, \bibinfo{person}{Eiichi
  Matsumoto}, {and} \bibinfo{person}{Vincent Sitzmann}.}
  \bibinfo{year}{2022}\natexlab{}.
\newblock \showarticletitle{Decomposing nerf for editing via feature field
  distillation}. In \bibinfo{booktitle}{\emph{Advances in Neural Information
  Processing Systems}}, Vol.~\bibinfo{volume}{35}.
  \bibinfo{pages}{23311--23330}.
\newblock


\bibitem[Lee et~al\mbox{.}(2024)]%
        {lee_rethinking_2024}
\bibfield{author}{\bibinfo{person}{Hyunjee Lee}, \bibinfo{person}{Youngsik
  Yun}, \bibinfo{person}{Jeongmin Bae}, \bibinfo{person}{Seoha Kim}, {and}
  \bibinfo{person}{Youngjung Uh}.} \bibinfo{year}{2024}\natexlab{}.
\newblock \bibinfo{title}{Rethinking {Open}-{Vocabulary} {Segmentation} of
  {Radiance} {Fields} in {3D} {Space}}.
\newblock
\urldef\tempurl%
\url{http://arxiv.org/abs/2408.07416}
\showURL{%
\tempurl}
\newblock
\shownote{arXiv:2408.07416 [cs]}.


\bibitem[Lepetit et~al\mbox{.}(2009)]%
        {lepetit_epnp_2009}
\bibfield{author}{\bibinfo{person}{Vincent Lepetit}, \bibinfo{person}{Francesc
  Moreno-Noguer}, {and} \bibinfo{person}{Pascal Fua}.}
  \bibinfo{year}{2009}\natexlab{}.
\newblock \showarticletitle{EPnP: An accurate O(n) solution to the PnP
  problem}. In \bibinfo{booktitle}{\emph{International Journal of Computer
  Vision}}, Vol.~\bibinfo{volume}{81}.
\newblock
\href{https://doi.org/10.1007/s11263-008-0152-6}{doi:\nolinkurl{10.1007/s11263-008-0152-6}}


\bibitem[Leroy et~al\mbox{.}(2024)]%
        {leroy_grounding_2024}
\bibfield{author}{\bibinfo{person}{Vincent Leroy}, \bibinfo{person}{Yohann
  Cabon}, {and} \bibinfo{person}{Jérôme Revaud}.}
  \bibinfo{year}{2024}\natexlab{}.
\newblock \bibinfo{title}{Grounding {Image} {Matching} in {3D} with {MASt3R}}.
\newblock
\urldef\tempurl%
\url{http://arxiv.org/abs/2406.09756}
\showURL{%
\tempurl}
\newblock
\shownote{arXiv:2406.09756 [cs]}.


\bibitem[Ling et~al\mbox{.}(2024)]%
        {ling2024dl3dv}
\bibfield{author}{\bibinfo{person}{Lu Ling}, \bibinfo{person}{Yichen Sheng},
  \bibinfo{person}{Zhi Tu}, \bibinfo{person}{Wentian Zhao},
  \bibinfo{person}{Cheng Xin}, \bibinfo{person}{Kun Wan},
  \bibinfo{person}{Lantao Yu}, \bibinfo{person}{Qianyu Guo},
  \bibinfo{person}{Zixun Yu}, \bibinfo{person}{Yawen Lu}, {et~al\mbox{.}}}
  \bibinfo{year}{2024}\natexlab{}.
\newblock \showarticletitle{Dl3dv-10k: A large-scale scene dataset for deep
  learning-based 3d vision}. In \bibinfo{booktitle}{\emph{Proceedings of the
  IEEE/CVF Conference on Computer Vision and Pattern Recognition}}.
  \bibinfo{pages}{22160--22169}.
\newblock


\bibitem[Liu et~al\mbox{.}(2024)]%
        {liu_groundingdino_2024}
\bibfield{author}{\bibinfo{person}{Shilong Liu}, \bibinfo{person}{Zhaoyang
  Zeng}, \bibinfo{person}{Tianhe Ren}, \bibinfo{person}{Feng Li},
  \bibinfo{person}{Hao Zhang}, \bibinfo{person}{Jie Yang},
  \bibinfo{person}{Chunyuan Li}, \bibinfo{person}{Jianwei Yang},
  \bibinfo{person}{Hang Su}, \bibinfo{person}{Jun Zhu}, {et~al\mbox{.}}}
  \bibinfo{year}{2024}\natexlab{}.
\newblock \showarticletitle{Grounding dino: Marrying dino with grounded
  pre-training for open-set object detection}. In
  \bibinfo{booktitle}{\emph{ECCV}}.
\newblock


\bibitem[Lowe(2004)]%
        {lowe_distinctive_2004}
\bibfield{author}{\bibinfo{person}{David~G. Lowe}.}
  \bibinfo{year}{2004}\natexlab{}.
\newblock \showarticletitle{Distinctive {Image} {Features} from
  {Scale}-{Invariant} {Keypoints}}.
\newblock \bibinfo{journal}{\emph{International Journal of Computer Vision}}
  \bibinfo{volume}{60}, \bibinfo{number}{2} (\bibinfo{date}{Nov.}
  \bibinfo{year}{2004}), \bibinfo{pages}{91--110}.
\newblock
\showISSN{0920-5691}
\href{https://doi.org/10.1023/B:VISI.0000029664.99615.94}{doi:\nolinkurl{10.1023/B:VISI.0000029664.99615.94}}


\bibitem[Mallick et~al\mbox{.}(2024)]%
        {mallick2024taming}
\bibfield{author}{\bibinfo{person}{Saswat~Subhajyoti Mallick},
  \bibinfo{person}{Rahul Goel}, \bibinfo{person}{Bernhard Kerbl},
  \bibinfo{person}{Markus Steinberger}, \bibinfo{person}{Francisco~Vicente
  Carrasco}, {and} \bibinfo{person}{Fernando De~La~Torre}.}
  \bibinfo{year}{2024}\natexlab{}.
\newblock \showarticletitle{Taming 3dgs: High-quality radiance fields with
  limited resources}. In \bibinfo{booktitle}{\emph{SIGGRAPH Asia 2024
  Conference Papers}}. \bibinfo{pages}{1--11}.
\newblock


\bibitem[Martin-Brualla et~al\mbox{.}(2021)]%
        {martinbrualla_nerfw_2020}
\bibfield{author}{\bibinfo{person}{Ricardo Martin-Brualla},
  \bibinfo{person}{Noha Radwan}, \bibinfo{person}{Mehdi S.~M. Sajjadi},
  \bibinfo{person}{Jonathan~T. Barron}, \bibinfo{person}{Alexey Dosovitskiy},
  {and} \bibinfo{person}{Daniel Duckworth}.} \bibinfo{year}{2021}\natexlab{}.
\newblock \showarticletitle{{NeRF in the Wild: Neural Radiance Fields for
  Unconstrained Photo Collections}}. In \bibinfo{booktitle}{\emph{CVPR}}.
\newblock


\bibitem[Mildenhall et~al\mbox{.}(2020)]%
        {mildenhall_nerf_2020}
\bibfield{author}{\bibinfo{person}{Ben Mildenhall}, \bibinfo{person}{Pratul~P.
  Srinivasan}, \bibinfo{person}{Matthew Tancik}, \bibinfo{person}{Jonathan~T.
  Barron}, \bibinfo{person}{Ravi Ramamoorthi}, {and} \bibinfo{person}{Ren Ng}.}
  \bibinfo{year}{2020}\natexlab{}.
\newblock \showarticletitle{NeRF: Representing Scenes as Neural Radiance Fields
  for View Synthesis}. In \bibinfo{booktitle}{\emph{ECCV}}.
\newblock


\bibitem[Mitra et~al\mbox{.}(2006)]%
        {mitra_partial_2006}
\bibfield{author}{\bibinfo{person}{Niloy~J. Mitra},
  \bibinfo{person}{Leonidas~J. Guibas}, {and} \bibinfo{person}{Mark Pauly}.}
  \bibinfo{year}{2006}\natexlab{}.
\newblock \showarticletitle{Partial and approximate symmetry detection for {3D}
  geometry}. In \bibinfo{booktitle}{\emph{{ACM} {SIGGRAPH} 2006 {Papers}}}
  \emph{(\bibinfo{series}{{SIGGRAPH} '06})}. \bibinfo{publisher}{Association
  for Computing Machinery}, \bibinfo{address}{New York, NY, USA},
  \bibinfo{pages}{560--568}.
\newblock
\showISBNx{978-1-59593-364-5}
\href{https://doi.org/10.1145/1179352.1141924}{doi:\nolinkurl{10.1145/1179352.1141924}}


\bibitem[Mitra et~al\mbox{.}(2013)]%
        {mitra_symmetry_2013}
\bibfield{author}{\bibinfo{person}{Niloy~J. Mitra}, \bibinfo{person}{Mark
  Pauly}, \bibinfo{person}{Michael Wand}, {and} \bibinfo{person}{Duygu
  Ceylan}.} \bibinfo{year}{2013}\natexlab{}.
\newblock \showarticletitle{Symmetry in {3D} {Geometry}: {Extraction} and
  {Applications}}.
\newblock \bibinfo{journal}{\emph{Computer Graphics Forum}}
  \bibinfo{volume}{32}, \bibinfo{number}{6} (\bibinfo{year}{2013}),
  \bibinfo{pages}{1--23}.
\newblock
\showISSN{1467-8659}
\href{https://doi.org/10.1111/cgf.12010}{doi:\nolinkurl{10.1111/cgf.12010}}
\newblock
\shownote{\_eprint: https://onlinelibrary.wiley.com/doi/pdf/10.1111/cgf.12010}.


\bibitem[M\"uller et~al\mbox{.}(2022)]%
        {mueller2022instant}
\bibfield{author}{\bibinfo{person}{Thomas M\"uller}, \bibinfo{person}{Alex
  Evans}, \bibinfo{person}{Christoph Schied}, {and} \bibinfo{person}{Alexander
  Keller}.} \bibinfo{year}{2022}\natexlab{}.
\newblock \showarticletitle{Instant Neural Graphics Primitives with a
  Multiresolution Hash Encoding}.
\newblock \bibinfo{journal}{\emph{ACM Transactions on Graphics}}
  (\bibinfo{year}{2022}).
\newblock


\bibitem[Nguyen et~al\mbox{.}(2024)]%
        {Nguyen_gigapose_2024}
\bibfield{author}{\bibinfo{person}{Van~Nguyen Nguyen},
  \bibinfo{person}{Thibault Groueix}, \bibinfo{person}{Mathieu Salzmann}, {and}
  \bibinfo{person}{Vincent Lepetit}.} \bibinfo{year}{2024}\natexlab{}.
\newblock \showarticletitle{GigaPose: Fast and Robust Novel Object Pose
  Estimation via One Correspondence}. In \bibinfo{booktitle}{\emph{Proceedings
  of the IEEE/CVF Conference on Computer Vision and Pattern Recognition
  (CVPR)}}. \bibinfo{pages}{9903--9913}.
\newblock


\bibitem[Podolak et~al\mbox{.}(2007)]%
        {podolak_symmetry_2007}
\bibfield{author}{\bibinfo{person}{Joshua Podolak}, \bibinfo{person}{Aleksey
  Golovinskiy}, {and} \bibinfo{person}{Szymon Rusinkiewicz}.}
  \bibinfo{year}{2007}\natexlab{}.
\newblock \showarticletitle{Symmetry-enhanced remeshing of surfaces}. In
  \bibinfo{booktitle}{\emph{Proceedings of the fifth Eurographics symposium on
  Geometry processing}}. \bibinfo{pages}{235--242}.
\newblock


\bibitem[Poirier-Ginter et~al\mbox{.}(2024)]%
        {poirier_GSrelighting_2024}
\bibfield{author}{\bibinfo{person}{Yohan Poirier-Ginter},
  \bibinfo{person}{Alban Gauthier}, \bibinfo{person}{Julien Phillip},
  \bibinfo{person}{J-F Lalonde}, {and} \bibinfo{person}{George Drettakis}.}
  \bibinfo{year}{2024}\natexlab{}.
\newblock \showarticletitle{A Diffusion Approach to Radiance Field Relighting
  using Multi-Illumination Synthesis}. In \bibinfo{booktitle}{\emph{Computer
  Graphics Forum}}, Vol.~\bibinfo{volume}{43}. Wiley Online Library,
  \bibinfo{pages}{e15147}.
\newblock


\bibitem[Potje et~al\mbox{.}(2024)]%
        {potje_xfeat_2024}
\bibfield{author}{\bibinfo{person}{Guilherme Potje}, \bibinfo{person}{Felipe
  Cadar}, \bibinfo{person}{André Araujo}, \bibinfo{person}{Renato Martins},
  {and} \bibinfo{person}{Erickson~R. Nascimento}.}
  \bibinfo{year}{2024}\natexlab{}.
\newblock \showarticletitle{{XFeat}: {Accelerated} {Features} for {Lightweight}
  {Image} {Matching}}. \bibinfo{pages}{2682--2691}.
\newblock
\urldef\tempurl%
\url{https://openaccess.thecvf.com/content/CVPR2024/html/Potje_XFeat_Accelerated_Features_for_Lightweight_Image_Matching_CVPR_2024_paper.html}
\showURL{%
\tempurl}


\bibitem[Qi et~al\mbox{.}(2017)]%
        {qi_pointnet_2017}
\bibfield{author}{\bibinfo{person}{Charles~Ruizhongtai Qi}, \bibinfo{person}{Li
  Yi}, \bibinfo{person}{Hao Su}, {and} \bibinfo{person}{Leonidas~J Guibas}.}
  \bibinfo{year}{2017}\natexlab{}.
\newblock \showarticletitle{{PointNet}++: {Deep} {Hierarchical} {Feature}
  {Learning} on {Point} {Sets} in a {Metric} {Space}}. In
  \bibinfo{booktitle}{\emph{31st {Conference} on {Neural} {Information}
  {Processing} {Systems} ({NIPS} 2017)}}. \bibinfo{address}{Long Beach, CA,
  USA.}
\newblock


\bibitem[Qin et~al\mbox{.}(2023)]%
        {qin_langsplat_2023}
\bibfield{author}{\bibinfo{person}{Minghan Qin}, \bibinfo{person}{Wanhua Li},
  \bibinfo{person}{Jiawei Zhou}, \bibinfo{person}{Haoqian Wang}, {and}
  \bibinfo{person}{Hanspeter Pfister}.} \bibinfo{year}{2023}\natexlab{}.
\newblock \bibinfo{title}{{LangSplat}: {3D} {Language} {Gaussian} {Splatting}}.
\newblock
\href{https://doi.org/10.48550/arXiv.2312.16084}{doi:\nolinkurl{10.48550/arXiv.2312.16084}}
\newblock
\shownote{arXiv:2312.16084 [cs]}.


\bibitem[Qiu et~al\mbox{.}(2024)]%
        {qiu_featuresplatting_2024}
\bibfield{author}{\bibinfo{person}{Ri-Zhao Qiu}, \bibinfo{person}{Ge Yang},
  \bibinfo{person}{Weijia Zeng}, {and} \bibinfo{person}{Xiaolong Wang}.}
  \bibinfo{year}{2024}\natexlab{}.
\newblock \showarticletitle{Language-Driven Physics-Based Scene Synthesis and
  Editing via Feature Splatting}. In \bibinfo{booktitle}{\emph{European
  Conference on Computer Vision (ECCV)}}.
\newblock


\bibitem[Radford et~al\mbox{.}(2021)]%
        {radford_clip_2021}
\bibfield{author}{\bibinfo{person}{Alec Radford}, \bibinfo{person}{Jong~Wook
  Kim}, \bibinfo{person}{Chris Hallacy}, \bibinfo{person}{Aditya Ramesh},
  \bibinfo{person}{Gabriel Goh}, \bibinfo{person}{Sandhini Agarwal},
  \bibinfo{person}{Girish Sastry}, \bibinfo{person}{Amanda Askell},
  \bibinfo{person}{Pamela Mishkin}, \bibinfo{person}{Jack Clark},
  {et~al\mbox{.}}} \bibinfo{year}{2021}\natexlab{}.
\newblock \showarticletitle{Learning transferable visual models from natural
  language supervision}. In \bibinfo{booktitle}{\emph{International conference
  on machine learning}}. PMLR, \bibinfo{pages}{8748--8763}.
\newblock


\bibitem[Ravi et~al\mbox{.}(2024)]%
        {ravi_sam2_2024}
\bibfield{author}{\bibinfo{person}{Nikhila Ravi}, \bibinfo{person}{Valentin
  Gabeur}, \bibinfo{person}{Yuan-Ting Hu}, \bibinfo{person}{Ronghang Hu},
  \bibinfo{person}{Chaitanya Ryali}, \bibinfo{person}{Tengyu Ma},
  \bibinfo{person}{Haitham Khedr}, \bibinfo{person}{Roman R{\"a}dle},
  \bibinfo{person}{Chloe Rolland}, \bibinfo{person}{Laura Gustafson},
  \bibinfo{person}{Eric Mintun}, \bibinfo{person}{Junting Pan},
  \bibinfo{person}{Kalyan~Vasudev Alwala}, \bibinfo{person}{Nicolas Carion},
  \bibinfo{person}{Chao-Yuan Wu}, \bibinfo{person}{Ross Girshick},
  \bibinfo{person}{Piotr Doll{\'a}r}, {and} \bibinfo{person}{Christoph
  Feichtenhofer}.} \bibinfo{year}{2024}\natexlab{}.
\newblock \showarticletitle{SAM 2: Segment Anything in Images and Videos}.
\newblock \bibinfo{journal}{\emph{arXiv preprint arXiv:2408.00714}}
  (\bibinfo{year}{2024}).
\newblock
\urldef\tempurl%
\url{https://arxiv.org/abs/2408.00714}
\showURL{%
\tempurl}


\bibitem[Rodriguez et~al\mbox{.}(2018)]%
        {rodriguez_exploiting_2018}
\bibfield{author}{\bibinfo{person}{Simon Rodriguez}, \bibinfo{person}{Adrien
  Bousseau}, \bibinfo{person}{Fredo Durand}, {and} \bibinfo{person}{George
  Drettakis}.} \bibinfo{year}{2018}\natexlab{}.
\newblock \showarticletitle{Exploiting {Repetitions} for {Image}-{Based}
  {Rendering} of {Facades}}.
\newblock \bibinfo{journal}{\emph{Computer Graphics Forum}}
  \bibinfo{volume}{37}, \bibinfo{number}{4} (\bibinfo{year}{2018}),
  \bibinfo{pages}{119--131}.
\newblock
\showISSN{1467-8659}
\href{https://doi.org/10.1111/cgf.13480}{doi:\nolinkurl{10.1111/cgf.13480}}
\newblock
\shownote{\_eprint: https://onlinelibrary.wiley.com/doi/pdf/10.1111/cgf.13480}.


\bibitem[Rusu et~al\mbox{.}(2009)]%
        {Rusu_histogram3DFPFH_2009}
\bibfield{author}{\bibinfo{person}{Radu~Bogdan Rusu}, \bibinfo{person}{Nico
  Blodow}, {and} \bibinfo{person}{Michael Beetz}.}
  \bibinfo{year}{2009}\natexlab{}.
\newblock \showarticletitle{Fast Point Feature Histograms (FPFH) for 3D
  registration}. In \bibinfo{booktitle}{\emph{2009 IEEE International
  Conference on Robotics and Automation}}. \bibinfo{pages}{3212--3217}.
\newblock
\href{https://doi.org/10.1109/ROBOT.2009.5152473}{doi:\nolinkurl{10.1109/ROBOT.2009.5152473}}


\bibitem[Schonberger and Frahm(2016)]%
        {schonberger_structure--motion_2016}
\bibfield{author}{\bibinfo{person}{Johannes~L. Schonberger} {and}
  \bibinfo{person}{Jan-Michael Frahm}.} \bibinfo{year}{2016}\natexlab{}.
\newblock \showarticletitle{Structure-from-{Motion} {Revisited}}. In
  \bibinfo{booktitle}{\emph{2016 {IEEE} {Conference} on {Computer} {Vision} and
  {Pattern} {Recognition} ({CVPR})}}. \bibinfo{publisher}{IEEE},
  \bibinfo{address}{Las Vegas, NV, USA}, \bibinfo{pages}{4104--4113}.
\newblock
\showISBNx{978-1-4673-8851-1}
\href{https://doi.org/10.1109/CVPR.2016.445}{doi:\nolinkurl{10.1109/CVPR.2016.445}}


\bibitem[Siddiqui et~al\mbox{.}(2023)]%
        {siddiqui_panoptic_2023}
\bibfield{author}{\bibinfo{person}{Yawar Siddiqui}, \bibinfo{person}{Lorenzo
  Porzi}, \bibinfo{person}{Samuel~Rota Bulò}, \bibinfo{person}{Norman
  Müller}, \bibinfo{person}{Matthias Nießner}, \bibinfo{person}{Angela Dai},
  {and} \bibinfo{person}{Peter Kontschieder}.} \bibinfo{year}{2023}\natexlab{}.
\newblock \showarticletitle{Panoptic {Lifting} for {3D} {Scene} {Understanding}
  with {Neural} {Fields}}. In \bibinfo{booktitle}{\emph{2023 {IEEE}/{CVF}
  {Conference} on {Computer} {Vision} and {Pattern} {Recognition} ({CVPR})}}.
  \bibinfo{publisher}{IEEE}, \bibinfo{address}{Vancouver, BC, Canada},
  \bibinfo{pages}{9043--9052}.
\newblock
\showISBNx{9798350301298}
\href{https://doi.org/10.1109/CVPR52729.2023.00873}{doi:\nolinkurl{10.1109/CVPR52729.2023.00873}}


\bibitem[Tancik et~al\mbox{.}(2023)]%
        {tancik2023nerfstudio}
\bibfield{author}{\bibinfo{person}{Matthew Tancik}, \bibinfo{person}{Ethan
  Weber}, \bibinfo{person}{Evonne Ng}, \bibinfo{person}{Ruilong Li},
  \bibinfo{person}{Brent Yi}, \bibinfo{person}{Terrance Wang},
  \bibinfo{person}{Alexander Kristoffersen}, \bibinfo{person}{Jake Austin},
  \bibinfo{person}{Kamyar Salahi}, \bibinfo{person}{Abhik Ahuja},
  {et~al\mbox{.}}} \bibinfo{year}{2023}\natexlab{}.
\newblock \showarticletitle{Nerfstudio: A modular framework for neural radiance
  field development}. In \bibinfo{booktitle}{\emph{ACM SIGGRAPH 2023 Conference
  Proceedings}}. \bibinfo{pages}{1--12}.
\newblock


\bibitem[Thrun and Wegbreit(2005)]%
        {thrun_completion_2005}
\bibfield{author}{\bibinfo{person}{S. Thrun} {and} \bibinfo{person}{B.
  Wegbreit}.} \bibinfo{year}{2005}\natexlab{}.
\newblock \showarticletitle{Shape from symmetry}. In
  \bibinfo{booktitle}{\emph{Tenth IEEE International Conference on Computer
  Vision (ICCV'05) Volume 1}}, Vol.~\bibinfo{volume}{2}.
  \bibinfo{pages}{1824--1831 Vol. 2}.
\newblock
\href{https://doi.org/10.1109/ICCV.2005.221}{doi:\nolinkurl{10.1109/ICCV.2005.221}}


\bibitem[Tschernezki et~al\mbox{.}(2022)]%
        {tschernezki_N3F_2022}
\bibfield{author}{\bibinfo{person}{Vadim Tschernezki}, \bibinfo{person}{Iro
  Laina}, \bibinfo{person}{Diane Larlus}, {and} \bibinfo{person}{Andrea
  Vedaldi}.} \bibinfo{year}{2022}\natexlab{}.
\newblock \showarticletitle{Neural Feature Fusion Fields: 3D Distillation of
  Self-Supervised 2D Image Representations}. In \bibinfo{booktitle}{\emph{2022
  International Conference on 3D Vision (3DV)}}. \bibinfo{pages}{443--453}.
\newblock
\href{https://doi.org/10.1109/3DV57658.2022.00056}{doi:\nolinkurl{10.1109/3DV57658.2022.00056}}


\bibitem[Wang et~al\mbox{.}(2023)]%
        {wang_dust3r_2023}
\bibfield{author}{\bibinfo{person}{Shuzhe Wang}, \bibinfo{person}{Vincent
  Leroy}, \bibinfo{person}{Yohann Cabon}, \bibinfo{person}{Boris Chidlovskii},
  {and} \bibinfo{person}{Jerome Revaud}.} \bibinfo{year}{2023}\natexlab{}.
\newblock \bibinfo{title}{{DUSt3R}: {Geometric} {3D} {Vision} {Made} {Easy}}.
\newblock
\href{https://doi.org/10.48550/arXiv.2312.14132}{doi:\nolinkurl{10.48550/arXiv.2312.14132}}
\newblock
\shownote{arXiv:2312.14132 [cs]}.


\bibitem[Wang and Solomon(2019)]%
        {wang2019deep}
\bibfield{author}{\bibinfo{person}{Yue Wang} {and} \bibinfo{person}{Justin~M
  Solomon}.} \bibinfo{year}{2019}\natexlab{}.
\newblock \showarticletitle{Deep closest point: Learning representations for
  point cloud registration}. In \bibinfo{booktitle}{\emph{Proceedings of the
  IEEE/CVF international conference on computer vision}}.
  \bibinfo{pages}{3523--3532}.
\newblock


\bibitem[Wang et~al\mbox{.}(2004)]%
        {wang_ssim_2003}
\bibfield{author}{\bibinfo{person}{Zhou Wang}, \bibinfo{person}{A.C. Bovik},
  \bibinfo{person}{H.R. Sheikh}, {and} \bibinfo{person}{E.P. Simoncelli}.}
  \bibinfo{year}{2004}\natexlab{}.
\newblock \showarticletitle{Image quality assessment: from error visibility to
  structural similarity}.
\newblock \bibinfo{journal}{\emph{IEEE Transactions on Image Processing}}
  \bibinfo{volume}{13}, \bibinfo{number}{4}, \bibinfo{pages}{600--612}.
\newblock
\href{https://doi.org/10.1109/TIP.2003.819861}{doi:\nolinkurl{10.1109/TIP.2003.819861}}


\bibitem[Warburg et~al\mbox{.}(2023)]%
        {warburg_nerfbusters_2023}
\bibfield{author}{\bibinfo{person}{Frederik Warburg}, \bibinfo{person}{Ethan
  Weber}, \bibinfo{person}{Matthew Tancik}, \bibinfo{person}{Aleksander
  Holynski}, {and} \bibinfo{person}{Angjoo Kanazawa}.}
  \bibinfo{year}{2023}\natexlab{}.
\newblock \bibinfo{title}{Nerfbusters: {Removing} {Ghostly} {Artifacts} from
  {Casually} {Captured} {NeRFs}}.
\newblock
\urldef\tempurl%
\url{http://arxiv.org/abs/2304.10532}
\showURL{%
\tempurl}
\newblock
\shownote{arXiv:2304.10532 [cs]}.


\bibitem[Wu et~al\mbox{.}(2024)]%
        {wu_reconfusion_2024}
\bibfield{author}{\bibinfo{person}{Rundi Wu}, \bibinfo{person}{Ben Mildenhall},
  \bibinfo{person}{Philipp Henzler}, \bibinfo{person}{Keunhong Park},
  \bibinfo{person}{Ruiqi Gao}, \bibinfo{person}{Daniel Watson},
  \bibinfo{person}{Pratul~P Srinivasan}, \bibinfo{person}{Dor Verbin},
  \bibinfo{person}{Jonathan~T Barron}, \bibinfo{person}{Ben Poole},
  {et~al\mbox{.}}} \bibinfo{year}{2024}\natexlab{}.
\newblock \showarticletitle{Reconfusion: 3d reconstruction with diffusion
  priors}. In \bibinfo{booktitle}{\emph{Proceedings of the IEEE/CVF Conference
  on Computer Vision and Pattern Recognition}}. \bibinfo{pages}{21551--21561}.
\newblock


\bibitem[Yang et~al\mbox{.}(2024)]%
        {Yang_depthanythingv2_2024}
\bibfield{author}{\bibinfo{person}{Lihe Yang}, \bibinfo{person}{Bingyi Kang},
  \bibinfo{person}{Zilong Huang}, \bibinfo{person}{Zhen Zhao},
  \bibinfo{person}{Xiaogang Xu}, \bibinfo{person}{Jiashi Feng}, {and}
  \bibinfo{person}{Hengshuang Zhao}.} \bibinfo{year}{2024}\natexlab{}.
\newblock \showarticletitle{Depth Anything V2}. In
  \bibinfo{booktitle}{\emph{NeurIPS}}.
\newblock


\bibitem[Ye et~al\mbox{.}(2023)]%
        {ye_gaussian_2023}
\bibfield{author}{\bibinfo{person}{Mingqiao Ye}, \bibinfo{person}{Martin
  Danelljan}, \bibinfo{person}{Fisher Yu}, {and} \bibinfo{person}{Lei Ke}.}
  \bibinfo{year}{2023}\natexlab{}.
\newblock \bibinfo{title}{Gaussian {Grouping}: {Segment} and {Edit} {Anything}
  in {3D} {Scenes}}.
\newblock
\href{https://doi.org/10.48550/arXiv.2312.00732}{doi:\nolinkurl{10.48550/arXiv.2312.00732}}
\newblock
\shownote{arXiv:2312.00732 [cs]}.


\bibitem[Yeshwanth et~al\mbox{.}(2023)]%
        {yeshwanth2023scannet++}
\bibfield{author}{\bibinfo{person}{Chandan Yeshwanth},
  \bibinfo{person}{Yueh-Cheng Liu}, \bibinfo{person}{Matthias Nie{\ss}ner},
  {and} \bibinfo{person}{Angela Dai}.} \bibinfo{year}{2023}\natexlab{}.
\newblock \showarticletitle{Scannet++: A high-fidelity dataset of 3d indoor
  scenes}. In \bibinfo{booktitle}{\emph{Proceedings of the IEEE/CVF
  International Conference on Computer Vision}}. \bibinfo{pages}{12--22}.
\newblock


\bibitem[Ying et~al\mbox{.}(2024)]%
        {ying_omniseg3d_2024}
\bibfield{author}{\bibinfo{person}{Haiyang Ying}, \bibinfo{person}{Yixuan Yin},
  \bibinfo{person}{Jinzhi Zhang}, \bibinfo{person}{Fan Wang},
  \bibinfo{person}{Tao Yu}, \bibinfo{person}{Ruqi Huang}, {and}
  \bibinfo{person}{Lu Fang}.} \bibinfo{year}{2024}\natexlab{}.
\newblock \showarticletitle{OmniSeg3D: Omniversal 3D Segmentation via
  Hierarchical Contrastive Learning}. In \bibinfo{booktitle}{\emph{Proceedings
  of the IEEE/CVF Conference on Computer Vision and Pattern Recognition
  (CVPR)}}.
\newblock


\bibitem[Zhang et~al\mbox{.}(2018)]%
        {zhang_lpips_2018}
\bibfield{author}{\bibinfo{person}{Richard Zhang}, \bibinfo{person}{Phillip
  Isola}, \bibinfo{person}{Alexei~A. Efros}, \bibinfo{person}{Eli Shechtman},
  {and} \bibinfo{person}{Oliver Wang}.} \bibinfo{year}{2018}\natexlab{}.
\newblock \showarticletitle{The Unreasonable Effectiveness of Deep Features as
  a Perceptual Metric}. In \bibinfo{booktitle}{\emph{2018 IEEE/CVF Conference
  on Computer Vision and Pattern Recognition}}. \bibinfo{pages}{586--595}.
\newblock
\href{https://doi.org/10.1109/CVPR.2018.00068}{doi:\nolinkurl{10.1109/CVPR.2018.00068}}


\bibitem[Zhang et~al\mbox{.}(2023)]%
        {zhang2023seeing}
\bibfield{author}{\bibinfo{person}{Yunzhi Zhang}, \bibinfo{person}{Shangzhe
  Wu}, \bibinfo{person}{Noah Snavely}, {and} \bibinfo{person}{Jiajun Wu}.}
  \bibinfo{year}{2023}\natexlab{}.
\newblock \showarticletitle{Seeing a rose in five thousand ways}. In
  \bibinfo{booktitle}{\emph{Proceedings of the IEEE/CVF Conference on Computer
  Vision and Pattern Recognition}}. \bibinfo{pages}{962--971}.
\newblock


\bibitem[Zhang(1994)]%
        {zhang_icp_1994}
\bibfield{author}{\bibinfo{person}{Zhengyou Zhang}.}
  \bibinfo{year}{1994}\natexlab{}.
\newblock \showarticletitle{Iterative point matching for registration of
  free-form curves and surfaces}.
\newblock \bibinfo{journal}{\emph{International journal of computer vision}}
  \bibinfo{volume}{13}, \bibinfo{number}{2} (\bibinfo{year}{1994}),
  \bibinfo{pages}{119--152}.
\newblock


\bibitem[Zheng et~al\mbox{.}(2010)]%
        {zheng_consolidation_2010}
\bibfield{author}{\bibinfo{person}{Qian Zheng}, \bibinfo{person}{Andrei Sharf},
  \bibinfo{person}{Guowei Wan}, \bibinfo{person}{Yangyan Li},
  \bibinfo{person}{Niloy~J Mitra}, \bibinfo{person}{Daniel Cohen-Or}, {and}
  \bibinfo{person}{Baoquan Chen}.} \bibinfo{year}{2010}\natexlab{}.
\newblock \showarticletitle{Non-local scan consolidation for 3D urban scenes}.
  In \bibinfo{booktitle}{\emph{ACM Trans. Graph.}}, Vol.~\bibinfo{volume}{29}.
  \bibinfo{pages}{94--1}.
\newblock


\bibitem[Zhi et~al\mbox{.}(2021)]%
        {Shuaifeng_semanticnerf_2021}
\bibfield{author}{\bibinfo{person}{Shuaifeng Zhi}, \bibinfo{person}{Tristan
  Laidlow}, \bibinfo{person}{Stefan Leutenegger}, {and}
  \bibinfo{person}{Andrew~J. Davison}.} \bibinfo{year}{2021}\natexlab{}.
\newblock \showarticletitle{In-Place Scene Labelling and Understanding With
  Implicit Scene Representation}. In \bibinfo{booktitle}{\emph{Proceedings of
  the IEEE/CVF International Conference on Computer Vision (ICCV)}}.
  \bibinfo{pages}{15838--15847}.
\newblock


\bibitem[Zhou et~al\mbox{.}(2024)]%
        {Zhou_Feature3DGS_2024}
\bibfield{author}{\bibinfo{person}{Shijie Zhou}, \bibinfo{person}{Haoran
  Chang}, \bibinfo{person}{Sicheng Jiang}, \bibinfo{person}{Zhiwen Fan},
  \bibinfo{person}{Zehao Zhu}, \bibinfo{person}{Dejia Xu},
  \bibinfo{person}{Pradyumna Chari}, \bibinfo{person}{Suya You},
  \bibinfo{person}{Zhangyang Wang}, {and} \bibinfo{person}{Achuta Kadambi}.}
  \bibinfo{year}{2024}\natexlab{}.
\newblock \showarticletitle{Feature 3DGS: Supercharging 3D Gaussian Splatting
  to Enable Distilled Feature Fields}. In \bibinfo{booktitle}{\emph{Proceedings
  of the IEEE/CVF Conference on Computer Vision and Pattern Recognition
  (CVPR)}}. \bibinfo{pages}{21676--21685}.
\newblock


\bibitem[Ziwen et~al\mbox{.}(2024)]%
        {ziwen_longlrm_2024}
\bibfield{author}{\bibinfo{person}{Chen Ziwen}, \bibinfo{person}{Hao Tan},
  \bibinfo{person}{Kai Zhang}, \bibinfo{person}{Sai Bi}, \bibinfo{person}{Fujun
  Luan}, \bibinfo{person}{Yicong Hong}, \bibinfo{person}{Li Fuxin}, {and}
  \bibinfo{person}{Zexiang Xu}.} \bibinfo{year}{2024}\natexlab{}.
\newblock \showarticletitle{Long-LRM: Long-sequence Large Reconstruction Model
  for Wide-coverage Gaussian Splats}.
\newblock \bibinfo{journal}{\emph{arXiv preprint 2410.12781}}
  (\bibinfo{year}{2024}).
\newblock


\end{thebibliography}

\end{document}